\documentclass[journal,draftclsnofoot,onecolumn]{IEEEtran}
\usepackage{graphicx}
\usepackage{amssymb}
\usepackage{color}
\usepackage{caption}
\renewcommand{\theequation}{\arabic{section}.\arabic{equation}}
\newtheorem{theorem}{Theorem}

\newtheorem{definition}{Definition}

\newtheorem{remark}{Remark}

\begin{document}

\title{Relay Broadcast Channel with Confidential Messages}

\author{Bin~Dai,
        Linman~Yu,
        and~Zheng~Ma
\thanks{B. Dai is with the
School of Information Science and Technology,
Southwest JiaoTong University, Chengdu 610031, China, and with
the National Mobile Communications Research Laboratory, Southeast University, Nanjing 210096, China,
e-mail: daibin@home.swjtu.edu.cn.}
\thanks{L. Yu is with the School of Economics and Management, Chengdu Textile College, Chengdu 611731, China,
Email: yulinmanylm@163.com.}
\thanks{
Z. Ma is with the
School of Information Science and Technology,
Southwest JiaoTong University, Chengdu 610031, China,
e-mail: zma@home.swjtu.edu.cn.}
}

\maketitle

\begin{abstract}

In this paper, we investigate the effects of a trusted relay node on the secrecy of the broadcast channel by considering the
model of relay broadcast channel with confidential messages (RBC-CM).
Inner and outer bounds on the capacity-equivocation region of the RBC-CM are provided, and the capacity results are further explained via
a degraded Gaussian example, which we call the degraded Gaussian relay broadcast channel with one common and one confidential messages.
Numerical results show that this trusted relay node
helps to enhance the security of the Gaussian broadcast channel with one common and one confidential messages.

\end{abstract}

\begin{IEEEkeywords}
Capacity-equivocation region, confidential messages, relay broadcast channel, secrecy capacity region.
\end{IEEEkeywords}

\section{Introduction \label{secI}}

The secure communication over broadcast channel was first studied by Wyner \cite{Wy}, where a transmitter wished to send a confidential message
to a legitimate receiver through a broadcast channel, while he wished to keep a wiretapper as ignorant of the message
as possible. This model is called the wiretap channel.
Measuring the wiretapper's uncertainty about the confidential message by
equivocation, Wyner \cite{Wy} determined the capacity-equivocation region of the discrete memoryless wiretap channel.
Based on Wyner's work, Leung-Yan-Cheong and Hellman \cite{CH} determined the capacity-equivocation region of the
Gaussian wiretap channel.
Wyner's work was generalized by Csisz$\acute{a}$r and K\"{o}rner \cite{CK}, where common and confidential messages were
sent through the broadcast channel. The common message was assumed to be decoded correctly by both the
legitimate receiver and the wiretapper, while the confidential message was only allowed to be obtained by the
legitimate receiver. This model is called the broadcast channel with confidential messages (BCC).
The capacity-equivocation region of the discrete memoryless BCC was determined in \cite{CK}, and the capacity-equivocation region of the
Gaussian BCC was determined in \cite{LPS}.
By using the approach of \cite{Wy} and \cite{CK}, Liu et al. \cite{LMSY} studied the broadcast channel with two confidential messages (no
common message), and Xu et al. \cite{XCC}
studied the broadcast channel with two confidential messages and one common message. Both of them provided inner
and outer bounds on the capacity-equivocation regions.

Due to the open nature of the wireless media, the wireless communication is susceptible to eavesdropping.
Recently, the security of the wireless networks receives a lot attention.
For the multiple-access channel (MAC), Liang and Poor \cite{LP} studied the MAC with confidential messages, where
the degraded version of the MAC output is available at the transmitters. Each transmitter
treats the other one as a wiretapper, and wishes to
keep its confidential message as secret as possible from the wiretapper. Inner and outer bounds on
capacity-equivocation region is provided for this model, and the results are further explained via a Gaussian example.
Other related works on the MAC with secrecy constraint are in \cite{TY2, TY1, AZV1}.
For the interference channel, Liu et al. \cite{LMSY} studied the interference channel with two confidential messages,
and provided inner and outer bounds on the secrecy capacity region. Liang et al. studied the
cognitive interference channel with one common message and one confidential message \cite{LP2}, and Zaidi et al. investigated the
secure communication over the multi-input multi-output (MIMO) X-channel with asymmetric output feedback and delayed CSI at the transmitters \cite{AZV2}.
For the relay channel,
Lai and El Gamal \cite{LG} studied the effects of an
additional trusted relay node on the secrecy of the wiretap channel, where a source wished
to send messages to a destination while leveraging the help of a trusted relay node to hide those messages from
the wiretapper. Three inner bounds (with respect to decode-and-forward (DF), noise-and-forward (NF) and compress-and-forward (CF)
strategies) and one outer bound on the capacity-equivocation region were provided in \cite{LG}. Of particular interest is
the NF strategy, where the relay node sends codewords independent of the message to confuse the wiretapper.
Lai and El Gamal \cite{LG} showed that this NF strategy significantly improved the secrecy capacity of the wiretap channel.
Based on the work of \cite{LG}, Tang et. al. \cite{tang} improved Lai and El Gamal's NF strategy by considering an additional case
that both the legitimate receiver
and the wiretapper could not decode the relay codeword, and in this case, the relay codeword was served as
interference for both the legitimate receiver and the wiretapper.
Other related works in the relay channel with secrecy constraint include Oohama's relay channel with confidential messages \cite{Oo},
where an un-trusted relay helps the transmission of messages from
one sender to one receiver, and
Awan et al.'s secure communication over the parallel relay channel \cite{AZV3}.
Recently, Ekrem and Ulukus \cite{EU1} investigated the effects of user cooperation on the secrecy of broadcast
channels by considering a cooperative relay broadcast channel. They showed that user cooperation can increase
the achievable secrecy rate region of \cite{LMSY}.

\begin{figure}[htb]
\centering
\centerline{\includegraphics[scale=0.6]{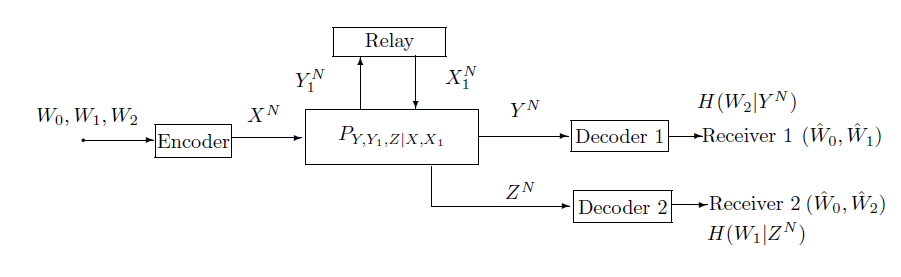}}
\caption{Relay broadcast channels with confidential messages}
\label{f1}
\end{figure}

\textcolor[rgb]{1.00,0.00,0.00}{In cellular and WiFi data networks, security is a critical issue when people wish to transmit
important/private information, such as credit card transactions or banking related data communications.
Another important issue in cellular and WiFi data networks is that the mobile users have been
demanding increasingly higher down-link data rate. Combining these two issues, how to achieve higher down-link secrecy data rate}
motivates us to study the down-link or broadcast
channel that exploits the techniques of relaying and user cooperation
to achieve higher secrecy rate.
In this paper, we study the model of
relay broadcast channel with confidential messages (RBC-CM), see Figure \ref{f1}.
We want to know whether the secrecy capacity region of the broadcast channel can be enhanced
by using a trusted relay node.

Inner and outer bounds on the
capacity-equivocation region of the model of Figure \ref{f1} are provided.
The inner bounds are constructed according to the decode-and-forward (DF) and noise-and-forward (NF) strategies.
Here note that in \cite{LG},
Lai and El Gamal has already shown that the compress-and-forward (CF) strategy is
a combination of the NF strategy and the classical Cover-El Gamal's CF strategy \cite{CG}, and it performs no better than the NF strategy.
Therefore, the CF inner bound on the
capacity-equivocation region of Figure \ref{f1} is not considered in this paper. The capacity results of the model of Figure \ref{f1} are further explained via
a degraded Gaussian example, which we call
the degraded Gaussian relay broadcast channel with one common and one confidential messages. The numerical results show that a trusted relay node
helps to enhance the security of the Gaussian BCC \cite{LPS}.

In this paper, random variab1es, sample values and
alphabets are denoted by capital letters, lower case letters and calligraphic letters, respectively.
A similar convention is applied to the random vectors and their sample values.
For example, $U^{N}$ denotes a random $N$-vector $(U_{1},...,U_{N})$,
and $u^{N}=(u_{1},...,u_{N})$ is a specific vector value in $\mathcal{U}^{N}$
that is the $N$th Cartesian power of $\mathcal{U}$.
$U_{i}^{N}$ denotes a random $N-i+1$-vector $(U_{i},...,U_{N})$,
and $u_{i}^{N}=(u_{i},...,u_{N})$ is a specific vector value in $\mathcal{U}_{i}^{N}$.
Let $p_{V}(v)$ denote the probability mass function $Pr\{V=v\}$. Throughout the paper,
the logarithmic function is to the base 2.

The organization of this paper is as follows. Section \ref{secII} provides the inner and outer bounds on the capacity-equivocation
region of the model of Figure \ref{f1}. The capacity results in Section \ref{secII} are further explained via a degraded Gaussian example,
which is shown in Section \ref{secIII}.
Final conclusions are
in Section \ref{secIV}.

\section{Problem formulation and the main results}\label{secII}

The model of Figure \ref{f1} is a four-terminal discrete channel consisting of finite sets $X$, $X_{1}$, $Y$, $Y_{1}$, $Z$ and a transition
probability distribution $p_{Y,Y_{1},Z|X_{1},X}(y,y_{1},z|x_{1},x)$. $X^{N}$ and $X^{N}_{1}$ are the channel inputs from the transmitter and
the relay respectively, while $Y^{N}$, $Y_{1}^{N}$, $Z^{N}$ are the channel outputs at receiver 1, relay and receiver 2, respectively.
The channel is discrete memoryless, i.e., the channel outputs $(y_{i},y_{1,i},z_{i})$ at time $i$ only depend on the channel
inputs $(x_{i}, x_{1,i})$ at time $i$.

\begin{definition}(\textbf{Channel encoder})\label{def1}
The confidential messages $W_{1}$ and $W_{2}$ take values in $\mathcal{W}_{1}$, $\mathcal{W}_{2}$, respectively.
The common message $W_{0}$ takes values in $\mathcal{W}_{0}$. $W_{1}$, $W_{2}$ and $W_{0}$ are independent and uniformly distributed over
their ranges. The channel encoder is a stochastic encoder $f_{E}$ that maps the messages $w_{1}$, $w_{2}$ and $w_{0}$ into a codeword
$x^{N}\in \mathcal{X}^{N}$. The transmission rates of the confidential messages ($W_{1}$, $W_{2}$) and the common message ($W_{0}$) are
$\frac{\log\|\mathcal{W}_{1}\|}{N}$, $\frac{\log\|\mathcal{W}_{2}\|}{N}$ and $\frac{\log\|\mathcal{W}_{0}\|}{N}$, respectively.
\end{definition}

\begin{definition}(\textbf{Relay encoder})\label{def2}
The relay encoder is also a stochastic encoder $\varphi_{i}$ that maps the signals $(y_{1,1},y_{1,2},...,y_{1,i-1})$
received before time $i$ to the channel input $x_{1,i}$.

\end{definition}

\begin{definition}(\textbf{Decoder})\label{def3}
The decoder for receiver 1 is a mapping $f_{D1}: \mathcal{Y}^{N}\rightarrow \mathcal{W}_{0}\times \mathcal{W}_{1}$,
with input $Y^{N}$ and outputs $\breve{W}_{0}$, $\breve{W}_{1}$. Let $P_{e1}$ be the error probability of receiver 1,
and it is defined as $Pr\{(W_{0},W_{1})\neq (\breve{W}_{0}, \breve{W}_{1})\}$.

The decoder for receiver 2 is a mapping $f_{D2}: \mathcal{Z}^{N}\rightarrow \mathcal{W}_{0}\times \mathcal{W}_{2}$,
with input $Z^{N}$ and outputs $\widehat{W}_{0}$, $\widehat{W}_{2}$. Let $P_{e2}$ be the error probability of receiver 2,
and it is defined as $Pr\{(W_{0},W_{2})\neq (\widehat{W}_{0}, \widehat{W}_{2})\}$.
\end{definition}

The equivocation rate at receiver 2 is defined as
\begin{equation}\label{e201}
\Delta_{1}=\frac{1}{N}H(W_{1}|Z^{N}).
\end{equation}
Analogously, the equivocation rate at receiver 1 is defined as
\begin{equation}\label{e202}
\Delta_{2}=\frac{1}{N}H(W_{2}|Y^{N}).
\end{equation}

A rate quintuple $(R_{0}, R_{1}, R_{2}, R_{e1},R_{e2})$ (where $R_{0}, R_{1}, R_{2}, R_{e1}, R_{e2}>0$) is called
achievable if, for any $\epsilon>0$ (where $\epsilon$ is an arbitrary small positive real number
and $\epsilon\rightarrow 0$), there exists a channel
encoder-decoder $(N, \Delta_{1}, \Delta_{2}, P_{e1}, P_{e2})$ such that
\begin{eqnarray}\label{e203}
&&\lim_{N\rightarrow \infty}\frac{\log\parallel \mathcal{W}_{0}\parallel}{N}= R_{0},
\lim_{N\rightarrow \infty}\frac{\log\parallel \mathcal{W}_{1}\parallel}{N}= R_{1},
\lim_{N\rightarrow \infty}\frac{\log\parallel \mathcal{W}_{2}\parallel}{N}= R_{2}, \nonumber\\
&&\lim_{N\rightarrow \infty}\Delta_{1}\geq R_{e1}, \lim_{N\rightarrow \infty}\Delta_{2}\geq R_{e2}, P_{e1}\leq \epsilon, P_{e2}\leq \epsilon.
\end{eqnarray}

The capacity-equivocation region $\mathcal{R}^{(A)}$ is a set composed of all achievable $(R_{0}, R_{1}, R_{2}, R_{e1},R_{e2})$ quintuples.
The inner and outer bounds on the capacity-equivocation region $\mathcal{R}^{(A)}$ are provided from Theorem \ref{T1} to
Theorem \ref{T3}, and they are proved in Appendix \ref{appen1}, Appendix \ref{appen2} and Appendix \ref{appen3}, respectively.
Our first result establishes an outer-bound on the capacity-equivocation region of the model of Figure \ref{f1}.

\begin{theorem}\label{T1}
\textbf{(Outer bound)} A single-letter characterization of the region $\mathcal{R}^{(Ao)}$ ($\mathcal{R}^{(A)}\subseteq \mathcal{R}^{(Ao)}$) is as follows,
\begin{eqnarray*}
&&\mathcal{R}^{(Ao)}=\{(R_{0}, R_{1}, R_{2}, R_{e1},R_{e2}): R_{e1}\leq R_{1},R_{e2}\leq R_{2},\\
&&R_{0}\leq \min\{I(U,U_{1};Y),I(U;Y,Y_{1}|U_{1})\},\\
&&R_{0}\leq \min\{I(U,U_{2};Z),I(U;Z,Y_{1}|U_{2})\},\\
&&R_{0}+R_{1}\leq \min\{I(U,U_{1},V_{1};Y),I(U,V_{1};Y,Y_{1}|U_{1})\},\\
&&R_{0}+R_{2}\leq \min\{I(U,U_{2},V_{2};Z),I(U,V_{2};Z,Y_{1}|U_{2})\},\\
&&R_{0}+R_{1}+R_{2}\leq I(U,U_{2},V_{1};Y,Y_{1}|U_{1})+I(V_{2};Z,Y_{1}|U,U_{1},U_{2},V_{1}),\\
&&R_{0}+R_{1}+R_{2}\leq I(U,U_{1},V_{2};Z,Y_{1}|U_{2})+I(V_{1};Y,Y_{1}|U,U_{1},U_{2},V_{2}),\\
&&R_{e1}\leq \min\{I(V_{1};Y|U,V_{2})-I(V_{1};Z|U,V_{2}), I(V_{1};Y|U)-I(V_{1};Z|U)\},\\
&&R_{e2}\leq \min\{I(V_{2};Z|U,V_{1})-I(V_{2};Y|U,V_{1}), I(V_{2};Z|U)-I(V_{2};Y|U)\}\},
\end{eqnarray*}
where
$U\rightarrow (U_{1},U_{2}, V_{1}, V_{2})\rightarrow (X, X_{1})\rightarrow (Y,Y_{1},Z).$
Moreover, note that the outer bound on the secrecy capacity region of Figure \ref{f1} is the set of triples $(R_{0}, R_{1}, R_{2})$
such that $(R_{0}, R_{1}, R_{2}, R_{e1}=R_{1},R_{e2}=R_{2})\in \mathcal{R}^{(Ao)}$.
\end{theorem}

\begin{IEEEproof}

The auxiliary random variables in $\mathcal{R}^{(Ao)}$ are defined by
\begin{eqnarray}\label{emk1}
&&U_{1}\triangleq Y_{1}^{J-1}, U_{2}\triangleq Y_{1,J+1}^{N}, U\triangleq (Y^{J-1}, W_{0}, Z_{J+1}^{N}, J)\nonumber\\
&&V_{1}\triangleq (U, W_{1}), V_{2}\triangleq (U, W_{2}), Y\triangleq Y_{J}, Y_{1}\triangleq Y_{1,J}, Z\triangleq Z_{J},
\end{eqnarray}
where $J$ is a random variable (uniformly distributed over $\{1, 2, ,...,N\}$), and it is independent of
$X^{N}$, $X_{1}^{N}$, $Y^{N}$, $Y_{1}^{N}$, $Z^{N}$, $W_{0}$, $W_{1}$ and $W_{2}$.
From the above definitions, it is easy to see that the relay $X_{1}$ is represented by two auxiliary random variables $U_{1}$ and $U_{2}$. The common message $W_{0}$ is represented
by $U$, and the confidential messages $W_{1}$, $W_{2}$ are represented by $V_{1}$ and $V_{2}$, respectively.
The proof of Theorem \ref{T1} combines Csisz$\acute{a}$r-K\"{o}rner's equality \cite{CK} for the equivocation analysis, Nair-El Gamal's technique \cite{NG}
for the bounds on the sum rate $R_{0}+R_{1}+R_{2}$, and Cover-El Gamal's technique \cite{CG} for introducing the relay input and output into the bounds
on $R_{0}$, $R_{0}+R_{1}$, $R_{0}+R_{2}$ and $R_{0}+R_{1}+R_{2}$. The details of the proof are in Appendix \ref{appen1}.

\end{IEEEproof}

\begin{remark}\label{R1}
There are some notes on Theorem \ref{T1}, see the following.
\begin{itemize}

\item If we allow the input $X_{1}^{N}$ and output $Y_{1}^{N}$ of the relay to be constants, from the above definitions
in (\ref{emk1}),
it is easy to see that the auxiliary random variables $U_{1}$, $U_{2}$ and $Y_{1}$ all are constants.
Substituting $U_{1}=U_{2}=Y_{1}=const$ into the region $\mathcal{R}^{(Ao)}$, we obtain an outer bound on the capacity-equivocation region of the broadcast channel
with two confidential messages and one common message, and it is in accordance with the outer bound in \cite{XCC}.

\item Define a triple
$(R_{0}, R_{1}, R_{2})$ is
achievable if, for any $\epsilon>0$, there exists a channel
encoder-decoder $(N, P_{e1}, P_{e2})$ such that
\begin{eqnarray}\label{e203.xxx}
&&\lim_{N\rightarrow \infty}\frac{\log\parallel \mathcal{W}_{0}\parallel}{N}= R_{0},
\lim_{N\rightarrow \infty}\frac{\log\parallel \mathcal{W}_{1}\parallel}{N}= R_{1},
\lim_{N\rightarrow \infty}\frac{\log\parallel \mathcal{W}_{2}\parallel}{N}= R_{2}, P_{e1}\leq \epsilon, P_{e2}\leq \epsilon.
\end{eqnarray}
The region $\mathcal{R}^{rbc}$, which is composed of all achievable $(R_{0}, R_{1}, R_{2})$ triples defined in (\ref{e203.xxx}),
is the capacity region of the general relay broadcast channel.
To the best of our knowledge, there is no outer bound on $\mathcal{R}^{rbc}$, and the inner bounds on $\mathcal{R}^{rbc}$ are studied in
\cite{KGG}. However, we find that the region $\mathcal{R}^{(Ao)}$ without the bounds on $R_{e1}$ and $R_{e2}$ can be served as
an outer bound on $\mathcal{R}^{rbc}$, and this is because without the consideration of the equivocation rates, our model
reduces to the general relay broadcast channel.

\end{itemize}
\end{remark}

We now turn our attention to constructing cooperation strategies for the model of Figure \ref{f1}. Our first step is
to characterize the inner bound on the capacity-equivocation region by using Cover-El Gamal's Decode and Forward (DF)
strategy \cite{CG}. In our DF strategy, the relay node will first decode the common message, and then re-encode the
common message to cooperate with the transmitter. The superposition coding and random binning techniques
used in \cite{XCC} will be combined with the DF cooperation strategy to characterize the inner bound.
Note that the DF inner bound of \cite{LG} is obtained by
allowing the relay to decode both the confidential and common messages. In general, for these two DF strategies,
we do not know which one is better. In Section \ref{secIII}, we show that for the degraded Gaussian relay-eavesdropper
\footnote{Here ``eavesdropper'' is another name for ``wiretapper''} channel,
in some particular cases, our DF strategy performs better than that of \cite{LG}.

\begin{theorem}\label{T2}
\textbf{(Inner bound 1: DF strategy)} A single-letter characterization of the region $\mathcal{R}^{(Ai1)}$
($\mathcal{R}^{(Ai1)}\subseteq \mathcal{R}^{(A)}$) is as follows,
\begin{eqnarray*}
&&\mathcal{R}^{(Ai1)}=\{(R_{0}, R_{1}, R_{2}, R_{e1},R_{e2}): R_{e1}\leq R_{1},R_{e2}\leq R_{2},\\
&&R_{0}\leq \min\{I(U;Y_{1}|X_{1}),I(U,X_{1};Y),I(U,X_{1};Z)\},\\
&&R_{0}+R_{1}\leq \min\{I(U;Y_{1}|X_{1}),I(U,X_{1};Y),I(U,X_{1};Z)\}+I(V_{1};Y|U,X_{1}),\\
&&R_{0}+R_{2}\leq \min\{I(U;Y_{1}|X_{1}),I(U,X_{1};Y),I(U,X_{1};Z)\}+I(V_{2};Z|U,X_{1}),\\
&&R_{0}+R_{1}+R_{2}\leq \min\{I(U;Y_{1}|X_{1}),I(U,X_{1};Y),I(U,X_{1};Z)\}+I(V_{1};Y|U,X_{1})+I(V_{2};Z|U,X_{1})-I(V_{1};V_{2}|U,X_{1}),\\
&&R_{e1}\leq I(V_{1};Y|U,X_{1})-I(V_{1};V_{2}|U,X_{1})-I(V_{1};Z|U,X_{1},V_{2}),\\
&&R_{e2}\leq I(V_{2};Z|U,X_{1})-I(V_{1};V_{2}|U,X_{1})-I(V_{2};Y|U,X_{1},V_{1})\},
\end{eqnarray*}
for some distribution
\begin{eqnarray*}
&&P_{Y,Z,Y_{1},X,X_{1},V_{1},V_{2},U}(y,z,y_{1},x,x_{1},v_{1},v_{2},u)=P_{Y,Z,Y_{1}|X,X_{1}}(y,z,y_{1}|x,x_{1})
P_{X,X_{1}|U,V_{1},V_{2}}(x,x_{1}|u,v_{1},v_{2})P_{U,V_{1},V_{2}}(u,v_{1},v_{2}).
\end{eqnarray*}
Moreover, note that the DF inner bound on the secrecy capacity region of Figure \ref{f1}
is the set of triples $(R_{0}, R_{1}, R_{2})$
such that $(R_{0}, R_{1}, R_{2}, R_{e1}=R_{1},R_{e2}=R_{2})\in \mathcal{R}^{(Ai1)}$.
\end{theorem}

\begin{IEEEproof}

The coding scheme of $\mathcal{R}^{Ai1}$ combines Cover-El Gamal's decode-and-forward (DF) strategy
and block Markov coding scheme for the relay channel \cite{CG}, Marton's double binning
and superposition coding techniques for the general broadcast channel \cite{Ma}, and Csisz$\acute{a}$r-K\"{o}rner's random binning technique
for the BCC \cite{CK}. The total transmission
is formed by $B$ blocks, in which $B-1$ messages will be sent.
The coding structure for block $b$ ($2\leq b\leq B$) is depicted in the following Figure \ref{fx1}.
In Figure \ref{fx1}, at block $b$, the new confidential messages $(w_{1,b},w_{2,b})$ are
first split into four sub-messages $(w_{10,b},w_{11,b},w_{20,b},w_{22,b})$, where $w_{10,b}$ and $w_{20,b}$
are common messages decoded by
both receivers, and $w_{11,b}$ and $w_{22,b}$ are confidential messages
for receiver 1 and receiver 2, respectively. Define $w_{0,b}^{*}=(w_{0,b},w_{10,b},w_{20,b})$. For block $b$, the relay sends
$x_{1}^{N}(w_{r,b})$, where $w_{r,b}$ is a deterministic function of $w_{0,b-1}^{*}$. The codeword
$u^{N}$ represents the superposition code in which the new common message $w_{0,b}^{*}$ is superimposed on the relay message $w_{r,b}$.
The codeword $v_{1}^{N}$ (or $v_{2}^{N}$) represents the superposition code
in which the private message $w_{11,b}$ (or $w_{22,b}$) is superimposed on $w_{0,b}^{*}$
and $w_{r,b}$. The input of the channel $x^{N}$ is i.i.d. generated according to $P_{X|U,V_{1},V_{2},X_{1}}(x|u,v_{1},v_{2},x_{1})$.
The details of the proof are in Appendix \ref{appen2}.

\begin{figure}[htb]
\centering
\centerline{\includegraphics[scale=0.7]{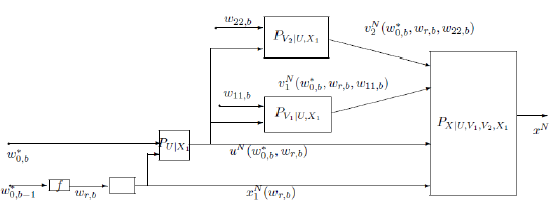}}
\caption{The encoder structure of the DF strategy in block $b$ ($2\leq b\leq B$)}
\label{fx1}
\end{figure}

\end{IEEEproof}

The second step is to characterize the inner bound on the capacity-equivocation region by using the generalized noise
and forward (GNF) strategy. 
\textcolor[rgb]{1.00,0.00,0.00}{In the GNF strategy, the relay node does not attempt to decode the messages
but sends codewords that are independent of the transmitters messages, and these codewords aid in confusing the less capable
receiver. Specifically, if the channel from the relay to receiver 1 is more capable than the channel from the relay to
receiver 2, only receiver 1 is allowed to decode the relay codeword, and vice versa.
Here note that the NF strategies of \cite{LG} and \cite{tang} only consider the case that
there is no confidential message for the receiver 2, and thus confusing receiver 1 is not needed, i.e.,
if the channel from the relay to receiver 2 is more capable than the channel from the relay to
receiver 1, we allow both or none of the receivers to decode the relay codeword. It is easy to see that
our GNF strategy is a generalization of the NF strategies \cite{LG,tang} for the single confidential message case.}

\begin{theorem}\label{T3}
\textbf{(Inner bound 2: GNF strategy)} A single-letter characterization of the region $\mathcal{R}^{(Ai2)}$
($\mathcal{R}^{(Ai2)}\subseteq \mathcal{R}^{(A)}$) is as follows,
\begin{eqnarray*}
&&\mathcal{R}^{(Ai2)}=\mbox{convex closure of}\quad(\mathcal{L}^{1}\bigcup \mathcal{L}^{2})
\end{eqnarray*}
where $\mathcal{L}^{1}$ is given by
\begin{eqnarray*}
&&\mathcal{L}^{1}=\bigcup_{\mbox{\tiny$\begin{array}{c}
P_{Y,Z,Y_{1},X,X_{1},V_{1},V_{2},U}\textbf{:}\\
I(X_{1};Y)>I(X_{1};Z|U,V_{2})\end{array}$}}
\left\{
\begin{array}{ll}
(R_{0}, R_{1}, R_{2}, R_{e1},R_{e2}): R_{e1}\leq R_{1},R_{e2}\leq R_{2},\\
R_{0}\leq \min\{I(U;Y|X_{1}),I(U;Z)\},\\
R_{0}+R_{1}\leq \min\{I(U;Y|X_{1}),I(U;Z)\}+I(V_{1};Y|U,X_{1}),\\
R_{0}+R_{2}\leq \min\{I(U;Y|X_{1}),I(U;Z)\}+I(V_{2};Z|U),\\
R_{0}+R_{1}+R_{2}\leq \min\{I(U;Y|X_{1}),I(U;Z)\}\\
+I(V_{1};Y|U,X_{1})+I(V_{2};Z|U)-I(V_{1};V_{2}|U),\\
R_{e1}\leq \min\{I(X_{1};Z|U,V_{1},V_{2}), I(X_{1};Y)\}+I(V_{1};Y|U,X_{1})\\
-I(V_{1};V_{2}|U)-I(X_{1},V_{1};Z|U,V_{2}),\\
R_{e2}\leq I(V_{2};Z|U)-I(V_{1};V_{2}|U)-I(V_{2};Y|U,X_{1},V_{1}).
\end{array}
\right\},
\end{eqnarray*}
$\mathcal{L}^{2}$ is given by
\begin{eqnarray*}
&&\mathcal{L}^{2}=\bigcup_{\mbox{\tiny$\begin{array}{c}
P_{Y,Z,Y_{1},X,X_{1},V_{1},V_{2},U}\textbf{:}\\
I(X_{1};Z)>I(X_{1};Y|U,V_{1})\end{array}$}}
\left\{
\begin{array}{ll}
(R_{0}, R_{1}, R_{2}, R_{e1},R_{e2}): R_{e1}\leq R_{1},R_{e2}\leq R_{2},\\
R_{0}\leq \min\{I(U;Z|X_{1}),I(U;Y)\},\\
R_{0}+R_{1}\leq \min\{I(U;Z|X_{1}),I(U;Y)\}+I(V_{1};Y|U),\\
R_{0}+R_{2}\leq \min\{I(U;Z|X_{1}),I(U;Y)\}+I(V_{2};Z|U,X_{1}),\\
R_{0}+R_{1}+R_{2}\leq \min\{I(U;Z|X_{1}),I(U;Y)\}\\
+I(V_{1};Y|U)+I(V_{2};Z|U,X_{1})-I(V_{1};V_{2}|U),\\
R_{e1}\leq I(V_{1};Y|U)-I(V_{1};V_{2}|U)-I(V_{1};Z|U,V_{2},X_{1}),\\
R_{e2}\leq \min\{I(X_{1};Y|U,V_{1},V_{2}), I(X_{1};Z)\}+I(V_{2};Z|U,X_{1})\\
-I(V_{1};V_{2}|U)-I(X_{1},V_{2};Y|U,V_{1}).
\end{array}
\right\},
\end{eqnarray*}
and $P_{Y,Z,Y_{1},X,X_{1},V_{1},V_{2},U}(y,z,y_{1},x,x_{1},v_{1},v_{2},u)$ satisfies
\begin{eqnarray*}
&&P_{Y,Z,Y_{1},X,X_{1},V_{1},V_{2},U}(y,z,y_{1},x,x_{1},v_{1},v_{2},u)=P_{Y,Z,Y_{1}|X,X_{1}}(y,z,y_{1}|x,x_{1})
P_{X|U,V_{1},V_{2}}(x|u,v_{1},v_{2})P_{U,V_{1},V_{2}}(u,v_{1},v_{2})P_{X_{1}}(x_{1}).
\end{eqnarray*}
Moreover, note that the GNF inner bound on the secrecy capacity region of Figure \ref{f1} is the set of triples $(R_{0}, R_{1}, R_{2})$
such that $(R_{0}, R_{1}, R_{2}, R_{e1}=R_{1},R_{e2}=R_{2})\in \mathcal{R}^{(Ai2)}$.
\end{theorem}

\begin{IEEEproof}

The region $\mathcal{L}^{1}$ is characterized under the condition that
the channel from the relay to receiver 1 is more capable than the channel from the relay to
receiver 2 (here note that $I(X_{1};Y)>I(X_{1};Z|U,V_{2})$ and $X_{1}$ is independent of $U$, $V_{2}$ imply that $I(X_{1};Y)>I(X_{1};Z)$).
Then, in this case, receiver 1 is allowed to decode the relay codeword, and receiver 2 is not allowed to
decode it. The rate of the relay is defined as $\min\{I(X_{1};Z|U,V_{1},V_{2}), I(X_{1};Y)\}$, and the
relay codeword is viewed as pure noise for receiver 2.
Analogously, the region $\mathcal{L}^{2}$ is characterized
under the condition that the channel from the relay to receiver 2 is more capable than the channel from the relay to
receiver 1. Then, in this case, receiver 2 is allowed to decode the relay codeword, and receiver 1 is not allowed to
decode it. In this case, the relay codeword is viewed as pure noise for receiver 1.
Combining the proof of \cite[Theorem 3]{LG} with the double binning
technique of the broadcast channel with one common and two confidential messages \cite{XCC},
the achievable regions $\mathcal{L}^{1}$ and $\mathcal{L}^{2}$ are obtained.
The details of the proof are in Appendix \ref{appen3}.

\end{IEEEproof}

\section{Degraded relay broadcast channel with one common and one confidential messages}\label{secIII}

In this section, the capacity results on the discrete memoryless degraded relay broadcast channel with one common and one confidential messages
are provided in Subsection \ref{sub41}, the capacity results on the degraded Gaussian case are
shown in Subsection \ref{sub42}, the numerical results are in Subsection \ref{sub43}, and the comparison of our proposed relay strategies
and previous known strategies is shown in Subsection \ref{sub44}.

\subsection{Discrete memoryless degraded relay broadcast channel with one common and one confidential messages}\label{sub41}

The degraded relay broadcast channel with one common and one confidential messages is a special case of the model of Figure \ref{f1}.
and it implies the existence
of a Markov chain $X\rightarrow (X_{1},Y_{1})\rightarrow Y\rightarrow Z$.
Since the received symbols $Z^{N}$ of receiver 2 are degraded versions of those of receiver 1, there is no confidential message $W_{2}$. The
channel encoder is a stochastic encoder that maps the messages $w_{0}$ and $w_{1}$ into a codeword $x^{N}\in \mathcal{X}^{N}$. Moreover,
the decoder for receiver 1 is a mapping $f_{D1}: \mathcal{Y}^{N}\rightarrow \mathcal{W}_{0}\times\mathcal{W}_{1}$, with input $Y^{N}$ and outputs
$\check{W}_{0}$ and $\check{W}_{1}$. Let $P_{e1}$ be the error probability of receiver 1, and it is defined as
$Pr\{(\check{W}_{0},\check{W}_{1})\neq (W_{0},W_{1})\}$.
Analogously, the decoder for receiver 2 is a mapping $f_{D2}: \mathcal{Z}^{N}\rightarrow \mathcal{W}_{0}$, with input $Z^{N}$ and output
$\hat{W}_{0}$. Let $P_{e2}$ be the error probability of receiver 2, and it is defined as $Pr\{\hat{W}_{0}\neq W_{0}\}$.

A rate triple $(R_{0}, R_{1}, R_{e})$ (where $R_{0}, R_{1}, R_{e}>0$) is called
achievable if, for any $\epsilon>0$ (where $\epsilon$ is an arbitrary small positive real number
and $\epsilon\rightarrow 0$), there exists a channel
encoder-decoder $(N, \Delta, P_{e1}, P_{e2})$ such that
\begin{eqnarray}\label{e401}
&&\lim_{N\rightarrow \infty}\frac{\log\parallel \mathcal{W}_{0}\parallel}{N}= R_{0},
\lim_{N\rightarrow \infty}\frac{\log\parallel \mathcal{W}_{1}\parallel}{N}= R_{1}, \nonumber\\
&&\lim_{N\rightarrow \infty}\Delta\geq R_{e}, P_{e1}\leq \epsilon, P_{e2}\leq \epsilon.
\end{eqnarray}

The capacity-equivocation region $\mathcal{R}^{(C)}$ is a set composed of all achievable $(R_{0}, R_{1}, R_{e})$ triples.
The inner and outer bounds on the capacity-equivocation region $\mathcal{R}^{(C)}$ are provided from Theorem \ref{T9} to
Theorem \ref{T11}, see the remainder of this subsection.
The first result is an outer bound on the capacity-equivocation region $\mathcal{R}^{(C)}$.

\begin{theorem}\label{T9}
\textbf{(Outer bound)} A single-letter characterization of the region $\mathcal{R}^{(Co)}$ ($\mathcal{R}^{(C)}\subseteq \mathcal{R}^{(Co)}$) is as follows,
\begin{eqnarray*}
&&\mathcal{R}^{(Co)}=\{(R_{0}, R_{1}, R_{e}): R_{e}\leq R_{1},\\
&&R_{0}\leq \min\{I(U,Q;Z),I(U;Y_{1}|Q)\},\\
&&R_{0}+R_{1}\leq \min\{I(Q,U,V;Y),I(U,V;Y_{1}|Q)\},\\
&&R_{e}\leq \min\{I(V;Y_{1}|U,Q)-I(V;Z|U,Q),I(V;Y|U)-I(V;Z|U)\}\},
\end{eqnarray*}
where
$(Q,U,V,X)\rightarrow (X_{1}, Y_{1})\rightarrow Y\rightarrow Z.$
Here note that the outer bound on the secrecy capacity region of $\mathcal{R}^{(C)}$
is the set of pairs $(R_{0}, R_{1})$
such that $(R_{0}, R_{1}, R_{e}=R_{1})\in \mathcal{R}^{(Co)}$
\end{theorem}

\begin{IEEEproof}

See Appendix \ref{appen11.1}.

\end{IEEEproof}

\begin{remark}\label{Ra}
There are some notes on Theorem \ref{T9}, see the following.

\begin{itemize}

\item In fact, we can directly obtain an outer bound $\mathcal{R}^{(Co*)}$ on $\mathcal{R}^{(C)}$ by substituting $V_{2}=U$
(here $V_{2}=U$
is from the definition that $V_{2}=(U, W_{2})$ and $W_{2}=const$) and $R_{2}=R_{e2}=0$ into Theorem \ref{T1}, and $\mathcal{R}^{(Co*)}$ is given by
\begin{eqnarray*}
&&\mathcal{R}^{(Co*)}=\{(R_{0}, R_{1}, R_{e}): R_{e}\leq R_{1},\\
&&R_{0}\leq \min\{I(U,U_{1};Y),I(U;Y,Y_{1}|U_{1})\},\\
&&R_{0}\leq \min\{I(U,U_{2};Z),I(U;Z,Y_{1}|U_{2})\},\\
&&R_{0}+R_{1}\leq \min\{I(U,U_{1},V;Y),I(U,V;Y,Y_{1}|U_{1})\},\\
&&R_{0}+R_{1}\leq I(U,U_{2},V;Y,Y_{1}|U_{1}),\\
&&R_{0}+R_{1}\leq I(U,U_{1};Z,Y_{1}|U_{2})+I(V;Y,Y_{1}|U,U_{1},U_{2}),\\
&&R_{e}\leq I(V;Y|U)-I(V;Z|U)\},
\end{eqnarray*}
where $V$ and $R_{e}$ represent $V_{1}$ and $R_{e1}$ of Theorem \ref{T1}, respectively.
However, we
find that a tighter outer bound may be obtained by using the degradedness assumption $X\rightarrow (X_{1},Y_{1})\rightarrow Y\rightarrow Z$.
Specifically, in $\mathcal{R}^{(Co*)}$, with the help of the degradedness assumption, we can remove the auxiliary random variable $U_{2}$,
simplify the bounds on $R_{0}$ and $R_{0}+R_{1}$,
and obtain a new upper bound $R_{e}\leq I(V;Y_{1}|U,Q)-I(V;Z|U,Q)$ (here $Q=U_{1}$) on the equivocation $R_{e}$.
Since the equivocation $R_{e}$ of $\mathcal{R}^{(Co)}$ satisfies
$R_{e}\leq \min\{I(V;Y_{1}|U,Q)-I(V;Z|U,Q),I(V;Y|U)-I(V;Z|U)\}\leq I(V;Y|U)-I(V;Z|U)$, it is easy to see that the equivocation bound of
$\mathcal{R}^{(Co)}$
is tighter than that of $\mathcal{R}^{(Co*)}$.

\item Letting $R_{0}=0$ and $R_{e}=R_{1}$, and observing that $I(V;Y|U)-I(V;Z|U)\leq I(Q,U,V;Y)$ and
$I(V;Y_{1}|U,Q)-I(V;Z|U,Q)\leq I(U,V;Y_{1}|Q)$, an outer bound $C_{s}^{Co}$ on the secrecy capacity $C_{s}^{C}$ of the
discrete memoryless degraded relay broadcast channel with one common and one confidential messages
is given by
\begin{eqnarray*}
&&C_{s}^{Co}=\max_{P_{QUVXX_{1}Y_{1}}}\min\{I(V;Y_{1}|U,Q)-I(V;Z|U,Q),I(V;Y|U)-I(V;Z|U)\}.
\end{eqnarray*}
Note that the degraded relay broadcast channel with one common and one confidential messages reduces to the
degraded relay-eavesdropper channel \cite{LG} when there is no common message ($R_{0}=0$).
We also notice that in \cite[Theorem 5]{LG}, the secrecy capacity of a kind of physically degraded relay-eavesdropper channel
has been determined. However, we find that the degradedness assumption of \cite[Theorem 5]{LG} is
$(X, X_{1})\rightarrow Y_{1}\rightarrow Y$, which is different from that of this paper. The secrecy capacity of
the degraded relay-eavesdropper channel (with the degradedness assumption $X\rightarrow (X_{1},Y_{1})\rightarrow Y$) is still unknown.

\end{itemize}
\end{remark}

Now we turn to the inner bounds on the capacity-equivocation region $\mathcal{R}^{(C)}$. The following
 Theorem \ref{T10} and Theorem \ref{T11} provide the decode-and-forward (DF) and noise-and-forward (NF) inner bounds on $\mathcal{R}^{(C)}$, respectively.

\begin{theorem}\label{T10}
\textbf{(Inner bound 1: DF strategy)} A single-letter characterization of the region $\mathcal{R}^{(Ci1)}$
($\mathcal{R}^{(Ci1)}\subseteq \mathcal{R}^{(C)}$) is as follows,
\begin{eqnarray*}
&&\mathcal{R}^{(Ci1)}=\{(R_{0}, R_{1}, R_{e}): R_{e}\leq R_{1},\\
&&R_{0}\leq \min\{I(U;Y_{1}|X_{1}),I(U,X_{1};Z)\},\\
&&R_{0}+R_{1}\leq \min\{I(U;Y_{1}|X_{1}),I(U,X_{1};Z)\}+I(V;Y|U,X_{1}),\\
&&R_{e}\leq I(V;Y|U,X_{1})-I(V;Z|U,X_{1}),
\end{eqnarray*}
for some distribution
\begin{eqnarray*}
&&P_{Y,Z,Y_{1},X,X_{1},V,U}(y,z,y_{1},x,x_{1},v,u)=P_{Z|Y}(z|y)P_{Y|X_{1},Y_{1}}(y|x_{1},y_{1})
P_{Y_{1}|X,X_{1}}(y_{1}|x,x_{1})P_{X,X_{1},U,V}(x,x_{1},u,v).
\end{eqnarray*}
Note that the DF inner bound on the secrecy capacity region
is denoted as $\mathcal{C}_{s}^{Ci1}$, which is the set of pairs $(R_{0}, R_{1})$
such that $(R_{0}, R_{1}, R_{e}=R_{1})\in \mathcal{R}^{(Ci1)}$.
\end{theorem}

\begin{IEEEproof}
By letting $V_{2}=const$, $V_{1}=V$ and using the Markov chain $(U, V)\rightarrow (X_{1},Y_{1})\rightarrow Y\rightarrow Z$,
the DF inner bound on $\mathcal{R}^{(C)}$ is directly obtained from Theorem \ref{T2}. Thus, the proof is omitted here.
\end{IEEEproof}

\begin{remark}\label{Rb}

Letting $R_{0}=0$ and $R_{e}=R_{1}$, the DF inner bound $C_{s}^{Ci1}$ on the secrecy capacity $C_{s}^{C}$
is given by
\begin{eqnarray}\label{xman1}
&&C_{s}^{Ci1}=\max_{P_{UVXX_{1}}}(I(V;Y|U,X_{1})-I(V;Z|U,X_{1})).
\end{eqnarray}
Here note that $C_{s}^{Ci1}$ is obtained by allowing the relay to decode the common message which is represented by $U$,
and the relay is not allowed to decode the confidential message represented by $V$. We notice that Lai and El Gamal \cite[Theorem 2]{LG}
also presents an DF inner bound $C_{s}^{Ci*}$ on the secrecy capacity of the relay-eavesdropper channel, which can also be viewed as
an DF inner bound on $C_{s}^{C}$, and it is given by
\begin{eqnarray}\label{xman2}
&&C_{s}^{Ci1*}=\max_{P_{XX_{1}|V_{1}V_{2}}P_{UV_{1}V_{2}}}(\min \{I(V_{1},V_{2};Y|U),I(V_{1};Y_{1}|V_{2},U)\}-I(V_{1},V_{2};Z|U)).
\end{eqnarray}
The bound $C_{s}^{Ci1*}$ is obtained by allowing the relay to decode both the confidential message represented by $V_{1}$ and the common message
represented by $U$.
In general, for these two bounds ($C_{s}^{Ci1}$ and $C_{s}^{Ci1*}$), we do not know which one is larger.
In Subsection \ref{sub44}, we present a Gaussian example and show that in some particular cases, $C_{s}^{Ci1}$ is larger than $C_{s}^{Ci1*}$.

\end{remark}

\begin{theorem}\label{T11}
\textbf{(Inner bound 2: NF strategy)} A single-letter characterization of the region $\mathcal{R}^{(Ci2)}$
($\mathcal{R}^{(Ci2)}\subseteq \mathcal{R}^{(C)}$) is as follows,
\begin{eqnarray*}
&&\mathcal{R}^{(Ci2)}=\mbox{convex closure of}\quad(\mathcal{L}^{3}\bigcup \mathcal{L}^{4})
\end{eqnarray*}
where $\mathcal{L}^{3}$ is given by
\begin{eqnarray*}
&&\mathcal{L}^{3}=\bigcup_{\mbox{\tiny$\begin{array}{c}
P_{Y,Z,Y_{1},X,X_{1},V,U}\textbf{:}\\
I(X_{1};Y)\geq I(X_{1};Z|U)\end{array}$}}
\left\{
\begin{array}{ll}
(R_{0}, R_{1}, R_{e}): R_{e}\leq R_{1}, R_{0}\leq I(U;Z),\\
R_{0}+R_{1}\leq I(U;Z)+I(V;Y|U,X_{1}),\\
R_{e}\leq \min\{I(X_{1};Z|U,V), I(X_{1};Y)\}+I(V;Y|U,X_{1})-I(X_{1},V;Z|U)
\end{array}
\right\},
\end{eqnarray*}
$\mathcal{L}^{4}$ is given by
\begin{eqnarray*}
&&\mathcal{L}^{4}=\bigcup_{\mbox{\tiny$\begin{array}{c}
P_{Y,Z,Y_{1},X,X_{1},V,U}\textbf{:}\\
I(X_{1};Y)<I(X_{1};Z|U)\end{array}$}}
\left\{
\begin{array}{ll}
(R_{0}, R_{1}, R_{e}): R_{e}\leq R_{1}, R_{0}\leq I(U;Z|X_{1}),\\
R_{0}+R_{1}\leq I(U;Z|X_{1})+I(V;Y|U,X_{1}),\\
R_{e}\leq I(V;Y|U,X_{1})-I(V;Z|U,X_{1})
\end{array}
\right\},
\end{eqnarray*}
and $P_{Y,Z,Y_{1},X,X_{1},V,U}(y,z,y_{1},x,x_{1},v,u)$ satisfies
\begin{eqnarray*}
&&P_{Y,Z,Y_{1},X,X_{1},V,U}(y,z,y_{1},x,x_{1},v,u)=P_{Z|Y}(z|y)P_{Y|X_{1},Y_{1}}(y|x_{1},y_{1})
P_{Y_{1}|X,X_{1}}(y_{1}|x,x_{1})P_{X,U,V}(x,u,v)P_{X_{1}}(x_{1}).
\end{eqnarray*}
Note that the NF inner bound on the secrecy capacity region
is denoted as $\mathcal{C}_{s}^{Ci2}$, and it is the set of pairs $(R_{0}, R_{1})$
such that $(R_{0}, R_{1}, R_{e}=R_{1})\in \mathcal{R}^{(Ci2)}$.
\end{theorem}

\begin{IEEEproof}

\begin{itemize}

\item The region $\mathcal{L}^{3}$ implies that the rate of the relay is $R_{r}=\min\{I(X_{1};Z|U,V), I(X_{1};Y)\}\geq I(X_{1};Z)$,
which means that receiver 1 is able to
decode the relay codeword $x_{1}^{N}$, and receiver 2 can not decode it. Therefore, in this case, $x_{1}^{N}$ can be
viewed as a noise signal to confuse receiver 2.
Letting $V_{2}=const$ and $V_{1}=V$, and using the Markov chain $X\rightarrow (X_{1},Y_{1})\rightarrow Y\rightarrow Z$,
the region $\mathcal{L}^{3}$ is directly obtained from the region $\mathcal{L}^{1}$ of Theorem \ref{T2}. Thus, the detail of the proof is omitted here.

\item For the region $\mathcal{L}^{4}$, the rate of the relay is $R_{r}=I(X_{1};Z)\leq I(X_{1};Y)$, which implies
that both the receivers can decode the relay codeword $x_{1}^{N}$, and in this case, $x_{1}^{N}$ serves as a common message decoded by
both receivers. Thus,
the achievability proof of the region $\mathcal{L}^{4}$ is along the lines of the proof of the broadcast channel with confidential messages \cite{CK},
and we omit the proof here.

\end{itemize}

\end{IEEEproof}

\begin{remark}\label{Rc}
There are some notes on Theorem \ref{T11}, see the following.

\begin{itemize}

\item The auxiliary random variable $U$ of Theorem \ref{T11} represents the common message $W_{0}$. Letting $U=const$ and $R_{e}=R_{1}$,
Theorem \ref{T11} can be served as an achievable secrecy rate $C_{s}^{Ci2}$ for the general relay-eavesdropper channel \cite{LG}
(without the degradedness assumption), and it is given as follows.
\begin{itemize}

\item If $I(X_{1};Y)\geq I(X_{1};Z)$, the achievable secrecy rate $C_{s}^{Ci2}$ satisfies
\begin{eqnarray}\label{umg1.1}
&&C_{s}^{Ci2}=\max_{P_{X|V}P_{V}P_{X_{1}}}[I(V;Y|X_{1})+\min\{I(X_{1};Y),I(X_{1};Z|V)\}-I(V,X_{1};Z)].
\end{eqnarray}

\item If $I(X_{1};Y)<I(X_{1};Z)$,
the achievable secrecy rate $C_{s}^{Ci2}$ satisfies
\begin{eqnarray}\label{umg1.2}
&&C_{s}^{Ci2}=\max_{P_{X|V}P_{V}P_{X_{1}}}[I(V;Y|X_{1})-I(V;Z|X_{1})].
\end{eqnarray}

\end{itemize}
Here note that (\ref{umg1.1}) and (\ref{umg1.2}) can be combined as
\begin{eqnarray}\label{umg1}
C_{s}^{Ci2}&=&\max_{P_{X|V}P_{V}P_{X_{1}}}[I(V;Y|X_{1})+\min\{I(X_{1};Y),I(X_{1};Z|V)\}\nonumber\\
&-&\min\{I(X_{1};Y),I(X_{1};Z)\}-I(V;Z|X_{1})].
\end{eqnarray}

We also notice that Lai and El Gamal \cite[Theorem 2]{LG}
has already provided a NF achievable secrecy rate $C_{s}^{Ci2*}$ for the relay-eavesdropper channel, and it is also given by
(\ref{umg1}). Thus, we can conclude that our NF achievable secrecy rate $C_{s}^{Ci2}$ is equivalent to
Lai-El Gamal's NF achievable secrecy rate $C_{s}^{Ci2*}$.

\item In \cite[Theorem 1]{tang}, Tang et al. provide a new NF achievable secrecy rate $C_{s}^{Ci2**}$ for the relay-eavesdropper
channel, and it can be equivalently characterized by
the following cases:
\begin{itemize}

\item Case 1: if $I(X_{1};Y|X)\leq I(X_{1};Z)$,
\begin{eqnarray}\label{umg2.1}
&&C_{s}^{Ci2**}=\max\{I(X;Y|X_{1})-I(X;Z|X_{1}), I(X;Y)-I(X;Z)\}.
\end{eqnarray}

\item Case 2: if $I(X_{1};Y)\leq I(X_{1};Z)\leq I(X_{1};Y|X)\leq I(X_{1};Z|X)$,
\begin{eqnarray}\label{umg2.2}
&&C_{s}^{Ci2**}=\max\{I(X;Y|X_{1})-I(X;Z|X_{1}), I(X;Y)-I(X;Z)\}.
\end{eqnarray}

\item Case 3: if $I(X_{1};Y)\leq I(X_{1};Z)\leq I(X_{1};Z|X)\leq I(X_{1};Y|X)$,
\begin{eqnarray}\label{umg2.3}
&&C_{s}^{Ci2**}=\max\{I(X;Y|X_{1})-I(X;Z|X_{1}), I(X;Y)-I(X;Z)\}.
\end{eqnarray}

\item Case 4: if $I(X_{1};Z)\leq I(X_{1};Y)\leq I(X_{1};Y|X)\leq I(X_{1};Z|X)$,
\begin{eqnarray}\label{umg2.4}
&&C_{s}^{Ci2**}=\max[I(X;Y)-I(X;Z)].
\end{eqnarray}

\item Case 5: if $I(X_{1};Z)\leq I(X_{1};Y)\leq I(X_{1};Z|X)\leq I(X_{1};Y|X)$,
\begin{eqnarray}\label{umg2.5}
&&C_{s}^{Ci2**}=\max[I(X,X_{1};Y)-I(X,X_{1};Z)].
\end{eqnarray}

\item Case 6: if $I(X_{1};Z|X)\leq I(X_{1};Y)$,
\begin{eqnarray}\label{umg2.6}
&&C_{s}^{Ci2**}=\max[I(X;Y|X_{1})-I(X;Z)].
\end{eqnarray}
\end{itemize}
Comparing $C_{s}^{Ci2}$, $C_{s}^{Ci2*}$ (replacing $V$ by $X$) with $C_{s}^{Ci2**}$, it is easy to see that the
NF achievable secrecy rates of this paper and \cite[Theorem 2]{LG} are included in that of \cite[Theorem 1]{tang}, and
the achievable secrecy rate $\max[I(X;Y)-I(X;Z)]$, which is not considered in \cite{LG} and this paper, is studied in \cite[Theorem 1]{tang}.
Specifically, in \cite[Theorem 1]{tang}, Tang et al.
show that if we allow the relay to generate the ``artificial noise'' with the rate
$R_{r}>\min\{I(X_{1};Y), I(X_{1};Z)\}$, both the legitimate receiver and the eavesdropper can not decode the relay codeword.
The relay codeword is served as interference for both the legitimate receiver and the eavesdropper, and thus Wyner's secrecy rate \cite{Wy}
($\max[I(X;Y)-I(X;Z)]$) is also achievable for this case. However, in Subsection \ref{sub44}, we show that the
achievable secrecy rates $C_{s}^{Ci2}$, $C_{s}^{Ci2*}$ and $C_{s}^{Ci2**}$
are the same for the Gaussian case.

\item In \cite{tang}, the NF achievable secrecy rate $C_{s}^{Ci2**}$ for three special cases (weak interference/eavesdropping, strong interference/eavesdropping
and very strong eavesdropping) is studied. Comparing $C_{s}^{Ci2}$ ($C_{s}^{Ci2*}$)
with $C_{s}^{Ci2**}$ for these special cases, we have the following comments.
\begin{itemize}

\item For the weak interference/eavesdropping (which implies that
$I(X;Y|X_{1})\geq I(X;Z|X_{1})$ and $I(X_{1};Z|X)\geq I(X_{1};Y|X)$), $C_{s}^{Ci2**}$ is given by
\begin{equation}\label{umg3}
C_{s}^{Ci2**}=\max\max
\left\{
\begin{array}{ll}
I(X;Y|X_{1})-I(X;Z|X_{1})\\
I(X;Y)-I(X;Z)
\end{array}
\right\}.
\end{equation}
As stated above, we have shown that $C_{s}^{Ci2}=C_{s}^{Ci2*}$. For the weak interference/eavesdropping, we have
\begin{eqnarray}\label{umg3.1}
&&C_{s}^{Ci2}=C_{s}^{Ci2*}=\max[I(X;Y|X_{1})-I(X;Z|X_{1})].
\end{eqnarray}
Comparing (\ref{umg3}) with (\ref{umg3.1}), it is easy to see that $C_{s}^{Ci2}=C_{s}^{Ci2*}\leq C_{s}^{Ci2**}$.

\item For the strong interference/eavesdropping (which implies that
$I(X;Y|X_{1})\leq I(X;Z|X_{1})$ and $I(X_{1};Z|X)\leq I(X_{1};Y|X)$), $C_{s}^{Ci2**}$ is given by
\begin{equation}\label{umg3.2}
C_{s}^{Ci2**}=\max\left[\min
\left\{
\begin{array}{ll}
I(X,X_{1};Y)-I(X,X_{1};Z)\\
I(X;Y|X_{1})-I(X;Z)
\end{array}
\right\}\right]^{+}.
\end{equation}
From (\ref{umg1.1}), (\ref{umg1.2}) and the definition of the strong interference/eavesdropping, we also have
\begin{equation}\label{umg3.3}
C_{s}^{Ci2}=C_{s}^{Ci2*}=\max\left[\min
\left\{
\begin{array}{ll}
I(X,X_{1};Y)-I(X,X_{1};Z)\\
I(X;Y|X_{1})-I(X;Z)
\end{array}
\right\}\right]^{+}.
\end{equation}
It is easy to see that $C_{s}^{Ci2}=C_{s}^{Ci2*}=C_{s}^{Ci2**}$ for the strong interference/eavesdropping.

\item For the very strong interference/eavesdropping (which implies that
$I(X;Z)\geq I(X;Y|X_{1})$), it is easy to see that $C_{s}^{Ci2}=C_{s}^{Ci2*}=C_{s}^{Ci2**}=0$.

\end{itemize}

\end{itemize}
\end{remark}

\subsection{Degraded Gaussian relay broadcast channel with one common and one confidential messages}\label{sub42}

In this subsection, we investigate the bounds on the secrecy capacity region of the degraded Gaussian relay broadcast channel with
one common and one confidential messages. The signal received at each node
is given by
\begin{eqnarray}\label{u1}
&&Y_{1}=X+Z_{r}, \,\,Y=X+X_{1}+Z_{r}+Z_{1}, \,\,Z=X+X_{1}+Z_{r}+Z_{2},
\end{eqnarray}
where $Z_{r}\sim \mathcal{N}(0, N_{r})$, $Z_{1}\sim \mathcal{N}(0, N_{1})$, $Z_{2}\sim \mathcal{N}(0, N_{2})$, $N_{2}>N_{1}$,
and they are independent. The average power
constraints of $X^{N}$ and $X_{1}^{N}$ are  $\frac{1}{N}\sum_{i=1}^{N}E[X_{i}^{2}]\leq P_{1}$ and
$\frac{1}{N}\sum_{i=1}^{N}E[X_{1,i}^{2}]\leq P_{2}$, respectively. The secrecy capacity region of this degraded Gaussian model is
denoted by $\mathcal{C}_{s}^{Cg}$.

First, the DF inner bound on $\mathcal{C}_{s}^{Cg}$ is given by
\begin{equation}\label{e402}
\mathcal{C}_{s}^{Cgi1}=\bigcup_{0\leq \alpha\leq 1}
\left\{
\begin{array}{ll}
(R_{0}, R_{1}): \\
R_{0}\leq \min\{\frac{1}{2}\log\frac{P_{1}+N_{r}}{\alpha P_{1}+N_{r}}, \frac{1}{2}\log\frac{P_{1}+P_{2}+N_{2}+N_{r}}{\alpha P_{1}+N_{2}+N_{r}}\}\\
R_{1}\leq \frac{1}{2}\log\frac{\alpha P_{1}+N_{r}+N_{1}}{N_{1}+N_{r}}-\frac{1}{2}\log\frac{\alpha P_{1}+N_{r}+N_{2}}{N_{2}+N_{r}}
\end{array}
\right\}.
\end{equation}

\begin{IEEEproof}
The region $\mathcal{C}_{s}^{Cgi1}$ is obtained by substituting $X=U+V$, $U=c_{1}X_{1}+X_{10}$, (\ref{u1}) and $R_{e}=R_{1}$ into Theorem \ref{T10},
where $U\sim \mathcal{N}(0, (1-\alpha)P_{1})$,
$V\sim \mathcal{N}(0, \alpha P_{1})$,
$X_{10}\sim \mathcal{N}(0, (1-\alpha)\beta P_{1})$ ($0\leq \beta\leq 1$), and $c_{1}=\sqrt{\frac{P_{1}(1-\alpha)(1-\beta)}{P_{2}}}$. Here note that
$X_{10}$ is independent of $X_{1}$, and
$V$ is independent of $U$.
\end{IEEEproof}

Second, the NF inner bound on $\mathcal{C}_{s}^{Cg}$ is considered into the following two cases:

\begin{itemize}

\item Case 1: If $N_{2}\geq N_{1}+P_{1}$, the NF inner bound $\mathcal{C}_{s}^{Cgi2}$ is given by
\begin{equation}\label{e402.u1}
\mathcal{C}_{s}^{Cgi2}=\bigcup_{0\leq \alpha\leq 1}
\left\{
\begin{array}{ll}
(R_{0}, R_{1}): \\
R_{0}\leq \frac{1}{2}\log\frac{P_{1}+P_{2}+N_{r}+N_{2}}{(1-\alpha)P_{1}+P_{2}+N_{r}+N_{2}}\\
R_{0}+R_{1}\leq \frac{1}{2}\log\frac{P_{1}+P_{2}+N_{r}+N_{2}}{(1-\alpha)P_{1}+P_{2}+N_{r}+N_{2}}+
\frac{1}{2}\log\frac{(1-\alpha)P_{1}+N_{r}+N_{1}}{N_{r}+N_{1}}\\
R_{1}\leq \min\{\frac{1}{2}\log\frac{N_{r}+N_{2}+P_{2}}{N_{r}+N_{2}},\frac{1}{2}\log\frac{P_{1}+P_{2}+N_{r}+N_{1}}{P_{1}+N_{r}+N_{1}}\}\\
+\frac{1}{2}\log\frac{(1-\alpha)P_{1}+N_{r}+N_{1}}{N_{r}+N_{1}}-\frac{1}{2}\log\frac{P_{2}+N_{r}+N_{2}}{N_{r}+N_{2}}
\end{array}
\right\}.
\end{equation}
\begin{IEEEproof}
The region (\ref{e402.u1}) is obtained by substituting $X=U+V$, (\ref{u1}) and $R_{e}=R_{1}$ into $\mathcal{L}^{3}$, where $V\sim \mathcal{N}(0, (1-\alpha)P_{1})$,
$U\sim \mathcal{N}(0, \alpha P_{1})$, and
$X_{1}$, $U$ and $V$ are independent random variables.
Here note that $P_{1}+N_{1}\leq N_{2}$ implies that $I(X_{1};Y)\geq I(X_{1};Z|U)$ for all $0\leq \alpha\leq 1$.
The proof is completed.
\end{IEEEproof}

\item Case 2: If $N_{2}<N_{1}+P_{1}$, the NF inner bound $\mathcal{C}_{s}^{Cgi2}$ is given by
\begin{eqnarray}\label{e402.u2.1}
&&\mathcal{C}_{s}^{Cgi2}=A\bigcup B,
\end{eqnarray}
where $A$ is given by
\begin{equation}\label{e402.u2.2}
A=\bigcup_{0\leq \alpha\leq \frac{N_{2}-N_{1}}{P_{1}}}
\left\{
\begin{array}{ll}
(R_{0}, R_{1}): \\
R_{0}\leq \frac{1}{2}\log\frac{P_{1}+P_{2}+N_{r}+N_{2}}{(1-\alpha)P_{1}+P_{2}+N_{r}+N_{2}}\\
R_{0}+R_{1}\leq \frac{1}{2}\log\frac{P_{1}+P_{2}+N_{r}+N_{2}}{(1-\alpha)P_{1}+P_{2}+N_{r}+N_{2}}+
\frac{1}{2}\log\frac{(1-\alpha)P_{1}+N_{r}+N_{1}}{N_{r}+N_{1}}\\
R_{1}\leq \min\{\frac{1}{2}\log\frac{N_{r}+N_{2}+P_{2}}{N_{r}+N_{2}},\frac{1}{2}\log\frac{P_{1}+P_{2}+N_{r}+N_{1}}{P_{1}+N_{r}+N_{1}}\}\\
+\frac{1}{2}\log\frac{(1-\alpha)P_{1}+N_{r}+N_{1}}{N_{r}+N_{1}}-\frac{1}{2}\log\frac{P_{2}+N_{r}+N_{2}}{N_{r}+N_{2}}
\end{array}
\right\},
\end{equation}
and $B$ is given by
\begin{equation}\label{e402.u2.3}
B=\bigcup_{\frac{N_{2}-N_{1}}{P_{1}}<\alpha\leq 1}
\left\{
\begin{array}{ll}
(R_{0}, R_{1}): \\
R_{0}\leq \frac{1}{2}\log\frac{P_{1}+N_{r}+N_{2}}{(1-\alpha)P_{1}+N_{r}+N_{2}}\\
R_{0}+R_{1}\leq \frac{1}{2}\log\frac{P_{1}+N_{r}+N_{2}}{(1-\alpha)P_{1}+N_{r}+N_{2}}+
\frac{1}{2}\log\frac{(1-\alpha)P_{1}+N_{r}+N_{1}}{N_{r}+N_{1}}\\
R_{1}\leq \frac{1}{2}\log\frac{N_{r}+N_{1}+(1-\alpha)P_{1}}{N_{r}+N_{1}}-\frac{1}{2}\log\frac{N_{r}+N_{2}+(1-\alpha)P_{1}}{N_{r}+N_{2}}
\end{array}
\right\}.
\end{equation}

\begin{IEEEproof}
Note that $N_{2}<N_{1}+P_{1}$ implies that $I(X_{1};Y)\geq I(X_{1};Z|U)$ holds if $0\leq \alpha\leq \frac{N_{2}-N_{1}}{P_{1}}$,
and $I(X_{1};Y)<I(X_{1};Z|U)$ holds if $\frac{N_{2}-N_{1}}{P_{1}}<\alpha\leq 1$. The region $A$ is the same as that of case 1,
and the region $B$ is obtained by substituting $X=U+V$, (\ref{u1}) and $R_{e}=R_{1}$ into $\mathcal{L}^{4}$. Thus, the proof is completed.

\end{IEEEproof}

\end{itemize}

Third, the outer bound ($\mathcal{C}_{s}^{Cgo}$) on the secrecy capacity region $\mathcal{C}_{s}^{Cg}$ is given by
\begin{equation}\label{e402.u2.4}
\mathcal{C}_{s}^{Cgo}=\bigcup_{0\leq \delta\leq 1}
\left\{
\begin{array}{ll}
(R_{0}, R_{1}): \\
R_{0}\leq \min\{\frac{1}{2}\log\frac{P_{1}+P_{2}+N_{r}+N_{2}}{\delta P_{1}+N_{r}+N_{2}}, \frac{1}{2}\log\frac{P_{1}+N_{r}}{\delta P_{1}+N_{r}}\}\\
R_{0}+R_{1}\leq \min\{\frac{1}{2}\log\frac{P_{1}+P_{2}+N_{r}+N_{1}}{N_{r}+N_{1}},\frac{1}{2}\log\frac{P_{1}+N_{r}}{N_{r}}\}\\
R_{1}\leq \min\{\frac{1}{2}\log\frac{P_{1}+P_{2}+N_{r}+N_{1}}{N_{r}+N_{1}}-\frac{1}{2}\log\frac{P_{1}+P_{2}+N_{r}+N_{2}}{\delta P_{1}+N_{r}+N_{2}},
\frac{1}{2}\log\frac{\delta P_{1}+N_{r}}{N_{r}}\}
\end{array}
\right\}.
\end{equation}
\begin{IEEEproof}
First, note that $h(X+X_{1}+Z_{r}+Z_{2}|U,V,Q)\geq h(X+X_{1}+Z_{r}+Z_{2}|U,V,Q,X_{1},X)=h(Z_{r}+Z_{2})=\frac{1}{2}\log2\pi e(N_{r}+N_{2})$,
and $h(X+X_{1}+Z_{r}+Z_{2}|U,V,Q)\leq h(X+X_{1}+Z_{r}+Z_{2})\leq \frac{1}{2}\log2\pi e(N_{r}+N_{2}+P_{1}+P_{2})$, thus we can conclude that
\begin{equation}\label{xx1}
h(X+X_{1}+Z_{r}+Z_{2}|U,V,Q)=\frac{1}{2}\log2\pi e(N_{r}+N_{2}+\alpha(P_{1}+P_{2})),
\end{equation}
where $0\leq\alpha\leq 1$.
Analogously, we have
\begin{equation}\label{xx2}
h(X+X_{1}+Z_{r}+Z_{2}|U,V)=\frac{1}{2}\log2\pi e(N_{r}+N_{2}+(\alpha+\beta-\alpha\beta)(P_{1}+P_{2})),
\end{equation}
\begin{equation}\label{xx3}
h(X+X_{1}+Z_{r}+Z_{2}|U)=\frac{1}{2}\log2\pi e(N_{r}+N_{2}+(\alpha+\beta-\alpha\beta+\gamma-\alpha\gamma-\beta\gamma+\alpha\beta\gamma)(P_{1}+P_{2})),
\end{equation}
\begin{equation}\label{xx4}
h(X+Z_{r}|Q,U)=\frac{1}{2}\log2\pi e(N_{r}+\delta P_{1}),
\end{equation}
where $0\leq \beta,\gamma,\delta\leq 1$.
Substituting (\ref{u1}) and $R_{e}=R_{1}$ into Theorem \ref{T9}, using the above (\ref{xx1}), (\ref{xx2}), (\ref{xx3}), (\ref{xx4}) and the entropy power inequality,
and maximizing the parameters $\alpha$, $\beta$ and $\gamma$, we have the outer bound (\ref{e402.u2.4}).
Here note that (\ref{e402.u2.4}) is achieved if $\beta=0$, $\gamma=1$ and $\alpha=\frac{P_{1}\delta}{P_{1}+P_{2}}$.
Thus, the proof is completed.
\end{IEEEproof}

Finally, remember that \cite{LPS} provides the secrecy capacity region of the Gaussian broadcast channel with one confidential
message and one common message (GBCC), and it is given by
\begin{equation}
\mathcal{C}_{s}^{GBCC}=\bigcup_{0\leq \alpha\leq 1}
\left\{
\begin{array}{ll}
(R_{0}, R_{1}): \\
R_{0}\leq \frac{1}{2}\log\frac{P_{1}+N_{r}+N_{2}}{\alpha P_{1}+N_{r}+N_{2}},\\
R_{1}\leq \frac{1}{2}\log\frac{\alpha P_{1}+N_{r}+N_{1}}{N_{1}+N_{r}}-\frac{1}{2}\log\frac{\alpha P_{1}+N_{r}+N_{2}}{N_{2}+N_{r}}.
\end{array}
\right\}.
\end{equation}

\subsection{Numerical Results on the Gaussian Example}\label{sub43}

The following Figure \ref{f5} and Figure \ref{f5.r1} show the inner and outer bounds on $\mathcal{C}_{s}^{Cg}$ and the
secrecy capacity region of the Gaussian BCC for several values of $P_{1}$, $P_{2}$, $N_{r}$, $N_{1}$ and $N_{2}$.
Specifically, in Figure \ref{f5}, we choose $P_{1}=5$, $P_{2}=20$, $N_{1}=2$, $N_{2}=8$ and $N_{r}=2$,
which implies that $N_{2}\geq N_{1}+P_{1}$. For this case, the NF inner bound on $\mathcal{C}_{s}^{Cg}$ reduces to the region of case 1.
Compared with the secrecy capacity region of the Gaussian BCC,
it is easy to see that the maximum secrecy rate $R_{1}$ of $\mathcal{C}_{s}^{GBCC}$ is enhanced by using the NF
strategy. For the DF strategy, though it can not increase the maximum $R_{1}$ of $\mathcal{C}_{s}^{GBCC}$, the
maximum common rate $R_{0}$ and the entire secrecy capacity region $\mathcal{C}_{s}^{GBCC}$ are enhanced.

In Figure \ref{f5.r1}, we choose $P_{1}=10$, $P_{2}=20$, $N_{1}=2$, $N_{2}=8$ and $N_{r}=2$,
which implies that $N_{2}<N_{1}+P_{1}$. For this case, the NF inner bound on $\mathcal{C}_{s}^{Cg}$ reduces to the region of case 2.
Compared with the secrecy capacity region of the Gaussian BCC,
it is easy to see that the maximum secrecy rate $R_{1}$ of $\mathcal{C}_{s}^{GBCC}$ is enhanced by using the NF
strategy. However, when $R_{0}$ is larger than $0.26$,
the NF strategy makes no contribution to enhance the security of the Gaussian BCC.
For the DF strategy, it enhances the
maximum common rate $R_{0}$ and the entire secrecy capacity region $\mathcal{C}_{s}^{GBCC}$.
Moreover, from Figure \ref{f5} and Figure \ref{f5.r1}, we can see that there is a huge gap between the inner and outer bounds on $\mathcal{C}_{s}^{Cg}$.
Eliminating the gap (improving both the inner and outer bounds) is our future work.

\begin{figure}[htb]
\centering
\centerline{\includegraphics[scale=0.55]{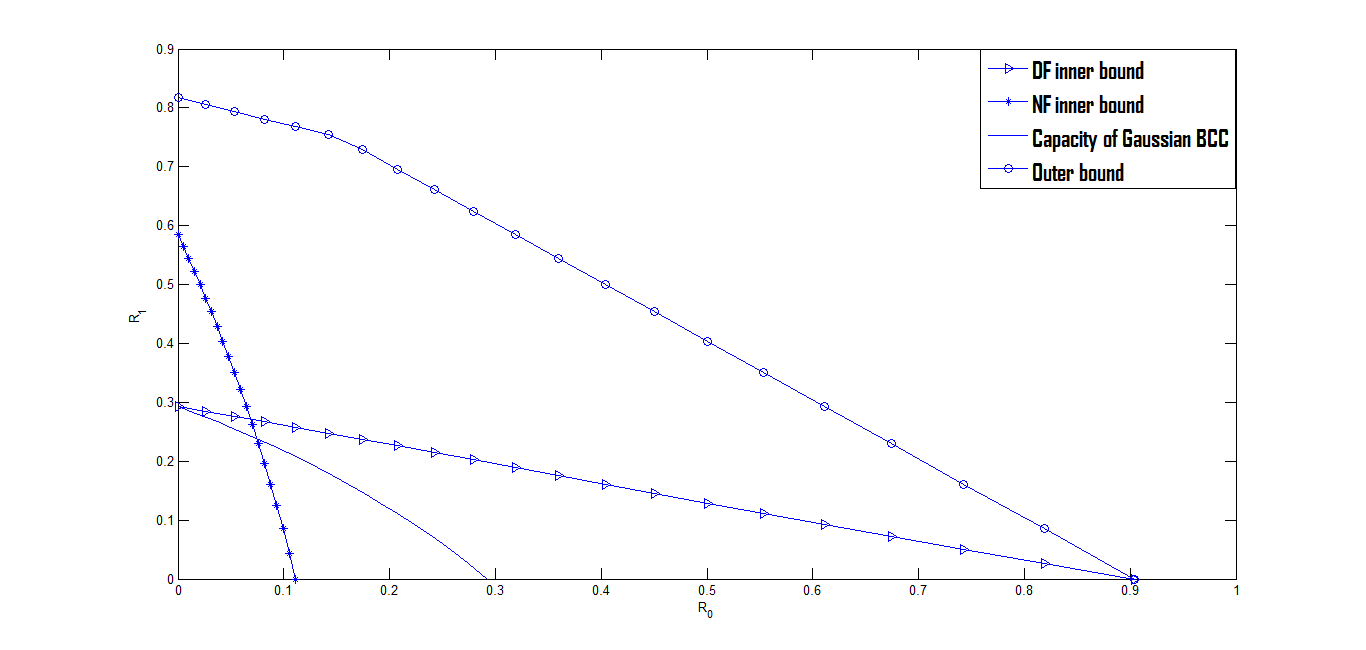}}
\caption{The inner and outer bounds on $\mathcal{C}_{s}^{Cg}$ and the secrecy
capacity region of the Gaussian BCC for $P_{1}=5$, $P_{2}=20$, $N_{1}=2$, $N_{2}=8$ and $N_{r}=2$}
\label{f5}
\end{figure}

\begin{figure}[htb]
\centering
\centerline{\includegraphics[scale=0.55]{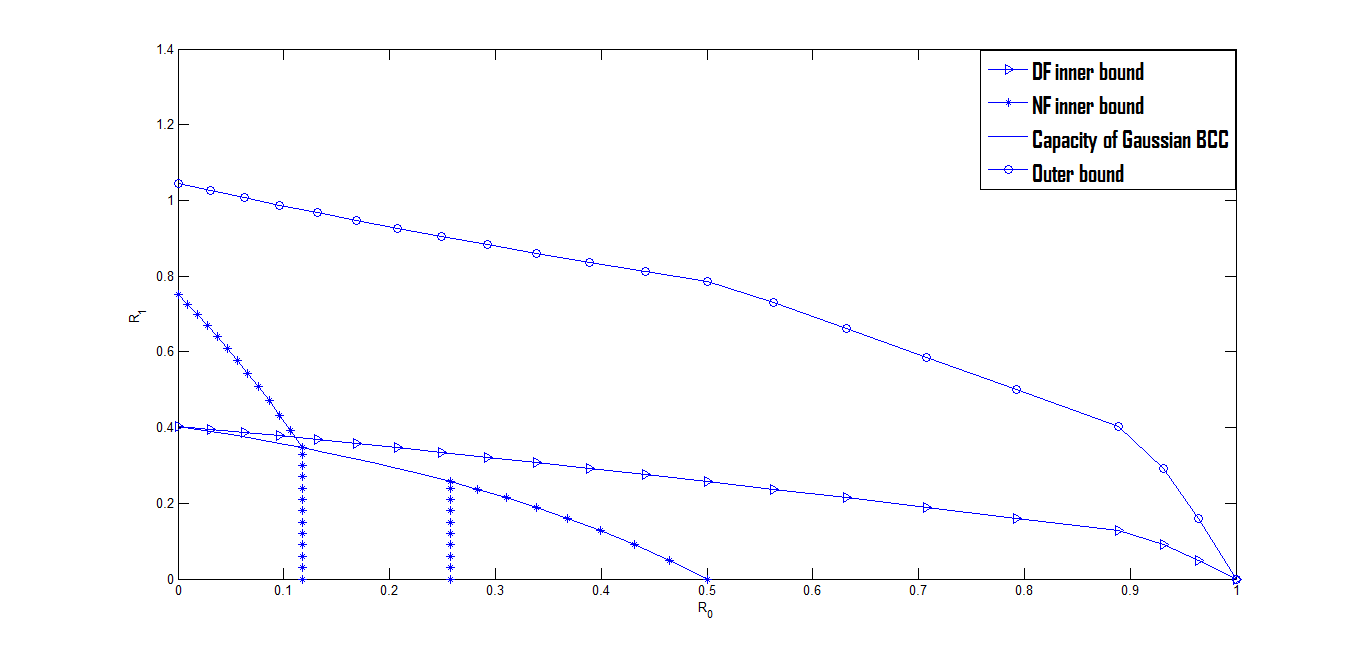}}
\caption{The inner and outer bounds on $\mathcal{C}_{s}^{Cg}$ and the secrecy
capacity region of the Gaussian BCC for $P_{1}=10$, $P_{2}=20$, $N_{1}=2$, $N_{2}=8$ and $N_{r}=2$}
\label{f5.r1}
\end{figure}

\subsection{The Comparison of the DF and NF Achievable Secrecy Rates with the Previous Known Results}\label{sub44}

In this subsection, we study the achievable secrecy rates of the degraded Gaussian relay-eavesdropper channel.
First, we show the DF and NF achievable secrecy rates of the degraded Gaussian relay-eavesdropper channel.
Then, we compare our DF secrecy rate with that of \cite[bottom of page 4009]{LG}, and show that in some particular cases, our result is better than
Lai and El Gamal's DF secrecy rate. Finally, we compare our NF secrecy rate with the Gaussian case of \cite[Theorem 1]{tang}, and show that
our NF secrecy rate is in accordance with the Gaussian case of \cite[Theorem 1]{tang}.

\subsubsection{Comparison of the DF Achievable Secrecy Rates of the Degraded Gaussian Relay-Eavesdropper Channel}\label{sub44.1}

In the previous subsection, we have already shown that the DF inner bound on the secrecy capacity region of the degraded Gaussian relay broadcast
channel with one common and one confidential messages is given by (\ref{e402}). Letting $\alpha=1$ (which implies that the auxiliary random variable $U=const$
and $R_{0}=0$), (\ref{e402}) can be served as our DF achievable secrecy rate $C_{s}^{Cgi1}$ for the degraded Gaussian relay-eavesdropper channel, and it is given by
\begin{equation}\label{xman3.1}
C_{s}^{Cgi1}=\frac{1}{2}\log\frac{P_{1}+N_{r}+N_{1}}{N_{r}+N_{1}}-\frac{1}{2}\log\frac{P_{1}+N_{r}+N_{2}}{N_{r}+N_{2}}.
\end{equation}

In the Remark \ref{Rb}, we have already shown that for the discrete memoryless case,
Lai-El Gamal's DF achievable secrecy rate is given by (\ref{xman2}). Letting $V_{1}=X$, $V_{2}=X_{1}$, $U=const$,
$X=cX_{1}+X_{10}$, $X_{1}\sim \mathcal{N}(0, P_{2})$, $X_{10}\sim \mathcal{N}(0, \alpha P_{1})$,
$c=\sqrt{\frac{P_{1}(1-\alpha)}{P_{2}}}$, straightforward calculations of (\ref{xman2}) result in the degraded Gaussian DF achievable secrecy rate $C_{s}^{Cgi1*}$,
and it is given by
\begin{equation}\label{xman3.2}
C_{s}^{Cgi1*}=\min\{\frac{1}{2}\log\frac{P_{1}+P_{2}+N_{r}+N_{1}}{N_{r}+N_{1}}, \frac{1}{2}\log\frac{P_{1}+N_{r}}{N_{r}}\}
-\frac{1}{2}\log\frac{P_{1}+P_{2}+N_{r}+N_{2}}{N_{r}+N_{2}}.
\end{equation}
Comparing (\ref{xman3.1}) with (\ref{xman3.2}), we can conclude that
\begin{itemize}

\item If $N_{1}\geq N_{2}$, $C_{s}^{Cgi1}=C_{s}^{Cgi1*}=0$.

\item If $N_{1}\leq N_{2}$ and $\frac{P_{1}}{P_{2}}\leq \frac{N_{r}}{N_{1}}\cdot \frac{P_{1}+N_{1}+N_{r}}{P_{1}+N_{2}+N_{r}}$,
$C_{s}^{Cgi1}$ is larger than $C_{s}^{Cgi1*}$.

\item If $N_{1}\leq N_{2}$ and $\frac{P_{1}}{P_{2}}\geq \frac{N_{r}}{N_{1}}\cdot \frac{P_{1}+N_{1}+N_{r}}{P_{1}+N_{2}+N_{r}}$,
$C_{s}^{Cgi1*}$ is larger than $C_{s}^{Cgi1}$.

\end{itemize}

\textcolor[rgb]{1.00,0.00,0.00}{The following Figure \ref{fxman1} shows $C_{s}^{Cgi1}$ and $C_{s}^{Cgi1*}$ for fixed $N_{1}$ and $N_{2}$.
As we can see, Lai-El Gamal's DF secrecy rate $C_{s}^{Cgi1*}$ dominates our DF secrecy rate  $C_{s}^{Cgi1}$ when $P_{2}$ and $N_{r}$ are small
($P_{2}=1$ and $N_{r}=0.5$). However, the gap between $C_{s}^{Cgi1}$ and $C_{s}^{Cgi1*}$ is decreasing while $P_{2}$ and $N_{r}$ are increasing, and
our DF secrecy rate  $C_{s}^{Cgi1}$ 
dominates Lai-El Gamal's DF secrecy rate $C_{s}^{Cgi1*}$ when $P_{2}$ and $N_{r}$ are large enough ($P_{2}=30$ and $N_{r}=4$).}

\begin{figure}[htb]
\centering
\centerline{\includegraphics[scale=0.55]{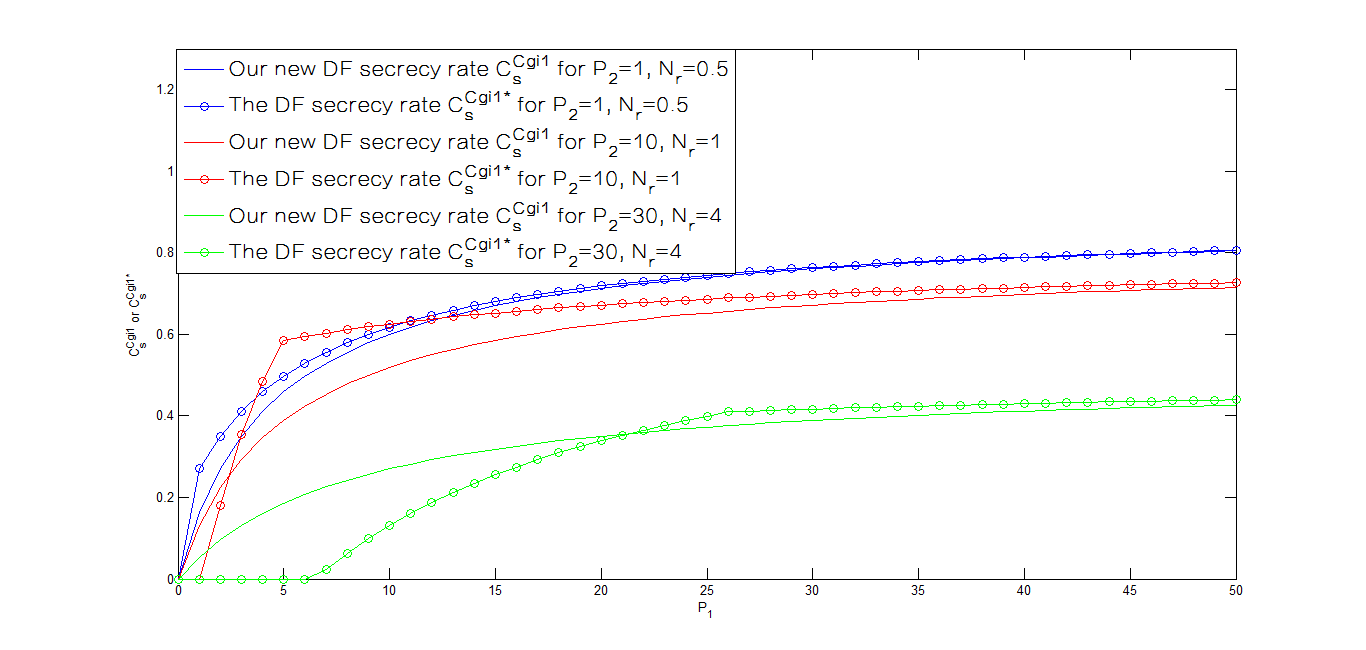}}
\caption{The curves $P_{1}-C_{s}^{Cgi1}$ and $P_{1}-C_{s}^{Cgi1*}$ for
$N_{1}=2$, $N_{2}=8$ and several values of $P_{2}$ and $N_{r}$}
\label{fxman1}
\end{figure}

\subsubsection{Comparison of the NF Achievable Secrecy Rates of the Degraded Gaussian Relay-Eavesdropper Channel}\label{sub44.2}

In the Remark \ref{Rc}, we have already shown that for the discrete memoryless case, our and Lai-El Gamal's NF achievable secrecy rates can be characterized by (\ref{umg1}).
Letting $V=X$,
$X\sim \mathcal{N}(0, P_{1})$ and $X_{1}\sim \mathcal{N}(0, P_{2})$, straightforward calculations of (\ref{umg1}) result in the degraded Gaussian
NF achievable secrecy rates $C_{s}^{Cgi2}$ and $C_{s}^{Cgi2*}$, and they are given by
\begin{equation}\label{xman4.1}
C_{s}^{Cgi2}=C_{s}^{Cgi2*}=\left\{
\begin{array}{ll}
\frac{1}{2}\log\frac{P_{1}+P_{2}+N_{1}}{N_{1}}-\frac{1}{2}\log\frac{P_{1}+P_{2}+N_{2}}{N_{2}}, & \mbox{if}\; N_{1}\leq N_{2}\leq N_{1}+P_{1},\\
\frac{1}{2}\log\frac{P_{1}+N_{1}}{N_{1}}-\frac{1}{2}\log\frac{P_{1}+P_{2}+N_{2}}{P_{2}+N_{2}}, & \mbox{if}\; N_{2}>N_{1}+P_{1},\\
0, & \mbox{if}\; N_{2}<N_{1}.
\end{array}
\right.
\end{equation}
Moreover, we have also shown that for the discrete memoryless case, Tang et al.'s NF achievable secrecy rate $C_{s}^{Ci2**}$ is characterized by
(\ref{umg2.1})-(\ref{umg2.6}). Letting
$X\sim \mathcal{N}(0, P_{1})$ and $X_{1}\sim \mathcal{N}(0, P_{2})$, straightforward calculations of (\ref{umg2.1})-(\ref{umg2.6}) also result in
(\ref{xman4.1}), i.e., the degraded Gaussian NF achievable secrecy rate $C_{s}^{Cgi2**}$ of \cite{tang} satisfies $C_{s}^{Cgi2**}=C_{s}^{Cgi2}=C_{s}^{Cgi2*}$.

The following Figure \ref{fxman2} shows $C_{s}^{Cgi2}$, $C_{s}^{Cgi2*}$ and $C_{s}^{Cgi2**}$ for $P_{1}=5$, $P_{2}=8$, $N_{1}=2$ and $N_{r}=4$.
It is easy to see that the secrecy rates are increasing while the noise variance $N_{2}$ of the eavesdropper's channel is increasing, and when
$N_{2}\leq N_{1}$, no positive secrecy rate can be achieved.

\begin{figure}[htb]
\centering
\centerline{\includegraphics[scale=0.55]{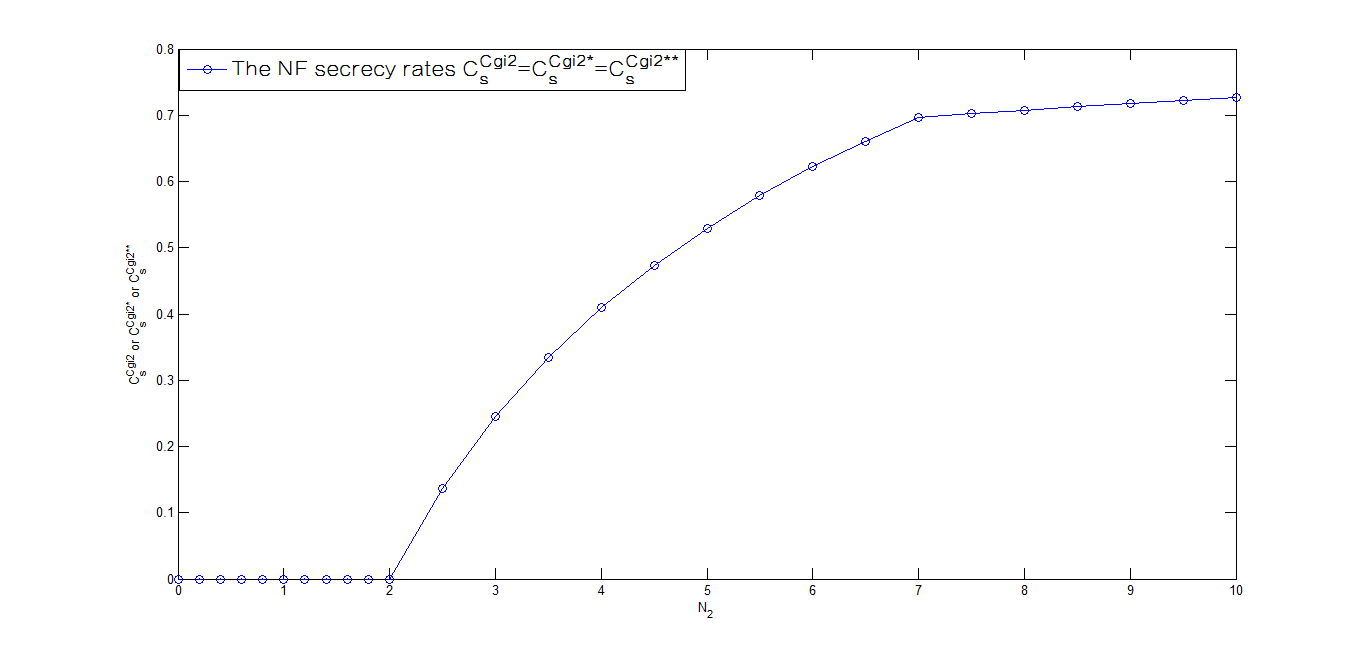}}
\caption{The curves $N_{2}-C_{s}^{Cgi2}$, $N_{2}-C_{s}^{Cgi2*}$ and $N_{2}-C_{s}^{Cgi2**}$ for $P_{1}=5$, $P_{2}=8$, $N_{1}=2$ and $N_{r}=4$}
\label{fxman2}
\end{figure}

\section{Conclusion}\label{secIV}

In this paper, we provide inner and outer bounds on the capacity-equivocation of RBC-CM. The capacity results
are further explained via the degraded Gaussian relay broadcast channel with one common and one confidential messages.
Numerical results show that
a trusted relay node
helps to enhance the security of the Gaussian BCC. Moreover, for the degraded Gaussian relay-eavesdropper channel,
we find that in some particular cases, our DF strategy is better than that of \cite{LG}. As for the NF strategies,
we find that the NF strategies of \cite{LG}, \cite{tang}
and this paper perform the same.

\section*{Acknowledgement}

The authors would like to thank Professor Ning Cai and the anonymous reviewers
for their valuable suggestions on improving this paper.
This work was supported by a
sub-project in National Basic Research Program of China under Grant 2012CB316100 on Broadband Mobile Communications at High Speeds,
the National Natural Science Foundation of China under Grants 61301121 and 61571373, the Fundamental Research Funds for the Central Universities
under Grant 2682014CX099, the National High-tech R\&D Program of
China (863 Program) under Grant 2014AA01A707, 
the Key Grant Project of Chinese Ministry of Education under Grant 311031100, the Young Innovative Research Team of Sichuan Province under Grant 2011JTD0007
and the Open Research Fund of National Mobile Communications Research Laboratory, Southeast University (No. 2014D01).

\renewcommand{\theequation}{A\arabic{equation}}
\appendices\section{Proof of Theorem \ref{T1}\label{appen1}}
\setcounter{equation}{0}

In this section, we will prove Theorem \ref{T1}: all the achievable $(R_{0},R_{1},R_{2},R_{e1},R_{e2})$ quintuples are contained in
the set $\mathcal{R}^{Ao}$. The inequalities of Theorem \ref{T1} are proved in the remainder of this section.

First, define the following auxiliary random variables,
\begin{eqnarray}\label{appen1.1}
&&U_{1}\triangleq Y_{1}^{J-1}, U_{2}\triangleq Y_{1,J+1}^{N}, U\triangleq (Y^{J-1}, W_{0}, Z_{J+1}^{N}, J)\nonumber\\
&&V_{1}\triangleq (U, W_{1}), V_{2}\triangleq (U, W_{2})\nonumber\\
&&Y\triangleq Y_{J}, Y_{1}\triangleq Y_{1,J}, Z\triangleq Z_{J},
\end{eqnarray}
where $J$ is a random variable (uniformly distributed over $\{1, 2, ,...,N\}$), and it is independent of
$Y^{N}$, $Y_{1}^{N}$, $Z^{N}$, $W_{0}$, $W_{1}$ and $W_{2}$.

\textbf{(Proof of $R_{0}\leq \min\{I(U,U_{1};Y),I(U;Y,Y_{1}|U_{1})\}$)}

The inequality $R_{0}\leq I(U,U_{1};Y)$ is proved as follows.
\begin{eqnarray}\label{appen1.2}
\frac{1}{N}H(W_{0})&\leq&\frac{1}{N}(I(W_{0};Y^{N})+H(W_{0}|Y^{N}))\nonumber\\
&\stackrel{(a)}\leq&\frac{1}{N}(I(W_{0};Y^{N})+\delta(P_{e1}))\nonumber\\
&=&\frac{1}{N}(\sum_{i=1}^{N}I(W_{0};Y_{i}|Y^{i-1})+\delta(P_{e1}))\nonumber\\
&=&\frac{1}{N}\sum_{i=1}^{N}(H(Y_{i}|Y^{i-1})-H(Y_{i}|Y^{i-1},W_{0}))+\frac{\delta(P_{e1})}{N}\nonumber\\
&\leq&\frac{1}{N}\sum_{i=1}^{N}(H(Y_{i})-H(Y_{i}|Y^{i-1},W_{0},Y_{1}^{i-1},Z_{i+1}^{N}))+\frac{\delta(P_{e1})}{N}\nonumber\\
&\stackrel{(b)}=&\frac{1}{N}\sum_{i=1}^{N}(H(Y_{i}|J=i)-H(Y_{i}|Y^{i-1},W_{0},Y_{1}^{i-1},Z_{i+1}^{N},J=i))+\frac{\delta(P_{e1})}{N}\nonumber\\
&\stackrel{(c)}\leq&H(Y_{J})-H(Y_{J}|Y^{J-1},W_{0},Y_{1}^{J-1},Z_{J+1}^{N},J)+\frac{\delta(P_{e1})}{N}\nonumber\\
&\stackrel{(d)}=&H(Y)-H(Y|U_{1},U)+\frac{\delta(P_{e1})}{N}\nonumber\\
&\stackrel{(e)}\leq&I(U_{1},U;Y)+\frac{\delta(\epsilon)}{N},
\end{eqnarray}
where (a) is from the Fano's inequality, (b) is from the fact that J is a random variable (uniformly distributed
over $\{1,2,...,N\}$), and it is independent of $Y^{N}$, $Y_{1}^{N}$, $Z^{N}$, $W_{0}$, $W_{1}$ and $W_{2}$, (c) is from $J$ is
uniformly distributed over
$\{1,2,...,N\}$, (d) is from the definitions of the auxiliary random variables (see (\ref{appen1.1})), and (e) is from $P_{e1}\leq \epsilon$.

By using $\epsilon\rightarrow 0$, $R_{0}=\lim_{N\rightarrow \infty}\frac{H(W_{0})}{N}$ and (\ref{appen1.2}), $R_{0}\leq I(U,U_{1};Y)$ is obtained.

The inequality $R_{0}\leq I(U;Y,Y_{1}|U_{1})$ is proved as follows.
\begin{eqnarray}\label{appen1.3}
\frac{1}{N}H(W_{0})&\leq&\frac{1}{N}(I(W_{0};Y_{1}^{N},Y^{N})+H(W_{0}|Y_{1}^{N},Y^{N}))\nonumber\\
&\leq&\frac{1}{N}(I(W_{0};Y_{1}^{N},Y^{N})+\delta(P_{e1}))\nonumber\\
&=&\frac{1}{N}(\sum_{i=1}^{N}I(W_{0};Y_{1,i},Y_{i}|Y_{1}^{i-1},Y^{i-1})+\delta(P_{e1}))\nonumber\\
&=&\frac{1}{N}\sum_{i=1}^{N}(H(Y_{1,i},Y_{i}|Y_{1}^{i-1},Y^{i-1})-H(Y_{1,i},Y_{i}|Y_{1}^{i-1},Y^{i-1},W_{0}))+\frac{\delta(P_{e1})}{N}\nonumber\\
&\leq&\frac{1}{N}\sum_{i=1}^{N}(H(Y_{1,i},Y_{i}|Y_{1}^{i-1})-H(Y_{1,i},Y_{i}|Y^{i-1},W_{0},Y_{1}^{i-1},Z_{i+1}^{N}))+\frac{\delta(P_{e1})}{N}\nonumber\\
&=&\frac{1}{N}\sum_{i=1}^{N}(H(Y_{1,i},Y_{i}|Y_{1}^{i-1},J=i)-H(Y_{1,i},Y_{i}|Y^{i-1},W_{0},Y_{1}^{i-1},Z_{i+1}^{N},J=i))+\frac{\delta(P_{e1})}{N}\nonumber\\
&\leq&H(Y_{J},Y_{1,J}|Y_{1}^{J-1})-H(Y_{J},Y_{1,J}|Y^{J-1},W_{0},Y_{1}^{J-1},Z_{J+1}^{N},J)+\frac{\delta(P_{e1})}{N}\nonumber\\
&\stackrel{(a)}=&H(Y,Y_{1}|U_{1})-H(Y,Y_{1}|U_{1},U)+\frac{\delta(P_{e1})}{N}\nonumber\\
&\leq&I(U;Y,Y_{1}|U_{1})+\frac{\delta(\epsilon)}{N},
\end{eqnarray}
where (a) is from (\ref{appen1.1}). By using $\epsilon\rightarrow 0$, $R_{0}=\lim_{N\rightarrow \infty}\frac{H(W_{0})}{N}$ and (\ref{appen1.3}),
$R_{0}\leq I(U;Y,Y_{1}|U_{1})$ is obtained.

Therefore, $R_{0}\leq \min\{I(U,U_{1};Y),I(U;Y,Y_{1}|U_{1})\}$ is proved.

\textbf{(Proof of $R_{0}\leq \min\{I(U,U_{2};Z),I(U;Z,Y_{1}|U_{2})\}$)}

The inequality $R_{0}\leq I(U,U_{2};Z)$ is proved as follows.
\begin{eqnarray}\label{appen1.4}
\frac{1}{N}H(W_{0})&\leq&\frac{1}{N}(I(W_{0};Z^{N})+H(W_{0}|Z^{N}))\nonumber\\
&\leq&\frac{1}{N}(I(W_{0};Z^{N})+\delta(P_{e2}))\nonumber\\
&=&\frac{1}{N}(\sum_{i=1}^{N}I(W_{0};Z_{i}|Z_{i+1}^{N})+\delta(P_{e2}))\nonumber\\
&=&\frac{1}{N}\sum_{i=1}^{N}(H(Z_{i}|Z_{i+1}^{N})-H(Z_{i}|Z_{i+1}^{N},W_{0}))+\frac{\delta(P_{e2})}{N}\nonumber\\
&\leq&\frac{1}{N}\sum_{i=1}^{N}(H(Z_{i})-H(Z_{i}|Y^{i-1},W_{0},Y_{1,i+1}^{N},Z_{i+1}^{N}))+\frac{\delta(P_{e2})}{N}\nonumber\\
&=&\frac{1}{N}\sum_{i=1}^{N}(H(Z_{i}|J=i)-H(Z_{i}|Y^{i-1},W_{0},Y_{1,i+1}^{N},Z_{i+1}^{N},J=i))+\frac{\delta(P_{e2})}{N}\nonumber\\
&\leq&H(Z_{J})-H(Z_{J}|Y^{J-1},W_{0},Y_{1,J+1}^{N},Z_{J+1}^{N},J)+\frac{\delta(P_{e2})}{N}\nonumber\\
&=&H(Z)-H(Z|U_{2},U)+\frac{\delta(P_{e2})}{N}\nonumber\\
&\leq&I(U_{2},U;Z)+\frac{\delta(\epsilon)}{N}.
\end{eqnarray}
By using $\epsilon\rightarrow 0$, $R_{0}=\lim_{N\rightarrow \infty}\frac{H(W_{0})}{N}$ and (\ref{appen1.4}),
$R_{0}\leq I(U_{2},U;Z)$ is obtained.

The inequality $R_{0}\leq I(U;Z,Y_{1}|U_{2})$ is proved as follows.
\begin{eqnarray}\label{appen1.5}
\frac{1}{N}H(W_{0})&\leq&\frac{1}{N}(I(W_{0};Y_{1}^{N},Z^{N})+H(W_{0}|Y_{1}^{N},Z^{N}))\nonumber\\
&\leq&\frac{1}{N}(I(W_{0};Y_{1}^{N},Z^{N})+\delta(P_{e2}))\nonumber\\
&=&\frac{1}{N}(\sum_{i=1}^{N}I(W_{0};Y_{1,i},Z_{i}|Y_{1,i+1}^{N},Z_{i+1}^{N})+\delta(P_{e2}))\nonumber\\
&=&\frac{1}{N}\sum_{i=1}^{N}(H(Y_{1,i},Z_{i}|Y_{1,i+1}^{N},Z_{i+1}^{N})-H(Y_{1,i},Z_{i}|Y_{1,i+1}^{N},Z_{i+1}^{N},W_{0}))+\frac{\delta(P_{e2})}{N}\nonumber\\
&\leq&\frac{1}{N}\sum_{i=1}^{N}(H(Y_{1,i},Z_{i}|Y_{1,i+1}^{N})-H(Y_{1,i},Z_{i}|Y^{i-1},W_{0},Y_{1,i+1}^{N},Z_{i+1}^{N}))+\frac{\delta(P_{e2})}{N}\nonumber\\
&=&\frac{1}{N}\sum_{i=1}^{N}(H(Y_{1,i},Z_{i}|Y_{1,i+1}^{N},J=i)-H(Y_{1,i},Z_{i}|Y^{i-1},W_{0},Y_{1,i+1}^{N},Z_{i+1}^{N},J=i))+\frac{\delta(P_{e2})}{N}\nonumber\\
&\leq&H(Z_{J},Y_{1,J}|Y_{1,J+1}^{N})-H(Z_{J},Y_{1,J}|Y^{J-1},W_{0},Y_{1,J+1}^{N},Z_{J+1}^{N},J)+\frac{\delta(P_{e2})}{N}\nonumber\\
&=&H(Z,Y_{1}|U_{2})-H(Z,Y_{1}|U_{2},U)+\frac{\delta(P_{e2})}{N}\nonumber\\
&\leq&I(U;Z,Y_{1}|U_{2})+\frac{\delta(\epsilon)}{N}.
\end{eqnarray}
By using $\epsilon\rightarrow 0$, $R_{0}=\lim_{N\rightarrow \infty}\frac{H(W_{0})}{N}$ and (\ref{appen1.5}),
$R_{0}\leq I(U;Z,Y_{1}|U_{2})$ is obtained.

Therefore, $R_{0}\leq \min\{I(U,U_{2};Z),I(U;Z,Y_{1}|U_{2})\}$ is proved.

\textbf{(Proof of $R_{0}+R_{1}\leq \min\{I(U,U_{1},V_{1};Y),I(U,V_{1};Y,Y_{1}|U_{1})\}$)}

The inequality $R_{0}+R_{1}\leq I(U,U_{1},V_{1};Y)$ is proved as follows.
\begin{eqnarray}\label{appen1.6}
\frac{1}{N}H(W_{0},W_{1})&\leq&\frac{1}{N}(I(W_{0},W_{1};Y^{N})+H(W_{0},W_{1}|Y^{N}))\nonumber\\
&\leq&\frac{1}{N}(I(W_{0},W_{1};Y^{N})+\delta(P_{e1}))\nonumber\\
&=&\frac{1}{N}(\sum_{i=1}^{N}I(W_{0},W_{1};Y_{i}|Y^{i-1})+\delta(P_{e1}))\nonumber\\
&=&\frac{1}{N}\sum_{i=1}^{N}(H(Y_{i}|Y^{i-1})-H(Y_{i}|Y^{i-1},W_{0},W_{1}))+\frac{\delta(P_{e1})}{N}\nonumber\\
&\leq&\frac{1}{N}\sum_{i=1}^{N}(H(Y_{i})-H(Y_{i}|Y^{i-1},W_{0},W_{1},Y_{1}^{i-1},Z_{i+1}^{N}))+\frac{\delta(P_{e1})}{N}\nonumber\\
&=&\frac{1}{N}\sum_{i=1}^{N}(H(Y_{i}|J=i)-H(Y_{i}|Y^{i-1},W_{0},W_{1},Y_{1}^{i-1},Z_{i+1}^{N},J=i))+\frac{\delta(P_{e1})}{N}\nonumber\\
&\leq&H(Y_{J})-H(Y_{J}|Y^{J-1},W_{0},W_{1},Y_{1}^{J-1},Z_{J+1}^{N},J)+\frac{\delta(P_{e1})}{N}\nonumber\\
&=&H(Y)-H(Y|U_{1},U,V_{1})+\frac{\delta(P_{e1})}{N}\nonumber\\
&\leq&I(U_{1},U,V_{1};Y)+\frac{\delta(\epsilon)}{N}.
\end{eqnarray}
By using $\epsilon\rightarrow 0$, $R_{0}+R_{1}=\lim_{N\rightarrow \infty}\frac{H(W_{0},W_{1})}{N}$ and (\ref{appen1.6}),
$R_{0}+R_{1}\leq I(U,U_{1},V_{1};Y)$ is obtained.

The inequality $R_{0}+R_{1}\leq I(U,V_{1};Y,Y_{1}|U_{1})$ is proved as follows.
\begin{eqnarray}\label{appen1.7}
\frac{1}{N}H(W_{0},W_{1})&\leq&\frac{1}{N}(I(W_{0},W_{1};Y_{1}^{N},Y^{N})+H(W_{0},W_{1}|Y_{1}^{N},Y^{N}))\nonumber\\
&\leq&\frac{1}{N}(I(W_{0},W_{1};Y_{1}^{N},Y^{N})+\delta(P_{e1}))\nonumber\\
&=&\frac{1}{N}(\sum_{i=1}^{N}I(W_{0},W_{1};Y_{1,i},Y_{i}|Y_{1}^{i-1},Y^{i-1})+\delta(P_{e1}))\nonumber\\
&=&\frac{1}{N}\sum_{i=1}^{N}(H(Y_{1,i},Y_{i}|Y_{1}^{i-1},Y^{i-1})-H(Y_{1,i},Y_{i}|Y_{1}^{i-1},Y^{i-1},W_{0},W_{1}))+\frac{\delta(P_{e1})}{N}\nonumber\\
&\leq&\frac{1}{N}\sum_{i=1}^{N}(H(Y_{1,i},Y_{i}|Y_{1}^{i-1})-H(Y_{1,i},Y_{i}|Y^{i-1},W_{0},W_{1},Y_{1}^{i-1},Z_{i+1}^{N}))+\frac{\delta(P_{e1})}{N}\nonumber\\
&=&\frac{1}{N}\sum_{i=1}^{N}(H(Y_{1,i},Y_{i}|Y_{1}^{i-1},J=i)-H(Y_{1,i},Y_{i}|Y^{i-1},W_{0},W_{1},Y_{1}^{i-1},Z_{i+1}^{N},J=i))+\frac{\delta(P_{e1})}{N}\nonumber\\
&\leq&H(Y_{J},Y_{1,J}|Y_{1}^{J-1})-H(Y_{J},Y_{1,J}|Y^{J-1},W_{0},W_{1},Y_{1}^{J-1},Z_{J+1}^{N},J)+\frac{\delta(P_{e1})}{N}\nonumber\\
&=&H(Y,Y_{1}|U_{1})-H(Y,Y_{1}|U_{1},U,V_{1})+\frac{\delta(P_{e1})}{N}\nonumber\\
&\leq&I(U,V_{1};Y,Y_{1}|U_{1})+\frac{\delta(\epsilon)}{N},
\end{eqnarray}
By using $\epsilon\rightarrow 0$, $R_{0}+R_{1}=\lim_{N\rightarrow \infty}\frac{H(W_{0},W_{1})}{N}$ and (\ref{appen1.7}),
$R_{0}+R_{1}\leq I(U,V_{1};Y,Y_{1}|U_{1})$ is obtained.

Therefore, $R_{0}+R_{1}\leq \min\{I(U,U_{1},V_{1};Y),I(U,V_{1};Y,Y_{1}|U_{1})\}$ is proved.

\textbf{(Proof of $R_{0}+R_{2}\leq \min\{I(U,U_{2},V_{2};Z),I(U,V_{2};Z,Y_{1}|U_{2})\}$)}

The proof of $R_{0}+R_{2}\leq \min\{I(U,U_{2},V_{2};Z),I(U,V_{2};Z,Y_{1}|U_{2})\}$ is analogous to the proof of
$R_{0}+R_{1}\leq \min\{I(U,U_{1},V_{1};Y),I(U,V_{1};Y,Y_{1}|U_{1})\}$, and it is omitted here.

\textbf{(Proof of $R_{0}+R_{1}+R_{2}\leq I(U,U_{2},V_{1};Y,Y_{1}|U_{1})+I(V_{2};Z,Y_{1}|U,U_{1},U_{2},V_{1})$)}

The inequality $R_{0}+R_{1}+R_{2}\leq I(U,U_{2},V_{1};Y,Y_{1}|U_{1})+I(V_{2};Z,Y_{1}|U,U_{1},U_{2},V_{1})$ is proved by the following
(\ref{appen1.8}), (\ref{appen1.9}), (\ref{appen1.10}) and (\ref{appen1.11}).

First, note that
\begin{eqnarray}\label{appen1.8}
\frac{1}{N}H(W_{0},W_{1},W_{2})&=&\frac{1}{N}(H(W_{0},W_{1})+H(W_{2}|W_{0},W_{1}))\nonumber\\
&=&\frac{1}{N}(I(W_{0},W_{1};Y_{1}^{N},Y^{N})+H(W_{0},W_{1}|Y_{1}^{N},Y^{N})+I(W_{2};Y_{1}^{N},Z^{N}|W_{0},W_{1})\nonumber\\
&&+H(W_{2}|W_{0},W_{1},Y_{1}^{N},Z^{N}))\nonumber\\
&\stackrel{(a)}\leq&\frac{1}{N}(I(W_{0},W_{1};Y_{1}^{N},Y^{N})+\delta(P_{e1})+I(W_{2};Y_{1}^{N},Z^{N}|W_{0},W_{1})+\delta(P_{e2})),
\end{eqnarray}
where (a) is from Fano's inequality.

The character $I(W_{0},W_{1};Y_{1}^{N},Y^{N})$ in (\ref{appen1.8}) is upper bounded by
\begin{eqnarray}\label{appen1.9}
&&I(W_{0},W_{1};Y_{1}^{N},Y^{N})\nonumber\\
&=&\sum_{i=1}^{N}I(W_{0},W_{1};Y_{1,i},Y_{i}|Y_{1}^{i-1},Y^{i-1})\nonumber\\
&=&\sum_{i=1}^{N}(H(Y_{1,i},Y_{i}|Y_{1}^{i-1},Y^{i-1})-H(Y_{1,i},Y_{i}|Y_{1}^{i-1},Y^{i-1},W_{0},W_{1})\nonumber\\
&&+H(Y_{1,i},Y_{i}|Y_{1}^{i-1},Y^{i-1},W_{0},W_{1},Y_{1,i+1}^{N},Z_{i+1}^{N})-H(Y_{1,i},Y_{i}|Y_{1}^{i-1},Y^{i-1},W_{0},W_{1},Y_{1,i+1}^{N},Z_{i+1}^{N}))\nonumber\\
&=&\sum_{i=1}^{N}(I(Y_{1,i},Y_{i};W_{0},W_{1},Y_{1,i+1}^{N},Z_{i+1}^{N}|Y_{1}^{i-1},Y^{i-1})\nonumber\\
&&-I(Y_{1,i},Y_{i};Y_{1,i+1}^{N},Z_{i+1}^{N}|Y_{1}^{i-1},Y^{i-1},W_{0},W_{1})),
\end{eqnarray}
and the character $I(W_{2};Y_{1}^{N},Z^{N}|W_{0},W_{1})$ in (\ref{appen1.8}) is upper bounded by
\begin{eqnarray}\label{appen1.10}
&&I(W_{2};Y_{1}^{N},Z^{N}|W_{0},W_{1})\nonumber\\
&=&\sum_{i=1}^{N}I(W_{2};Y_{1,i},Z_{i}|Y_{1,i+1}^{N},Z_{i+1}^{N},W_{0},W_{1})\nonumber\\
&\leq&\sum_{i=1}^{N}I(W_{2},Y^{i-1},Y_{1}^{i-1};Y_{1,i},Z_{i}|Y_{1,i+1}^{N},Z_{i+1}^{N},W_{0},W_{1})\nonumber\\
&=&\sum_{i=1}^{N}(H(Y_{1,i},Z_{i}|Y_{1,i+1}^{N},Z_{i+1}^{N},W_{0},W_{1})\nonumber\\
&&-H(Y_{1,i},Z_{i}|Y_{1,i+1}^{N},Z_{i+1}^{N},W_{0},W_{1},W_{2},Y^{i-1},Y_{1}^{i-1})\nonumber\\
&&+H(Y_{1,i},Z_{i}|Y_{1,i+1}^{N},Z_{i+1}^{N},W_{0},W_{1},Y^{i-1},Y_{1}^{i-1})-H(Y_{1,i},Z_{i}|Y_{1,i+1}^{N},Z_{i+1}^{N},W_{0},W_{1},Y^{i-1},Y_{1}^{i-1})\nonumber\\
&=&\sum_{i=1}^{N}(I(Y_{1,i},Z_{i};Y^{i-1},Y_{1}^{i-1}|Y_{1,i+1}^{N},Z_{i+1}^{N},W_{0},W_{1})\nonumber\\
&&+I(Y_{1,i},Z_{i};W_{2}|Y_{1,i+1}^{N},Z_{i+1}^{N},W_{0},W_{1},Y^{i-1},Y_{1}^{i-1})).
\end{eqnarray}

Here note that $\sum_{i=1}^{N}I(Y_{1,i},Y_{i};Y_{1,i+1}^{N},Z_{i+1}^{N}|Y_{1}^{i-1},Y^{i-1},W_{0},W_{1})$ appeared in the last step of (\ref{appen1.9})
is equal to $\sum_{i=1}^{N}I(Y_{1,i},Z_{i};Y^{i-1},Y_{1}^{i-1}|Y_{1,i+1}^{N},Z_{i+1}^{N},W_{0},W_{1})$ appeared in the last step of (\ref{appen1.10}),
i.e.,
\begin{eqnarray}\label{appen1.11}
&&\sum_{i=1}^{N}I(Y_{1,i},Y_{i};Y_{1,i+1}^{N},Z_{i+1}^{N}|Y_{1}^{i-1},Y^{i-1},W_{0},W_{1})\nonumber\\
&=&\sum_{i=1}^{N}I(Y_{1,i},Z_{i};Y^{i-1},Y_{1}^{i-1}|Y_{1,i+1}^{N},Z_{i+1}^{N},W_{0},W_{1}),
\end{eqnarray}
and it is proved by the following (\ref{appen1.12}) and (\ref{appen1.13}).

\begin{eqnarray}\label{appen1.12}
&&\sum_{i=1}^{N}I(Y_{1,i},Y_{i};Y_{1,i+1}^{N},Z_{i+1}^{N}|Y_{1}^{i-1},Y^{i-1},W_{0},W_{1})\nonumber\\
&=&\sum_{i=1}^{N}\sum_{j=i+1}^{N}I(Y_{1,i},Y_{i};Y_{1,j},Z_{j}|Y_{1}^{i-1},Y^{i-1},W_{0},W_{1},Y_{1,j+1}^{N},Z_{j+1}^{N}).
\end{eqnarray}

\begin{eqnarray}\label{appen1.13}
&&\sum_{i=1}^{N}I(Y_{1,i},Z_{i};Y^{i-1},Y_{1}^{i-1}|Y_{1,i+1}^{N},Z_{i+1}^{N},W_{0},W_{1})\nonumber\\
&=&\sum_{i=1}^{N}\sum_{j=1}^{i-1}I(Y_{1,i},Z_{i};Y_{1,j},Y_{j}|Y_{1,i+1}^{N},Z_{i+1}^{N},W_{0},W_{1},Y^{j-1},Y_{1}^{j-1})\nonumber\\
&=&\sum_{j=1}^{N}\sum_{i=1}^{j-1}I(Y_{1,j},Z_{j};Y_{1,i},Y_{i}|Y_{1,j+1}^{N},Z_{j+1}^{N},W_{0},W_{1},Y^{i-1},Y_{1}^{i-1})\nonumber\\
&=&\sum_{i=1}^{N}\sum_{j=i+1}^{N}I(Y_{1,j},Z_{j};Y_{1,i},Y_{i}|Y_{1,j+1}^{N},Z_{j+1}^{N},W_{0},W_{1},Y^{i-1},Y_{1}^{i-1}).
\end{eqnarray}

Finally, substituting (\ref{appen1.9}) and (\ref{appen1.10}) into (\ref{appen1.8}), and using the fact that (\ref{appen1.11}) holds,
then we have
\begin{eqnarray}\label{appen1.14}
&&\frac{1}{N}H(W_{0},W_{1},W_{2})\nonumber\\
&\leq&\frac{1}{N}\sum_{i=1}^{N}(I(Y_{1,i},Y_{i};W_{0},W_{1},Y_{1,i+1}^{N},Z_{i+1}^{N}|Y_{1}^{i-1},Y^{i-1})\nonumber\\
&&+I(Y_{1,i},Z_{i};W_{2}|Y_{1,i+1}^{N},Z_{i+1}^{N},W_{0},W_{1},Y^{i-1},Y_{1}^{i-1}))+\frac{\delta(P_{e1})+\delta(P_{e2})}{N}\nonumber\\
&\stackrel{(1)}\leq&\frac{1}{N}\sum_{i=1}^{N}(I(Y_{1,i},Y_{i};W_{0},W_{1},Y_{1,i+1}^{N},Z_{i+1}^{N}|Y_{1}^{i-1},Y^{i-1},J=i)\nonumber\\
&&+I(Y_{1,i},Z_{i};W_{2}|Y_{1,i+1}^{N},Z_{i+1}^{N},W_{0},W_{1},Y^{i-1},Y_{1}^{i-1},J=i))+\frac{2\delta(\epsilon)}{N}\nonumber\\
&\stackrel{(2)}=&I(Y_{1,J},Y_{J};W_{0},W_{1},Y_{1,J+1}^{N},Z_{J+1}^{N}|Y_{1}^{J-1},Y^{J-1},J)\nonumber\\
&&+I(Y_{1,J},Z_{J};W_{2}|Y_{1,J+1}^{N},Z_{J+1}^{N},W_{0},W_{1},Y^{J-1},Y_{1}^{J-1},J)+\frac{2\delta(\epsilon)}{N}\nonumber\\
&\leq&H(Y_{1,J},Y_{J}|Y_{1}^{J-1})-H(Y_{1,J},Y_{J}|W_{0},W_{1},Y_{1,J+1}^{N},Z_{J+1}^{N},Y_{1}^{J-1},Y^{J-1},J)\nonumber\\
&&+I(Y_{1,J},Z_{J};W_{2}|Y_{1,J+1}^{N},Z_{J+1}^{N},W_{0},W_{1},Y^{J-1},Y_{1}^{J-1},J)+\frac{2\delta(\epsilon)}{N}\nonumber\\
&\stackrel{(3)}=&I(Y_{1},Y|U_{1})-H(Y_{1},Y|U,V_{1},U_{1},U_{2})+H(Y_{1},Z|U,V_{1},U_{1},U_{2})\nonumber\\
&&-H(Y_{1},Z|U,V_{1},U_{1},U_{2},V_{2})+\frac{2\delta(\epsilon)}{N}\nonumber\\
&=&I(U,U_{2},V_{1};Y,Y_{1}|U_{1})+I(V_{2};Z,Y_{1}|U,U_{1},U_{2},V_{1})+\frac{2\delta(\epsilon)}{N},
\end{eqnarray}
where (1) is from $P_{e1}, P_{e2}\leq \epsilon$, $J$ is a random variable (uniformly distributed
over $\{1,2,...,N\}$), and it is independent of $Y^{N}$, $Y_{1}^{N}$, $Z^{N}$, $W_{0}$, $W_{1}$ and $W_{2}$, (2) is from $J$ is
uniformly distributed over
$\{1,2,...,N\}$, and (3) is from the definitions of the auxiliary random variables (see (\ref{appen1.1})).

By using $\epsilon\rightarrow 0$, $R_{0}+R_{1}+R_{2}=\lim_{N\rightarrow \infty}\frac{H(W_{0},W_{1},W_{2})}{N}$ and (\ref{appen1.14}),
$R_{0}+R_{1}+R_{2}\leq I(U,U_{2},V_{1};Y,Y_{1}|U_{1})+I(V_{2};Z,Y_{1}|U,U_{1},U_{2},V_{1})$ is proved.

\textbf{(Proof of $R_{0}+R_{1}+R_{2}\leq I(U,U_{1},V_{2};Z,Y_{1}|U_{2})+I(V_{1};Y,Y_{1}|U,U_{1},U_{2},V_{2})$)}

The inequality $R_{0}+R_{1}+R_{2}\leq I(U,U_{1},V_{2};Z,Y_{1}|U_{2})+I(V_{1};Y,Y_{1}|U,U_{1},U_{2},V_{2})$ is proved by letting
$H(W_{0},W_{1},W_{2})=H(W_{0},W_{2})+H(W_{1}|W_{0},W_{2})$, and the remainder of the proof is analogous to the proof of
$R_{0}+R_{1}+R_{2}\leq I(U,U_{2},V_{1};Y,Y_{1}|U_{1})+I(V_{2};Z,Y_{1}|U,U_{1},U_{2},V_{1})$. Thus ,we omit the proof here.

\textbf{(Proof of $R_{e1}\leq I(V_{1};Y|U,V_{2})-I(V_{1};Z|U,V_{2})$)}

The inequality $R_{e1}\leq I(V_{1};Y|U,V_{2})-I(V_{1};Z|U,V_{2})$ is proved by the following (\ref{appen1.15}), (\ref{appen1.16}),
(\ref{appen1.17}) and (\ref{appen1.20}). First note that
\begin{eqnarray}\label{appen1.15}
&&\frac{1}{N}H(W_{1}|Z^{N})\nonumber\\
&=&\frac{1}{N}(I(W_{1};W_{0},W_{2}|Z^{N})+H(W_{1}|Z^{N},W_{0},W_{2}))\nonumber\\
&\leq&\frac{1}{N}(H(W_{1}|Z^{N},W_{0},W_{2})+\delta(\epsilon))\nonumber\\
&=&\frac{1}{N}(H(W_{1}|W_{0},W_{2})-I(W_{1};Z^{N}|W_{0},W_{2})+\delta(\epsilon))\nonumber\\
&=&\frac{1}{N}(I(W_{1};Y^{N}|W_{0},W_{2})+H(W_{1}|Y^{N},W_{0},W_{2})-I(W_{1};Z^{N}|W_{0},W_{2})+\delta(\epsilon))\nonumber\\
&\leq&\frac{1}{N}(I(W_{1};Y^{N}|W_{0},W_{2})-I(W_{1};Z^{N}|W_{0},W_{2})+2\delta(\epsilon)).
\end{eqnarray}

Then, the character $I(W_{1};Y^{N}|W_{0},W_{2})$ in (\ref{appen1.15}) is upper bounded by
\begin{eqnarray}\label{appen1.16}
&&I(W_{1};Y^{N}|W_{0},W_{2})=\sum_{i=1}^{N}I(W_{1};Y_{i}|W_{0},W_{2},Y^{i-1})\nonumber\\
&=&\sum_{i=1}^{N}(H(Y_{i}|W_{0},W_{2},Y^{i-1})-H(Y_{i}|W_{0},W_{1},W_{2},Y^{i-1})\nonumber\\
&&+H(Y_{i}|W_{0},W_{2},Y^{i-1},W_{1},Z_{i+1}^{N})-H(Y_{i}|W_{0},W_{2},Y^{i-1},W_{1},Z_{i+1}^{N}))\nonumber\\
&=&\sum_{i=1}^{N}(I(Y_{i};W_{1},Z_{i+1}^{N}|W_{0},W_{2},Y^{i-1})-I(Y_{i};Z_{i+1}^{N}|W_{0},W_{1},W_{2},Y^{i-1}))\nonumber\\
&=&\sum_{i=1}^{N}(I(Y_{i};Z_{i+1}^{N}|W_{0},W_{2},Y^{i-1})+I(Y_{i};W_{1}|W_{0},W_{2},Y^{i-1},Z_{i+1}^{N})\nonumber\\
&&-I(Y_{i};Z_{i+1}^{N}|W_{0},W_{1},W_{2},Y^{i-1})),
\end{eqnarray}
and the character $I(W_{1};Z^{N}|W_{0},W_{2})$ in (\ref{appen1.15}) can be expressed as
\begin{eqnarray}\label{appen1.17}
&&I(W_{1};Z^{N}|W_{0},W_{2})=\sum_{i=1}^{N}I(W_{1};Z_{i}|W_{0},W_{2},Z_{i+1}^{N})\nonumber\\
&=&\sum_{i=1}^{N}(H(Z_{i}|W_{0},W_{2},Z_{i+1}^{N})-H(Z_{i}|W_{0},W_{1},W_{2},Z_{i+1}^{N})\nonumber\\
&&+H(Z_{i}|W_{0},W_{2},Y^{i-1},W_{1},Z_{i+1}^{N})-H(Z_{i}|W_{0},W_{2},Y^{i-1},W_{1},Z_{i+1}^{N}))\nonumber\\
&=&\sum_{i=1}^{N}(I(Z_{i};W_{1},Y^{i-1}|W_{0},W_{2},Z_{i+1}^{N})-I(Z_{i};Y^{i-1}|W_{0},W_{1},W_{2},Z_{i+1}^{N}))\nonumber\\
&=&\sum_{i=1}^{N}(I(Z_{i};Y^{i-1}|W_{0},W_{2},Z_{i+1}^{N})+I(Z_{i};W_{1}|W_{0},W_{2},Y^{i-1},Z_{i+1}^{N})\nonumber\\
&&-I(Z_{i};Y^{i-1}|W_{0},W_{1},W_{2},Z_{i+1}^{N})).
\end{eqnarray}
Note that
\begin{equation}\label{appen1.18}
\sum_{i=1}^{N}I(Y_{i};Z_{i+1}^{N}|W_{0},W_{2},Y^{i-1})=\sum_{i=1}^{N}I(Z_{i};Y^{i-1}|W_{0},W_{2},Z_{i+1}^{N}),
\end{equation}
and
\begin{equation}\label{appen1.19}
\sum_{i=1}^{N}I(Y_{i};Z_{i+1}^{N}|W_{0},W_{1},W_{2},Y^{i-1})=\sum_{i=1}^{N}I(Z_{i};Y^{i-1}|W_{0},W_{1},W_{2},Z_{i+1}^{N}),
\end{equation}
and these are from Csisz$\acute{a}$r's equality \cite{CK}.

Substituting (\ref{appen1.16}) and (\ref{appen1.17}) into (\ref{appen1.15}), and using the equalities (\ref{appen1.18}) and (\ref{appen1.19}), we have
\begin{eqnarray}\label{appen1.20}
&&\frac{1}{N}H(W_{1}|Z^{N})\nonumber\\
&\leq&\frac{1}{N}(I(W_{1};Y^{N}|W_{0},W_{2})-I(W_{1};Z^{N}|W_{0},W_{2})+2\delta(\epsilon))\nonumber\\
&=&\frac{1}{N}\sum_{i=1}^{N}(I(Y_{i};W_{1}|W_{0},W_{2},Y^{i-1},Z_{i+1}^{N})\nonumber\\
&&-(Z_{i};W_{1}|W_{0},W_{2},Y^{i-1},Z_{i+1}^{N}))+\frac{2\delta(\epsilon)}{N}\nonumber\\
&=&\frac{1}{N}\sum_{i=1}^{N}(I(Y_{i};W_{1}|W_{0},W_{2},Y^{i-1},Z_{i+1}^{N},J=i)\nonumber\\
&&-I(Z_{i};W_{1}|W_{0},W_{2},Y^{i-1},Z_{i+1}^{N},J=i))+\frac{2\delta(\epsilon)}{N}\nonumber\\
&=&I(Y_{J};W_{1}|W_{0},W_{2},Y^{J-1},Z_{J+1}^{N},J)\nonumber\\
&&-I(Z_{J};W_{1}|W_{0},W_{2},Y^{J-1},Z_{J+1}^{N},J)+\frac{2\delta(\epsilon)}{N}\nonumber\\
&=&I(Y;V_{1}|U,V_{2})-I(Z;V_{1}|U,V_{2})+\frac{2\delta(\epsilon)}{N}.
\end{eqnarray}

By using $\epsilon\rightarrow 0$, $R_{e1}\leq\lim_{N\rightarrow \infty}\frac{H(W_{1}|Z^{N})}{N}$ and (\ref{appen1.20}),
$R_{e1}\leq I(V_{1};Y|U,V_{2})-I(V_{1};Z|U,V_{2})$ is proved.

\textbf{(Proof of $R_{e1}\leq I(V_{1};Y|U)-I(V_{1};Z|U)$)}

The inequality $R_{e1}\leq I(V_{1};Y|U)-I(V_{1};Z|U)$ is proved by the following (\ref{appen1.21}), (\ref{appen1.22}),
(\ref{appen1.23}) and (\ref{appen1.26}). First note that
\begin{eqnarray}\label{appen1.21}
&&\frac{1}{N}H(W_{1}|Z^{N})\nonumber\\
&=&\frac{1}{N}(I(W_{1};W_{0}|Z^{N})+H(W_{1}|Z^{N},W_{0}))\nonumber\\
&\leq&\frac{1}{N}(H(W_{1}|Z^{N},W_{0})+\delta(\epsilon))\nonumber\\
&=&\frac{1}{N}(H(W_{1}|W_{0})-I(W_{1};Z^{N}|W_{0})+\delta(\epsilon))\nonumber\\
&=&\frac{1}{N}(I(W_{1};Y^{N}|W_{0})+H(W_{1}|Y^{N},W_{0})-I(W_{1};Z^{N}|W_{0})+\delta(\epsilon))\nonumber\\
&\leq&\frac{1}{N}(I(W_{1};Y^{N}|W_{0})-I(W_{1};Z^{N}|W_{0})+2\delta(\epsilon)).
\end{eqnarray}

Then, the character $I(W_{1};Y^{N}|W_{0})$ in (\ref{appen1.21}) is upper bounded by
\begin{eqnarray}\label{appen1.22}
&&I(W_{1};Y^{N}|W_{0})=\sum_{i=1}^{N}I(W_{1};Y_{i}|W_{0},Y^{i-1})\nonumber\\
&=&\sum_{i=1}^{N}(H(Y_{i}|W_{0},Y^{i-1})-H(Y_{i}|W_{0},W_{1},Y^{i-1})\nonumber\\
&&+H(Y_{i}|W_{0},Y^{i-1},W_{1},Z_{i+1}^{N})-H(Y_{i}|W_{0},Y^{i-1},W_{1},Z_{i+1}^{N}))\nonumber\\
&=&\sum_{i=1}^{N}(I(Y_{i};W_{1},Z_{i+1}^{N}|W_{0},Y^{i-1})-I(Y_{i};Z_{i+1}^{N}|W_{0},W_{1},Y^{i-1}))\nonumber\\
&=&\sum_{i=1}^{N}(I(Y_{i};Z_{i+1}^{N}|W_{0},Y^{i-1})+I(Y_{i};W_{1}|W_{0},Y^{i-1},Z_{i+1}^{N})\nonumber\\
&&-I(Y_{i};Z_{i+1}^{N}|W_{0},W_{1},Y^{i-1})),
\end{eqnarray}
and the character $I(W_{1};Z^{N}|W_{0})$ in (\ref{appen1.21}) can be expressed as
\begin{eqnarray}\label{appen1.23}
&&I(W_{1};Z^{N}|W_{0})=\sum_{i=1}^{N}I(W_{1};Z_{i}|W_{0},Z_{i+1}^{N})\nonumber\\
&=&\sum_{i=1}^{N}(H(Z_{i}|W_{0},Z_{i+1}^{N})-H(Z_{i}|W_{0},W_{1},Z_{i+1}^{N})\nonumber\\
&&+H(Z_{i}|W_{0},Y^{i-1},W_{1},Z_{i+1}^{N})-H(Z_{i}|W_{0},Y^{i-1},W_{1},Z_{i+1}^{N}))\nonumber\\
&=&\sum_{i=1}^{N}(I(Z_{i};W_{1},Y^{i-1}|W_{0},Z_{i+1}^{N})-I(Z_{i};Y^{i-1}|W_{0},W_{1},Z_{i+1}^{N}))\nonumber\\
&=&\sum_{i=1}^{N}(I(Z_{i};Y^{i-1}|W_{0},Z_{i+1}^{N})+I(Z_{i};W_{1}|W_{0},Y^{i-1},Z_{i+1}^{N})\nonumber\\
&&-I(Z_{i};Y^{i-1}|W_{0},W_{1},Z_{i+1}^{N})).
\end{eqnarray}
Note that
\begin{equation}\label{appen1.24}
\sum_{i=1}^{N}I(Y_{i};Z_{i+1}^{N}|W_{0},Y^{i-1})=\sum_{i=1}^{N}I(Z_{i};Y^{i-1}|W_{0},Z_{i+1}^{N}),
\end{equation}
and
\begin{equation}\label{appen1.25}
\sum_{i=1}^{N}I(Y_{i};Z_{i+1}^{N}|W_{0},W_{1},Y^{i-1})=\sum_{i=1}^{N}I(Z_{i};Y^{i-1}|W_{0},W_{1},Z_{i+1}^{N}),
\end{equation}
and these are from Csisz$\acute{a}$r's equality \cite{CK}.

Substituting (\ref{appen1.22}) and (\ref{appen1.23}) into (\ref{appen1.21}), and using the equalities (\ref{appen1.24}) and (\ref{appen1.25}), we have
\begin{eqnarray}\label{appen1.26}
&&\frac{1}{N}H(W_{1}|Z^{N})\nonumber\\
&\leq&\frac{1}{N}(I(W_{1};Y^{N}|W_{0})-I(W_{1};Z^{N}|W_{0})+2\delta(\epsilon))\nonumber\\
&=&\frac{1}{N}\sum_{i=1}^{N}(I(Y_{i};W_{1}|W_{0},Y^{i-1},Z_{i+1}^{N})\nonumber\\
&&-(Z_{i};W_{1}|W_{0},Y^{i-1},Z_{i+1}^{N}))+\frac{2\delta(\epsilon)}{N}\nonumber\\
&=&\frac{1}{N}\sum_{i=1}^{N}(I(Y_{i};W_{1}|W_{0},Y^{i-1},Z_{i+1}^{N},J=i)\nonumber\\
&&-I(Z_{i};W_{1}|W_{0},Y^{i-1},Z_{i+1}^{N},J=i))+\frac{2\delta(\epsilon)}{N}\nonumber\\
&=&I(Y_{J};W_{1}|W_{0},Y^{J-1},Z_{J+1}^{N},J)\nonumber\\
&&-I(Z_{J};W_{1}|W_{0},Y^{J-1},Z_{J+1}^{N},J)+\frac{2\delta(\epsilon)}{N}\nonumber\\
&=&I(Y;V_{1}|U)-I(Z;V_{1}|U)+\frac{2\delta(\epsilon)}{N}.
\end{eqnarray}

By using $\epsilon\rightarrow 0$, $R_{e1}\leq\lim_{N\rightarrow \infty}\frac{H(W_{1}|Z^{N})}{N}$ and (\ref{appen1.26}),
$R_{e1}\leq I(V_{1};Y|U)-I(V_{1};Z|U)$ is proved.

\textbf{(Proof of $R_{e2}\leq \min\{I(V_{2};Z|U,V_{1})-I(V_{2};Y|U,V_{1}), I(V_{2};Z|U)-I(V_{2};Y|U)\}$)}

The proof of $R_{e2}\leq \min\{I(V_{2};Z|U,V_{1})-I(V_{2};Y|U,V_{1}), I(V_{2};Z|U)-I(V_{2};Y|U)\}$ is analogous to the
proof of $R_{e1}\leq \min\{I(V_{1};Y|U,V_{2})-I(V_{1};Z|U,V_{2}), I(V_{1};Y|U)-I(V_{1};Z|U)\}$, and therefore, we omit the proof here.

The Markov chain $U\rightarrow (U_{1},U_{2},V_{1},V_{2})\rightarrow (X,X_{1})\rightarrow (Y,Y_{1},Z)$ is directly proved by the definitions of the
auxiliary random variables. Thus, the proof of Theorem \ref{T1} is completed.

\section{Proof of Theorem \ref{T2}\label{appen2}}

Suppose $(R_{0},R_{1},R_{2},R_{e1},R_{e2})\in \mathcal{R}^{(Ai1)}$, we will show that $(R_{0},R_{1},R_{2},R_{e1},R_{e2})$ is achievable, i.e., there
exists encoder-decoder $(N,\Delta_{1},\Delta_{2},P_{e1},P_{e2})$ such that (\ref{e203}) is satisfied. The existence of the
encoder-decoder is under the sufficient conditions that
\begin{equation}\label{appen2.1}
R_{e1}=I(V_{1};Y|U,X_{1})-I(V_{1};V_{2}|U,X_{1})-I(V_{1};Z|U,X_{1},V_{2}),
\end{equation}
and
\begin{equation}\label{appen2.2}
R_{e2}=I(V_{2};Z|U,X_{1})-I(V_{1};V_{2}|U,X_{1})-I(V_{2};Y|U,X_{1},V_{1}).
\end{equation}

The coding scheme combines the decode and forward (DF) strategy \cite{CG}, random binning, superposition coding,
block Markov coding and rate splitting techniques. The rate splitting technique is typically used in the interference channels to
achieve a larger rate region as it enables interference cancellation at the receivers.
Now we use it to split the confidential message $W_{1}$ into $W_{10}$ and
$W_{11}$, and $W_{2}$ into $W_{20}$ and $W_{22}$, and the details are as follows.

Define the messages $W_{0}$, $W_{10}$, $W_{11}$, $W_{20}$, $W_{22}$ taken values in the alphabets
$\mathcal{W}_{0}$, $\mathcal{W}_{10}$, $\mathcal{W}_{11}$, $\mathcal{W}_{20}$, $\mathcal{W}_{22}$, respectively, where
$$\mathcal{W}_{0}=\{1,2,...,2^{NR_{0}}\},$$
$$\mathcal{W}_{10}=\{1,2,...,2^{NR_{10}}\},$$
$$\mathcal{W}_{11}=\{1,2,...,2^{NR_{11}}\},$$
$$\mathcal{W}_{20}=\{1,2,...,2^{NR_{20}}\},$$
$$\mathcal{W}_{22}=\{1,2,...,2^{NR_{22}}\},$$
and $R_{10}+R_{11}=R_{1}$, $R_{20}+R_{22}=R_{2}$. Here note that the formulas (\ref{appen2.1}) and (\ref{appen2.2}) combined with the rate
splitting and the fact that $W_{10}$ and $W_{20}$ are decoded by both receivers ensure that,
\begin{equation}\label{appen2.3}
R_{11}\geq R_{e1}=I(V_{1};Y|U,X_{1})-I(V_{1};V_{2}|U,X_{1})-I(V_{1};Z|U,X_{1},V_{2}),
\end{equation}
and
\begin{equation}\label{appen2.4}
R_{22}\geq R_{e2}=I(V_{2};Z|U,X_{1})-I(V_{1};V_{2}|U,X_{1})-I(V_{2};Y|U,X_{1},V_{1}).
\end{equation}

\textbf{Code Construction:} Fix the joint probability mass function
$P_{Y,Z,Y_{1},X,X_{1},V_{1},V_{2},U}(y,z,y_{1},x,x_{1},v_{1},v_{2},u)$. For
arbitrary $\epsilon> 0$, define
\begin{equation}\label{appen2.5}
L_{11}=I(V_{1};Y|U,X_{1})-I(V_{1};V_{2}|U,X_{1})-I(V_{1};Z|U,X_{1},V_{2}),
\end{equation}
\begin{equation}\label{appen2.6}
L_{12}=I(V_{1};Z|U,X_{1},V_{2}),
\end{equation}
\begin{equation}\label{appen2.7}
L_{21}=I(V_{2};Z|U,X_{1})-I(V_{1};V_{2}|U,X_{1})-I(V_{2};Y|U,X_{1},V_{1}),
\end{equation}
\begin{equation}\label{appen2.8}
L_{22}=I(V_{2};Y|U,X_{1},V_{1}),
\end{equation}
\begin{equation}\label{appen2.9}
L_{3}=I(V_{1};V_{2}|U,X_{1})-\epsilon.
\end{equation}
Note that
\begin{equation}\label{appen2.10}
L_{11}+L_{12}+L_{3}=I(V_{1};Y|U,X_{1})-\epsilon,
\end{equation}
\begin{equation}\label{appen2.11}
L_{21}+L_{22}+L_{3}=I(V_{2};Z|U,X_{1})-\epsilon.
\end{equation}

\begin{itemize}

\item First, generate at random $2^{NR_{r}}$ i.i.d. sequences at the relay node each drawn according to
$p_{X_{1}^{N}}(x_{1}^{N})=\prod_{i=1}^{N}p_{X_{1}}(x_{1,i})$, index them as $x_{1}^{N}(a)$, $a\in [1,2^{NR_{r}}]$, where
\begin{equation}\label{appen2.12}
R_{r}=\min\{I(X_{1};Y),I(X_{1};Z)\}-\epsilon.
\end{equation}

\item Generate at random $2^{N(R_{10}+R_{20}+R_{0})}$ i.i.d. sequences $u^{N}(b|a)$ ($b\in [1,2^{N(R_{10}+R_{20}+R_{0})}], a\in [1,2^{NR_{r}}]$)
according to $\prod_{i=1}^{N}p_{U|X_{1}}(u_{i}|x_{1,i})$. In addition, partition $2^{N(R_{10}+R_{20}+R_{0})}$
i.i.d. sequences $u^{N}$ into $2^{NR_{r}}$ bins. These bins are denoted as $\{S_{1}, S_{2},...,S_{2^{NR_{r}}}\}$, where
$S_{i}$ ($1\leq i\leq 2^{NR_{r}}$) contains $2^{N(R_{10}+R_{20}+R_{0}-R_{r})}$ sequences about $u^{N}$.

\item For the transmitted sequences $u^{N}$ and $x_{1}^{N}$, generate $2^{N(L_{11}+L_{12}+L_{3})}$ i.i.d. sequences
$v_{1}^{N}(i^{'},i^{''},i^{'''})$, with $i^{'}\in \mathcal{I}^{'}=[1,2^{NL_{11}}]$, $i^{''}\in \mathcal{I}^{''}=[1,2^{NL_{12}}]$
and $i^{'''}\in \mathcal{I}^{'''}=[1,2^{NL_{3}}]$, according to $\prod_{i=1}^{N}p_{V_{1}|U,X_{1}}(v_{1,i}|u_{i},x_{1,i})$.

\item Similarly, for the transmitted sequences $u^{N}$ and $x_{1}^{N}$, generate $2^{N(L_{21}+L_{22}+L_{3})}$ i.i.d. sequences
$v_{2}^{N}(j^{'},j^{''},j^{'''})$, with $j^{'}\in \mathcal{J}^{'}=[1,2^{NL_{21}}]$, $j^{''}\in \mathcal{J}^{''}=[1,2^{NL_{22}}]$
and $j^{'''}\in \mathcal{J}^{'''}=[1,2^{NL_{3}}]$, according to $\prod_{i=1}^{N}p_{V_{2}|U,X_{1}}(v_{2,i}|u_{i},x_{1,i})$.

\item The $x^{N}$ is generated according to a new discrete memoryless channel (DMC) with inputs $x_{1}^{N}$, $u^{N}$, $v_{1}^{N}$,
$v_{2}^{N}$ and output $x^{N}$. The transition probability of this new DMC is $p_{X|X_{1},U,V_{1},V_{2}}(x|x_{1},u,v_{1},v_{2})$.
The probability $p_{X^{N}|X_{1}^{N},U^{N},V_{1}^{N},V_{2}^{N}}(x^{N}|x_{1}^{N},u^{N},v_{1}^{N},v_{2}^{N})$ is calculated as follows.
\begin{equation}\label{appen2.13}
p_{X^{N}|X_{1}^{N},U^{N},V_{1}^{N},V_{2}^{N}}(x^{N}|x_{1}^{N},u^{N},v_{1}^{N},v_{2}^{N})=
\prod_{i=1}^{N}p_{X|X_{1},U,V_{1},V_{2}}(x_{i}|x_{1,i},u_{i},v_{1,i},v_{2,i}).
\end{equation}
Denote $x^{N}$ by $x^{N}(a,w_{0},w_{10},w_{20},w_{11},w_{22})$.

\end{itemize}

\textbf{Encoding:} Encoding involves the mapping of message indices to channel inputs, which are facilitated by the
sequences generated above. We exploit the block Markov coding scheme, as argued in \cite{CG}, the loss induced by
this scheme is negligible as the number of blocks $n\rightarrow \infty$. For block $i$ ($1\leq i\leq n$), encoding proceeds as follows.

First, for convenience, define $w^{*}_{0,i}=(w_{0,i},w_{10,i},w_{20,i})$, where $w_{0,i}$, $w_{10,i}$ and $w_{20,i}$
are the messages transmitted in the $i$-th block. The messages $w_{11}$ and $w_{22}$ transmitted in the $i$-th block are denoted
by $w_{11,i}$ and $w_{22,i}$, respectively.

\begin{itemize}

\item \textbf{(Channel encoder)}

1) The transmitter sends $(u^{N}(w^{*}_{0,1}|1), v_{1}^{N}(i_{1}^{'},i_{1}^{''},i_{1}^{'''}|1,w^{*}_{0,1}),
v_{2}^{N}(j_{1}^{'},j_{1}^{''},j_{1}^{'''}|1,w^{*}_{0,1}))$ at the first block, \\
$(u^{N}(w^{*}_{0,i}|a_{i-1}), v_{1}^{N}(i_{i}^{'},i_{i}^{''},i_{i}^{'''}|a_{i-1},w^{*}_{0,i}),
v_{2}^{N}(j_{i}^{'},j_{i}^{''},j_{i}^{'''}|a_{i-1},w^{*}_{0,i}))$ from block $2$ to $n-1$,
and \\ $(u^{N}(1|a_{n-1}), v_{1}^{N}(1,1,1|a_{n-1},1),
v_{2}^{N}(1,1,1|a_{n-1},1))$ at block $n$. Here $i_{i}^{'}$, $i_{i}^{''}$, $i_{i}^{'''}$, $j_{i}^{'}$, $j_{i}^{''}$ and $j_{i}^{'''}$
are the indexes for block $i$.

2) In the $i$-th block ($1\leq i\leq n$), the indexes $i_{i}^{'}$, $i_{i}^{''}$, $j_{i}^{'}$ and $j_{i}^{''}$ are determined by the following methods.

\begin{itemize}

\item If $R_{11}\leq L_{11}+L_{12}$, define $\mathcal{W}_{11}=\mathcal{I}^{'}\times \mathcal{K}_{1}$. Thus the index $i_{i}^{'}$
is determined by a given message $w_{11,i}$. Evenly partition $\mathcal{I}^{''}$ into $\mathcal{K}_{1}$ bins, and the index $i_{i}^{''}$
is drawn at random (with uniform distribution) from the bin $k_{1}$.

Analogously, if $R_{22}\leq L_{21}+L_{22}$, define $\mathcal{W}_{22}=\mathcal{J}^{'}\times \mathcal{K}_{2}$. Thus the index $j_{i}^{'}$
is determined by a given message $w_{22,i}$. Evenly partition $\mathcal{J}^{''}$ into $\mathcal{K}_{2}$ bins, and the index $j_{i}^{''}$
is drawn at random (with uniform distribution) from the bin $k_{2}$.

\item If $L_{11}+L_{12}\leq R_{11}\leq L_{11}+L_{12}+L_{3}$, define $\mathcal{W}_{11}=\mathcal{I}^{'}\times\mathcal{I}^{''}\times \mathcal{K}_{1}$.
Thus the indexes $i_{i}^{'}$ and $i_{i}^{''}$
are determined by a given message $w_{11,i}$. Evenly partition $\mathcal{I}^{'''}$ into $\mathcal{K}_{1}$ bins, and the codeword
$v_{1}^{N}(i_{i}^{'},i_{i}^{''},i_{i}^{'''}|a_{i-1},w^{*}_{0,i})$ will be drawn from the bin $k_{1}$.

Analogously, if $L_{21}+L_{22}\leq R_{22}\leq L_{21}+L_{22}+L_{3}$, define $\mathcal{W}_{22}=\mathcal{J}^{'}\times\mathcal{J}^{''}\times \mathcal{K}_{2}$.
Thus the indexes $j_{i}^{'}$ and $j_{i}^{''}$
are determined by a given message $w_{22,i}$. Evenly partition $\mathcal{J}^{'''}$ into $\mathcal{K}_{2}$ bins, and the codeword
$v_{2}^{N}(j_{i}^{'},j_{i}^{''},j_{i}^{'''}|a_{i-1},w^{*}_{0,i})$ will be drawn from the bin $k_{2}$.

\end{itemize}

3) In the $i$-th block ($1\leq i\leq n$), the indexes $i_{i}^{'''}$ and $j_{i}^{'''}$ are determined as follows.

After the determination of $i_{i}^{'}$, $i_{i}^{''}$, $j_{i}^{'}$ and $j_{i}^{''}$, the transmitter tries to find a pair
$$(v_{1}^{N}(i_{i}^{'},i_{i}^{''},i_{i}^{'''}|a_{i-1},w^{*}_{0,i}), v_{2}^{N}(j_{i}^{'},j_{i}^{''},j_{i}^{'''}|a_{i-1},w^{*}_{0,i}))$$
such that $(u^{N}(w^{*}_{0,i}|a_{i-1}), x_{1}^{N}(a_{i-1}), v_{1}^{N}(i_{i}^{'},i_{i}^{''},
i_{i}^{'''}|a_{i-1},w^{*}_{0,i}), v_{2}^{N}(j_{i}^{'},j_{i}^{''},j_{i}^{'''}|a_{i-1},w^{*}_{0,i}))$ are jointly typical. If there
are more than one such pair, randomly choose one; if there is no such pair, an error is declared. Thus, all the indexes of
$v_{1}^{N}$ and $v_{2}^{N}$ (in block $i$) are determined. One can show that such a pair exists with high probability for sufficiently large $N$ if
(see \cite{Ma})
\begin{equation}\label{appen2.14}
I(V_{1};Y|U,X_{1})-\epsilon-R_{11}+I(V_{2};Z|U,X_{1})-\epsilon-R_{22}\geq I(V_{1};V_{2}|U,X_{1}).
\end{equation}

4) In the $i$-th block ($1\leq i\leq n$), the transmitter finally sends $x^{N}(a_{i-1},w_{0,i},w_{10,i},w_{20,i},w_{11,i},w_{22,i})$.

\item \textbf{(Relay encoder)}

The relay sends $x_{1}^{N}(1)$ at the first block, and $x_{1}^{N}(\hat{a}_{i-1})$ from block $2$ to $n$.

\end{itemize}

\textbf{Decoding:} Decoding proceeds as follows.

1) (At the relay) At the end of block $i$ ($1\leq i\leq n$), the relay already has an estimation of the $a_{i-1}$
(denoted as $\hat{a}_{i-1}$), which was sent at block $i-1$, and will declare that it receives $\hat{a}_{i}$, if this is the only triple such that
$(u^{N}(\hat{w}^{*}_{0,i}|\hat{a}_{i-1}),x_{1}^{N}(\hat{a}_{i-1}),y_{1}^{N}(i))$ are jointly typical. Here note that
$y_{1}^{N}(i)$ indicates the output sequence $y_{1}^{N}$ in block $i$, and $\hat{a}_{i}$ is the index of the bin
that $\hat{w}^{*}_{0,i}$ belongs to. Based on the AEP, the probability $Pr\{\hat{a}_{i}=a_{i}\}$ goes to $1$ if
\begin{equation}\label{appen2.15}
R_{0}+R_{10}+R_{20}\leq I(U;Y_{1}|X_{1}).
\end{equation}

2) (At receiver 1) Receiver 1 decodes from the last block, i.e., block $n$. Suppose that at the end of block $n-1$,
the relay decodes successfully, then receiver 1 will declare that $\check{a}_{n-1}$ is received if
$(x_{1}^{N}(\check{a}_{n-1}),y^{N}(n))$ jointly typical. By using (\ref{appen2.12}) and the AEP, it is easy to see that the probability
$Pr\{\check{a}_{n-1}=a_{n-1}\}$ goes to $1$. After getting $\check{a}_{n-1}$, receiver 1 can get an estimation of
$a_{i}$ ($1\leq i\leq n-2$) in a similar way.

Having $\check{a}_{i-1}$, receiver 1 can get the estimation of the message $w^{*}_{0,i}=(w_{0,i},w_{10,i},w_{20,i})$ by finding a unique triple
such that $(u^{N}(\check{w}^{*}_{0,i}|\check{a}_{i-1}),x_{1}^{N}(\check{a}_{i-1}),y^{N}(i))$ are jointly typical. Based on the AEP, the probability
$Pr\{\check{w}^{*}_{0,i}=w^{*}_{0,i}\}$ goes to $1$ if
\begin{equation}\label{appen2.16}
R_{0}+R_{10}+R_{20}-R_{r}\leq I(U;Y|X_{1}).
\end{equation}
After decoding $\check{w}^{*}_{0,i}$, receiver 1 tries to find a quadruple such that \\
$(v_{1}^{N}(\check{i}_{i}^{'},\check{i}_{i}^{''},\check{i}_{i}^{'''}|\check{a}_{i-1},\check{w}^{*}_{0,i}),
u^{N}(\check{w}^{*}_{0,i}|\check{a}_{i-1}),x_{1}^{N}(\check{a}_{i-1}),y^{N}(i))$ are jointly typical. Based on the AEP, the probability
$Pr\{\check{w}_{11,i}=w_{11,i}\}$ goes to $1$ if
\begin{equation}\label{appen2.17}
R_{11}\leq I(V_{1};Y|U,X_{1}).
\end{equation}
If such $v_{1}^{N}(\check{i}_{i}^{'},\check{i}_{i}^{''},\check{i}_{i}^{'''}|\check{a}_{i-1},\check{w}^{*}_{0,i})$ exists and is unique, set
$\check{i}_{i}^{'}=i_{i}^{'}$, $\check{i}_{i}^{''}=i_{i}^{''}$ and $\check{i}_{i}^{'''}=i_{i}^{'''}$; otherwise, declare an error.
From the values of $\check{i}_{i}^{'}$, $\check{i}_{i}^{''}$, $\check{i}_{i}^{'''}$, and the above encoding schemes, receiver 1 can calculate the message
$\check{w}_{11,i}$.

(At receiver 2) The decoding scheme for receiver 2 is symmetric, and it is omitted here. Analogously, we have
\begin{equation}\label{appen2.18}
R_{0}+R_{10}+R_{20}-R_{r}\leq I(U;Z|X_{1}),
\end{equation}
and
\begin{equation}\label{appen2.19}
R_{22}\leq I(V_{2};Z|U,X_{1}).
\end{equation}

By using (\ref{appen2.12}), (\ref{appen2.14}), (\ref{appen2.15}), (\ref{appen2.16}), (\ref{appen2.17}), (\ref{appen2.18})
and (\ref{appen2.19}), it is easy to check that $P_{e1}\leq\epsilon$ and $P_{e2}\leq\epsilon$.
Moreover, applying Fourier-Motzkin elimination on (\ref{appen2.12}), (\ref{appen2.14}), (\ref{appen2.15}), (\ref{appen2.16}), (\ref{appen2.17}), (\ref{appen2.18})
and (\ref{appen2.19}) with the definitions $R_{1}=R_{10}+R_{11}$ and $R_{2}=R_{20}+R_{22}$, we get

$$R_{0}\leq \min\{I(U;Y_{1}|X_{1}),I(U,X_{1};Y),I(U,X_{1};Z)\},$$

$$R_{0}+R_{1}\leq \min\{I(U;Y_{1}|X_{1}),I(U,X_{1};Y),I(U,X_{1};Z)\}+I(V_{1};Y|U,X_{1}),$$

$$R_{0}+R_{2}\leq \min\{I(U;Y_{1}|X_{1}),I(U,X_{1};Y),I(U,X_{1};Z)\}+I(V_{2};Z|U,X_{1}),$$

$$R_{0}+R_{1}+R_{2}\leq \min\{I(U;Y_{1}|X_{1}),I(U,X_{1};Y),I(U,X_{1};Z)\}+I(V_{1};Y|U,X_{1})+I(V_{2};Z|U,X_{1})-I(V_{1};V_{2}|U,X_{1}).$$

Note that the above inequalities are the same as those in Theorem \ref{T2}.

\textbf{Equivocation Analysis:} Now, it remains to prove $\lim_{N\rightarrow \infty}\Delta_{1}\geq R_{e1}=I(V_{1};Y|U,X_{1})-I(V_{1};
V_{2}|U,X_{1})-I(V_{1};Z|U,X_{1},V_{2})$.
The bound $\lim_{N\rightarrow \infty}\Delta_{2}\geq R_{e2}=I(V_{2};Z|U,X_{1})-I(V_{1};V_{2}|U,X_{1})-I(V_{2};Y|U,X_{1},V_{1})$ follows by symmetry.

\begin{eqnarray}\label{appen2.20}
H(W_{1}|Z^{N})&\geq& H(W_{1}|Z^{N},V_{2}^{N},U^{N},X_{1}^{N})\nonumber\\
&=&H(W_{10},W_{11}|Z^{N},V_{2}^{N},U^{N},X_{1}^{N})\nonumber\\
&\stackrel{(a)}=&H(W_{11}|Z^{N},V_{2}^{N},U^{N},X_{1}^{N})\nonumber\\
&=&H(W_{11},Z^{N}|V_{2}^{N},U^{N},X_{1}^{N})-H(Z^{N}|V_{2}^{N},U^{N},X_{1}^{N})\nonumber\\
&=&H(W_{11},Z^{N},V_{1}^{N}|V_{2}^{N},U^{N},X_{1}^{N})-H(V_{1}^{N}|W_{11},Z^{N},V_{2}^{N},U^{N},X_{1}^{N})-H(Z^{N}|V_{2}^{N},U^{N},X_{1}^{N})\nonumber\\
&\geq&H(Z^{N},V_{1}^{N}|V_{2}^{N},U^{N},X_{1}^{N})-H(V_{1}^{N}|W_{11},Z^{N},V_{2}^{N},U^{N},X_{1}^{N})-H(Z^{N}|V_{2}^{N},U^{N},X_{1}^{N})\nonumber\\
&=&H(V_{1}^{N}|V_{2}^{N},U^{N},X_{1}^{N})+H(Z^{N}|V_{1}^{N},V_{2}^{N},U^{N},X_{1}^{N})-H(V_{1}^{N}|W_{11},Z^{N},V_{2}^{N},U^{N},X_{1}^{N})\nonumber\\
&&-H(Z^{N}|V_{2}^{N},U^{N},X_{1}^{N})\nonumber\\
&=&H(V_{1}^{N}|U^{N},X_{1}^{N})-I(V_{1}^{N};V_{2}^{N}|U^{N},X_{1}^{N})-I(Z^{N};V_{1}^{N}|V_{2}^{N},U^{N},X_{1}^{N})\nonumber\\
&&-H(V_{1}^{N}|W_{11},Z^{N},V_{2}^{N},U^{N},X_{1}^{N}),
\end{eqnarray}
where (a) follows from the fact that given $U^{N}$, $W_{10}$ is uniquely determined.

Consider the first term in (\ref{appen2.20}), the codeword generation ensures that
\begin{equation}\label{appen2.21}
H(V_{1}^{N}|U^{N},X_{1}^{N})\geq\log2^{N(L_{11}+L_{12}+L_{3})}-\delta=N(I(V_{1};Y|U,X_{1})-\epsilon)-\delta,
\end{equation}
where $\delta$ is small for sufficiently large $N$.

For the second and third terms in (\ref{appen2.20}), using the same approach as that in \cite[Lemma 3]{CK}, we get
\begin{equation}\label{appen2.22}
I(V_{1}^{N};V_{2}^{N}|U^{N},X_{1}^{N})\leq N(I(V_{1};V_{2}|U,X_{1})+\epsilon^{'}),
\end{equation}
and
\begin{equation}\label{appen2.23}
I(Z^{N};V_{1}^{N}|V_{2}^{N},U^{N},X_{1}^{N})\leq N(I(V_{1};Z|U,X_{1},V_{2})+\epsilon^{''}),
\end{equation}
where $\epsilon^{'}, \epsilon^{''}\rightarrow 0$ as $N\rightarrow\infty$.

Now, we consider the last term of (\ref{appen2.20}). For the case that $R_{11}\leq L_{11}+L_{12}$, given $U^{N}$, $X_{1}^{N}$,
$V_{2}^{N}$ and $W_{11}$, the total number of possible codewords of $V_{1}^{N}$ is
\begin{equation}\label{appen2.24}
N_{1}\leq 2^{NL_{12}}=2^{NI(V_{1};Z|U,X_{1},V_{2})}.
\end{equation}
By using the Fano's inequality and (\ref{appen2.24}), we have
\begin{equation}\label{appen2.25}
H(V_{1}^{N}|W_{11},Z^{N},V_{2}^{N},U^{N},X_{1}^{N})\leq N\epsilon^{'''},
\end{equation}
where $\epsilon^{'''}\rightarrow 0$.

For the case that $L_{11}+L_{12}\leq R_{11}\leq L_{11}+L_{12}+L_{3}$, given $U^{N}$, $X_{1}^{N}$,
$V_{2}^{N}$ and $W_{11}$, $V_{1}^{N}$ is totally determined, and therefore
\begin{equation}\label{appen2.26}
H(V_{1}^{N}|W_{11},Z^{N},V_{2}^{N},U^{N},X_{1}^{N})=0.
\end{equation}

Substituting (\ref{appen2.21}), (\ref{appen2.22}), (\ref{appen2.23}) and (\ref{appen2.25}) (or (\ref{appen2.26})) into (\ref{appen2.20}),
and using the definition (\ref{e203}), we have
$\lim_{N\rightarrow \infty}\Delta_{1}\geq R_{e1}=I(V_{1};Y|U,X_{1})-I(V_{1};V_{2}|U,X_{1})-I(V_{1};Z|U,X_{1},V_{2})$.
This completes the proof of Theorem \ref{T2}.

\section{Proof of Theorem \ref{T3}\label{appen3}}

We consider the proof of Theorem \ref{T3} for the case $I(X_{1};Y)>I(X_{1};Z|U,V_{2})$, and the proof for $I(X_{1};Z)>I(X_{1};Y|U,V_{1})$
follows by symmetry.

In Theorem \ref{T3}, the relay node does not attempt to decode the messages but sends codewords that are independent
of the transmitter's messages, and these codewords aid in confusing the receivers. Since the channel between the
relay and receiver 1 is better than the channel between the relay and receiver 2 ($I(X_{1};Y)>I(X_{1};Z|U,V_{2})\geq I(X_{1};Z)$), we allow
receiver 1 to decode the relay codeword, and receiver 2 can not decode it. Therefore, in this case, the relay codeword
can be viewed as a noise signal to confuse receiver 2.

Now we will prove that the quintuple $(R_{0},R_{1},R_{2},R_{e1},R_{e2})\in \mathcal{R}^{(Ai2)}$ with the conditions
\begin{equation}\label{appen3.1}
R_{e1}=\min\{I(X_{1};Z|U,V_{1},V_{2}), I(X_{1};Y)\}+I(V_{1};Y|U,X_{1})-I(V_{1};V_{2}|U)-I(X_{1},V_{1};Z|U,V_{2}),
\end{equation}
and
\begin{equation}\label{appen3.2}
R_{e2}=I(V_{2};Z|U)-I(V_{1};V_{2}|U)-I(V_{2};Y|U,X_{1},V_{1}),
\end{equation}
is achievable.

Similar to the proof of Theorem \ref{T2}, we split the confidential message $W_{1}$ into $W_{10}$ and $W_{11}$, and $W_{2}$ into $W_{20}$
and $W_{22}$, and the definitions of these messages are the same as those in Appendix \ref{appen2}. Here note that the formulas
(\ref{appen3.1}) and (\ref{appen3.2}) combined with the rate splitting and the fact that $W_{10}$ and $W_{20}$ are decoded by both receivers
ensure that,
\begin{equation}\label{appen3.3}
R_{11}\geq R_{e1}=\min\{I(X_{1};Z|U,V_{1},V_{2}), I(X_{1};Y)\}+I(V_{1};Y|U,X_{1})-I(V_{1};V_{2}|U)-I(X_{1},V_{1};Z|U,V_{2}),
\end{equation}
and
\begin{equation}\label{appen3.4}
R_{22}\geq R_{e2}=I(V_{2};Z|U)-I(V_{1};V_{2}|U)-I(V_{2};Y|U,X_{1},V_{1}).
\end{equation}

\textbf{Code Construction:} Fix the joint probability mass function
\begin{eqnarray*}
&&P_{Y,Z,Y_{1},X,X_{1},V_{1},V_{2},U}(y,z,y_{1},x,x_{1},v_{1},v_{2},u)=P_{Y,Z,Y_{1}|X,X_{1}}(y,z,y_{1}|x,x_{1})
P_{X|U,V_{1},V_{2}}(x|u,v_{1},v_{2})P_{U,V_{1},V_{2}}(u,v_{1},v_{2})P_{X_{1}}(x_{1}).
\end{eqnarray*}
For arbitrary $\epsilon> 0$, define
\begin{equation}\label{appen3.50}
L_{11}=I(V_{1};Y|U,X_{1})-I(V_{1};V_{2}|U)-I(V_{1};Z|U,V_{2}),
\end{equation}
\begin{equation}\label{appen3.5}
L_{12}=I(V_{1};Z|U,V_{2}),
\end{equation}
\begin{equation}\label{appen3.6}
L_{21}=I(V_{2};Z|U)-I(V_{1};V_{2}|U)-I(V_{2};Y|U,X_{1},V_{1}),
\end{equation}
\begin{equation}\label{appen3.7}
L_{22}=I(V_{2};Y|U,X_{1},V_{1}),
\end{equation}
\begin{equation}\label{appen3.8}
L_{3}=I(V_{1};V_{2}|U)-\epsilon.
\end{equation}
Note that
\begin{equation}\label{appen3.9}
L_{11}+L_{12}+L_{3}=I(V_{1};Y|U,X_{1})-\epsilon,
\end{equation}
\begin{equation}\label{appen3.10}
L_{21}+L_{22}+L_{3}=I(V_{2};Z|U)-\epsilon,
\end{equation}
\begin{equation}\label{appen3.10x}
L_{11}\geq R_{e1}.
\end{equation}

\begin{itemize}

\item First, generate at random $2^{NR_{r}}$ i.i.d. sequences at the relay node each drawn according to
$p_{X_{1}^{N}}(x_{1}^{N})=\prod_{i=1}^{N}p_{X_{1}}(x_{1,i})$, index them as $x_{1}^{N}(a)$, $a\in [1,2^{NR_{r}}]$, where
\begin{equation}\label{appen3.13}
R_{r}=\min\{I(X_{1};Z|U,V_{1},V_{2}), I(X_{1};Y)\}-\epsilon,
\end{equation}
and $\epsilon\rightarrow 0^{+}$.
Note that $I(X_{1};Z|U,V_{2})\leq I(X_{1};Z|U,V_{1},V_{2})$ and $I(X_{1};Z|U,V_{2})\leq I(X_{1};Y)$, and thus
\begin{equation}\label{appen3.13xx}
R_{r}\geq I(X_{1};Z|U,V_{2})-\epsilon,
\end{equation}
and
\begin{equation}\label{appen3.13xxx}
R_{r}\leq I(X_{1};Z|U,V_{1},V_{2})-\epsilon.
\end{equation}

\item Generate at random $2^{N(R_{10}+R_{20}+R_{0})}$ i.i.d. sequences $u^{N}(b)$ ($b\in [1,2^{N(R_{10}+R_{20}+R_{0})}]$)
according to $\prod_{i=1}^{N}p_{U}(u_{i})$.

\item For the transmitted sequence $u^{N}(b)$, generate $2^{N(L_{11}+L_{12}+L_{3})}$ i.i.d. sequences
$v_{1}^{N}(i^{'},i^{''},i^{'''})$, with $i^{'}\in \mathcal{I}^{'}=[1,2^{NL_{11}}]$, $i^{''}\in \mathcal{I}^{''}=[1,2^{NL_{12}}]$
and $i^{'''}\in \mathcal{I}^{'''}=[1,2^{NL_{3}}]$, according to $\prod_{i=1}^{N}p_{V_{1}|U}(v_{1,i}|u_{i})$.

\item Similarly, for the transmitted sequences $u^{N}$ and $x_{1}^{N}$, generate $2^{N(L_{21}+L_{22}+L_{3})}$ i.i.d. sequences
$v_{2}^{N}(j^{'},j^{''},j^{'''})$, with $j^{'}\in \mathcal{J}^{'}=[1,2^{NL_{21}}]$, $j^{''}\in \mathcal{J}^{''}=[1,2^{NL_{22}}]$
and $j^{'''}\in \mathcal{J}^{'''}=[1,2^{NL_{3}}]$, according to $\prod_{i=1}^{N}p_{V_{2}|U}(v_{2,i}|u_{i})$.

\item The $x^{N}$ is generated according to a new discrete memoryless channel (DMC) with inputs $u^{N}$, $v_{1}^{N}$,
$v_{2}^{N}$ and output $x^{N}$. The transition probability of this new DMC is $p_{X|U,V_{1},V_{2}}(x|u,v_{1},v_{2})$.
The probability $p_{X^{N}|U^{N},V_{1}^{N},V_{2}^{N}}(x^{N}|u^{N},v_{1}^{N},v_{2}^{N})$ is calculated as follows.
\begin{equation}\label{appen3.14}
p_{X^{N}|U^{N},V_{1}^{N},V_{2}^{N}}(x^{N}|u^{N},v_{1}^{N},v_{2}^{N})=
\prod_{i=1}^{N}p_{X|U,V_{1},V_{2}}(x_{i}|u_{i},v_{1,i},v_{2,i}).
\end{equation}
Denote $x^{N}$ by $x^{N}(w_{0},w_{10},w_{20},w_{11},w_{22})$.

\end{itemize}

\textbf{Encoding:} Similar to the definitions in Appendix \ref{appen2}, define $w^{*}_{0,i}=(w_{0,i},w_{10,i},w_{20,i})$,
where $w_{0,i}$, $w_{10,i}$ and $w_{20,i}$ are the messages transmitted in the $i$-th block.
The messages $w_{11}$ and $w_{22}$ transmitted in the $i$-th block are denoted
by $w_{11,i}$ and $w_{22,i}$, respectively.

\begin{itemize}

\item \textbf{(Channel encoder)}

1) The transmitter sends $(u^{N}(w^{*}_{0,i}),v_{1}^{N}(i_{i}^{'},i_{i}^{''},i_{i}^{'''}|w^{*}_{0,i}),
v_{2}^{N}(j_{i}^{'},j_{i}^{''},j_{i}^{'''}|w^{*}_{0,i}))$ for the $i$-th block ($1\leq i\leq n$).
Here $i_{i}^{'}$, $i_{i}^{''}$, $i_{i}^{'''}$, $j_{i}^{'}$, $j_{i}^{''}$ and $j_{i}^{'''}$
are the indexes for block $i$.

2) The indexes $i_{i}^{'}$, $i_{i}^{''}$, $j_{i}^{'}$ and $j_{i}^{''}$ are determined by the following methods.

\begin{itemize}

\item If $R_{11}\leq L_{11}$, evenly partition $\mathcal{I}^{'}$ into $\mathcal{W}_{11}$ bins, and the index $i_{i}^{'}$
is drawn at random (with uniform distribution) from the bin $w_{11}$. The index $i_{i}^{''}$
is drawn at random (with uniform distribution) from $\mathcal{I}^{''}$.

Note that $R_{22}$ always satisfies $R_{22}\geq L_{21}$.

\item If $L_{11}\leq R_{11}\leq L_{11}+L_{12}$, define $\mathcal{W}_{11}=\mathcal{I}^{'}\times \mathcal{K}_{1}$. Thus the index $i_{i}^{'}$
is determined by a given message $w_{11,i}$. Evenly partition $\mathcal{I}^{''}$ into $\mathcal{K}_{1}$ bins, and the index $i_{i}^{''}$
is drawn at random (with uniform distribution) from the bin $k_{1}$.

Analogously, if $R_{22}\leq L_{21}+L_{22}$, define $\mathcal{W}_{22}=\mathcal{J}^{'}\times \mathcal{K}_{2}$. Thus the index $j_{i}^{'}$
is determined by a given message $w_{22,i}$. Evenly partition $\mathcal{J}^{''}$ into $\mathcal{K}_{2}$ bins, and the index $j_{i}^{''}$
is drawn at random (with uniform distribution) from the bin $k_{2}$.

\item If $L_{11}+L_{12}\leq R_{11}\leq L_{11}+L_{12}+L_{3}$, define $\mathcal{W}_{11}=\mathcal{I}^{'}\times\mathcal{I}^{''}\times \mathcal{K}_{1}$.
Thus the indexes $i_{i}^{'}$ and $i_{i}^{''}$
are determined by a given message $w_{11,i}$. Evenly partition $\mathcal{I}^{'''}$ into $\mathcal{K}_{1}$ bins, and the codeword
$v_{1}^{N}(i_{i}^{'},i_{i}^{''},i_{i}^{'''}|w^{*}_{0,i})$ will be drawn from the bin $k_{1}$.

Analogously, if $L_{21}+L_{22}\leq R_{22}\leq L_{21}+L_{22}+L_{3}$, define $\mathcal{W}_{22}=\mathcal{J}^{'}\times\mathcal{J}^{''}\times \mathcal{K}_{2}$.
Thus the indexes $j_{i}^{'}$ and $j_{i}^{''}$
are determined by a given message $w_{22,i}$. Evenly partition $\mathcal{J}^{'''}$ into $\mathcal{K}_{2}$ bins, and the codeword
$v_{2}^{N}(j_{i}^{'},j_{i}^{''},j_{i}^{'''}|w^{*}_{0,i})$ will be drawn from the bin $k_{2}$.

\end{itemize}

3) The indexes $i_{i}^{'''}$ and $j_{i}^{'''}$ are determined as follows.

After the determination of $i_{i}^{'}$, $i_{i}^{''}$, $j_{i}^{'}$ and $j_{i}^{''}$, the transmitter tries to find a pair
$(v_{1}^{N}(i_{i}^{'},i_{i}^{''},i_{i}^{'''}|w^{*}_{0,i}), v_{2}^{N}(j_{i}^{'},j_{i}^{''},j_{i}^{'''}|w^{*}_{0,i}))$
such that $(u^{N}(w^{*}_{0,i}), v_{1}^{N}(i_{i}^{'},i_{i}^{''},
i_{i}^{'''}|w^{*}_{0,i}), v_{2}^{N}(j_{i}^{'},j_{i}^{''},j_{i}^{'''}|w^{*}_{0,i}))$ are jointly typical. If there
are more than one such pair, randomly choose one; if there is no such pair, an error is declared. Thus, all the indexes of
$v_{1}^{N}$ and $v_{2}^{N}$ (in block $i$) are determined. One can show that such a pair exists with high probability for sufficiently large $N$ if
(see \cite{Ma})
\begin{equation}\label{appen3.15}
I(V_{1};Y|U,X_{1})-\epsilon-R_{11}+I(V_{2};Z|U)-\epsilon-R_{22}\geq I(V_{1};V_{2}|U).
\end{equation}

4) The transmitter finally sends $x^{N}(w_{0,i},w_{10,i},w_{20,i},w_{11,i},w_{22,i})$.

\item \textbf{(Relay encoder)}

In the $i$-th block, the relay uniformly picks a codeword $x_{1}^{N}(a_{i})$ from $a_{i}\in [1,2^{NR_{r}}]$, and sends
$x_{1}^{N}(a_{i})$.

\end{itemize}

\textbf{Decoding:} Decoding proceeds as follows.

(At receiver 1) At the end of block $i$, receiver 1 will declare that $\check{a}_{i}$ is received if
$(x_{1}^{N}(\check{a}_{i}),y^{N}(i))$ are jointly typical. By using (\ref{appen3.13}) and the AEP,
it is easy to see that the probability $Pr\{\check{a}_{i}=a_{i}\}$ goes to $1$.

Having $\check{a}_{i}$, receiver 1 can get the estimation of the message $w^{*}_{0,i}=(w_{0,i},w_{10,i},w_{20,i})$ by finding a unique triple
such that $(u^{N}(\check{w}^{*}_{0,i}),x_{1}^{N}(\check{a}_{i}),y^{N}(i))$ are jointly typical. Based on the AEP, the probability
$Pr\{\check{w}^{*}_{0,i}=w^{*}_{0,i}\}$ goes to $1$ if
\begin{equation}\label{appen3.16}
R_{0}+R_{10}+R_{20}\leq I(U;Y|X_{1}).
\end{equation}
After decoding $\check{w}^{*}_{0,i}$, receiver 1 tries to find a quadruple such that \\
$(v_{1}^{N}(\check{i}_{i}^{'},\check{i}_{i}^{''},\check{i}_{i}^{'''}|\check{w}^{*}_{0,i}),
u^{N}(\check{w}^{*}_{0,i}),x_{1}^{N}(\check{a}_{i}),y^{N}(i))$ are jointly typical. Based on the AEP, the probability
$Pr\{\check{w}_{11,i}=w_{11,i}\}$ goes to $1$ if
\begin{equation}\label{appen3.17}
R_{11}\leq I(V_{1};Y|U,X_{1}).
\end{equation}
If such $v_{1}^{N}(\check{i}_{i}^{'},\check{i}_{i}^{''},\check{i}_{i}^{'''}|\check{w}^{*}_{0,i})$ exists and is unique, set
$\check{i}_{i}^{'}=i_{i}^{'}$, $\check{i}_{i}^{''}=i_{i}^{''}$ and $\check{i}_{i}^{'''}=i_{i}^{'''}$; otherwise, declare an error.
From the values of $\check{i}_{i}^{'}$, $\check{i}_{i}^{''}$, $\check{i}_{i}^{'''}$, and the above encoding schemes, receiver 1 can calculate the message
$\check{w}_{11,i}$.

(At receiver 2) The decoding scheme for receiver 2 is as follows.

Receiver 2 gets the estimation of the message $w^{*}_{0,i}$ by finding a unique pair such that $(u^{N}(\hat{w}^{*}_{0,i}),z^{N}(i))$ are
jointly typical. Based on the AEP, the probability $Pr\{\hat{w}^{*}_{0,i}=w^{*}_{0,i}\}$ goes to 1 if
\begin{equation}\label{appen3.18}
R_{0}+R_{10}+R_{20}\leq I(U;Z).
\end{equation}
After decoding $\hat{w}^{*}_{0,i}$, receiver 2 tries to find a triple such that
$(v_{2}^{N}(\hat{j}_{i}^{'},\hat{j}_{i}^{''},\hat{j}_{i}^{'''}|\hat{w}^{*}_{0,i}),
u^{N}(\hat{w}^{*}_{0,i}),z^{N}(i))$ are jointly typical. Based on the AEP, the probability
$Pr\{\hat{w}_{22,i}=w_{22,i}\}$ goes to $1$ if
\begin{equation}\label{appen3.19}
R_{22}\leq I(V_{2};Z|U).
\end{equation}
If such $v_{2}^{N}(\hat{j}_{i}^{'},\hat{j}_{i}^{''},\hat{j}_{i}^{'''}|\hat{w}^{*}_{0,i})$ exists and is unique, set
$\hat{j}_{i}^{'}=j_{i}^{'}$, $\hat{j}_{i}^{''}=j_{i}^{''}$ and $\hat{j}_{i}^{'''}=j_{i}^{'''}$; otherwise, declare an error.
From the values of $\hat{j}_{i}^{'}$, $\hat{j}_{i}^{''}$, $\hat{j}_{i}^{'''}$, and the above encoding schemes, receiver 2 can calculate the message
$\hat{w}_{22,i}$.

By using (\ref{appen3.13}), (\ref{appen3.15}), (\ref{appen3.16}), (\ref{appen3.17}), (\ref{appen3.18})
and (\ref{appen3.19}), it is easy to check that $P_{e1}\leq\epsilon$ and $P_{e2}\leq\epsilon$.
Moreover, applying Fourier-Motzkin elimination on (\ref{appen3.13}), (\ref{appen3.15}), (\ref{appen3.16}), (\ref{appen3.17}), (\ref{appen3.18})
and (\ref{appen3.19}) with the definitions $R_{1}=R_{10}+R_{11}$ and $R_{2}=R_{20}+R_{22}$, we get

$$R_{0}\leq \min\{I(U;Y|X_{1}),I(U;Z)\},$$

$$R_{0}+R_{1}\leq \min\{I(U;Y|X_{1}),I(U;Z)\}+I(V_{1};Y|U,X_{1}),$$

$$R_{0}+R_{2}\leq \min\{I(U;Y|X_{1}),I(U;Z)\}+I(V_{2};Z|U),$$

$$R_{0}+R_{1}+R_{2}\leq \min\{I(U;Y|X_{1}),I(U;Z)\}+I(V_{1};Y|U,X_{1})+I(V_{2};Z|U)-I(V_{1};V_{2}|U).$$

Note that the above inequalities are the same as those in Theorem \ref{T3}.

\textbf{Equivocation Analysis:} Now, it remains to prove $\lim_{N\rightarrow \infty}\Delta_{1}\geq R_{e1}
=\min\{I(X_{1};Z|U,V_{1},V_{2}), I(X_{1};Y)\}+
I(V_{1};Y|U,X_{1})-I(V_{1};V_{2}|U)-I(X_{1},V_{1};Z|U,V_{2})$
and $\lim_{N\rightarrow \infty}\Delta_{2}\geq R_{e2}=I(V_{2};Z|U)-I(V_{1};V_{2}|U)-I(V_{2};Y|U,X_{1},V_{1})$.

\textbf{Proof of} $\lim_{N\rightarrow \infty}\Delta_{1}\geq R_{e1}=\min\{I(X_{1};Z|U,V_{1},V_{2}), I(X_{1};Y)\}+
I(V_{1};Y|U,X_{1})-I(V_{1};V_{2}|U)-I(X_{1},V_{1};Z|U,V_{2})$:

\begin{eqnarray}\label{appen3.20}
H(W_{1}|Z^{N})&\geq& H(W_{1}|Z^{N},V_{2}^{N},U^{N})\nonumber\\
&=&H(W_{10},W_{11}|Z^{N},V_{2}^{N},U^{N})\nonumber\\
&\stackrel{(a)}=&H(W_{11}|Z^{N},V_{2}^{N},U^{N})\nonumber\\
&=&H(W_{11},Z^{N}|V_{2}^{N},U^{N})-H(Z^{N}|V_{2}^{N},U^{N})\nonumber\\
&=&H(W_{11},Z^{N},V_{1}^{N},X_{1}^{N}|V_{2}^{N},U^{N})-H(V_{1}^{N},X_{1}^{N}|W_{11},Z^{N},V_{2}^{N},U^{N})-H(Z^{N}|V_{2}^{N},U^{N})\nonumber\\
&\geq&H(Z^{N},V_{1}^{N},X_{1}^{N}|V_{2}^{N},U^{N})-H(V_{1}^{N},X_{1}^{N}|W_{11},Z^{N},V_{2}^{N},U^{N})-H(Z^{N}|V_{2}^{N},U^{N})\nonumber\\
&=&H(V_{1}^{N},X_{1}^{N}|V_{2}^{N},U^{N})+H(Z^{N}|V_{1}^{N},V_{2}^{N},U^{N},X_{1}^{N})-H(V_{1}^{N},X_{1}^{N}|W_{11},Z^{N},V_{2}^{N},U^{N})\nonumber\\
&&-H(Z^{N}|V_{2}^{N},U^{N})\nonumber\\
&\stackrel{(b)}=&H(X_{1}^{N})+H(V_{1}^{N}|V_{2}^{N},U^{N})+H(Z^{N}|V_{1}^{N},V_{2}^{N},U^{N},X_{1}^{N})-H(V_{1}^{N},X_{1}^{N}|W_{11},Z^{N},V_{2}^{N},U^{N})\nonumber\\
&&-H(Z^{N}|V_{2}^{N},U^{N})\nonumber\\
&=&H(X_{1}^{N})+H(V_{1}^{N}|U^{N})-I(V_{1}^{N};V_{2}^{N}|U^{N})+H(Z^{N}|V_{1}^{N},V_{2}^{N},U^{N},X_{1}^{N})\nonumber\\
&&-H(V_{1}^{N},X_{1}^{N}|W_{11},Z^{N},V_{2}^{N},U^{N})-H(Z^{N}|V_{2}^{N},U^{N})\nonumber\\
&=&H(X_{1}^{N})+H(V_{1}^{N}|U^{N})-I(V_{1}^{N};V_{2}^{N}|U^{N})-I(Z^{N};X_{1}^{N},V_{1}^{N}|V_{2}^{N},U^{N})\nonumber\\
&&-H(V_{1}^{N},X_{1}^{N}|W_{11},Z^{N},V_{2}^{N},U^{N}),
\end{eqnarray}
where (a) follows from the fact that given $U^{N}$, $W_{10}$ is uniquely determined, and (b) is from that $X_{1}^{N}$ is independent
of $V_{1}^{N}$, $V_{2}^{N}$ and $U^{N}$.

Consider the first term in (\ref{appen3.20}), the codeword generation ensures that
\begin{equation}\label{appen3.21}
H(X_{1}^{N})\geq NR_{r}-\delta=N(\min\{I(X_{1};Z|U,V_{1},V_{2}), I(X_{1};Y)\}-\epsilon)-\delta,
\end{equation}
where $\delta$ is small for sufficiently large $N$.

For the second term in (\ref{appen3.20}), similarly we have
\begin{equation}\label{appen3.22}
H(V_{1}^{N}|U^{N})\geq\log2^{N(L_{11}+L_{12}+L_{3})}-\delta_{1}=N(I(V_{1};Y|U,X_{1})-\epsilon)-\delta_{1},
\end{equation}
where $\delta_{1}$ is small for sufficiently large $N$.

For the third and fourth terms in (\ref{appen3.20}), using the same approach as that in \cite[Lemma 3]{CK}, we get
\begin{equation}\label{appen3.23}
I(V_{1}^{N};V_{2}^{N}|U^{N})\leq N(I(V_{1};V_{2}|U)+\epsilon^{'}),
\end{equation}
and
\begin{equation}\label{appen3.24}
I(Z^{N};X_{1}^{N},V_{1}^{N}|V_{2}^{N},U^{N})\leq N(I(X_{1},V_{1};Z|U,V_{2})+\epsilon^{''}),
\end{equation}
where $\epsilon^{'}, \epsilon^{''}\rightarrow 0$ as $N\rightarrow\infty$.

Now, we consider the last term of (\ref{appen3.20}).
Given $W_{11}$, receiver 2 can do joint decoding.

\begin{itemize}

\item For the case that $R_{11}\leq L_{11}$, given $U^{N}$,
$V_{2}^{N}$, $W_{11}$ and $\epsilon^{'''}\rightarrow 0^{+}$,
\begin{equation}\label{appen3.26xx}
H(V_{1}^{N},X_{1}^{N}|W_{11},Z^{N},V_{2}^{N},U^{N})\leq N\epsilon^{'''},
\end{equation}
is guaranteed if $R_{r}\leq I(X_{1};Z|V_{1},V_{2},U)-\epsilon$ and $R_{r}\geq I(X_{1};Z|U,V_{2})-\epsilon$ ($\epsilon\rightarrow 0^{+}$), and this
is from the properties of AEP (similar argument is used in the proof of Theorem 3 in \cite{LG}). By using
(\ref{appen3.13xx}) and (\ref{appen3.13xxx}), (\ref{appen3.26xx}) is obtained.

\item For the case that $L_{11}\leq R_{11}\leq L_{11}+L_{12}$,
given $U^{N}$,
$V_{2}^{N}$ and $W_{11}$, the total number of possible codewords of $V_{1}^{N}$ is
\begin{equation}\label{appen3.25}
N_{1}\leq 2^{NL_{12}}=2^{NI(V_{1};Z|U,V_{2})}.
\end{equation}
By using the Fano's inequality and (\ref{appen3.25}), we have
\begin{equation}\label{appen3.26}
H(V_{1}^{N}|W_{11},Z^{N},V_{2}^{N},U^{N})\leq N\epsilon^{'''},
\end{equation}
where $\epsilon^{'''}\rightarrow 0$.

Given $U^{N}$, $V_{1}^{N}$, $V_{2}^{N}$ and $W_{11}$, the total number of possible codewords of $X_{1}^{N}$ is
\begin{equation}\label{appen3.27}
N_{2}\leq 2^{NR_{r}}=2^{N(\min\{I(X_{1};Y),I(X_{1};Z|V_{1},V_{2},U)\}-\epsilon)}.
\end{equation}
By using the Fano's inequality and (\ref{appen3.27}), we have
\begin{equation}\label{appen3.28}
H(X_{1}^{N}|W_{11},Z^{N},V_{1}^{N},V_{2}^{N},U^{N})\leq N\epsilon^{''''},
\end{equation}
where $\epsilon^{''''}\rightarrow 0$.

By using (\ref{appen3.26}) and (\ref{appen3.28}),
\begin{equation}\label{appen3.28xx}
\frac{1}{N}H(V_{1}^{N},X_{1}^{N}|W_{11},Z^{N},V_{2}^{N},U^{N})\leq \epsilon\rightarrow 0,
\end{equation}
is guaranteed.

\item For the case that $L_{11}+L_{12}\leq R_{11}\leq L_{11}+L_{12}+L_{3}$, given $U^{N}$,
$V_{2}^{N}$ and $W_{11}$, $V_{1}^{N}$ is totally determined, and therefore
\begin{equation}\label{appen3.29}
H(V_{1}^{N}|W_{11},Z^{N},V_{2}^{N},U^{N})=0.
\end{equation}
Similarly, note that $R_{r}=\min\{I(X_{1};Z|U,V_{1},V_{2}), I(X_{1};Y)\}-\epsilon$, by using the Fano's inequality, we have
(\ref{appen3.28}). Thus
\begin{equation}\label{appen3.28xxx}
\frac{1}{N}H(V_{1}^{N},X_{1}^{N}|W_{11},Z^{N},V_{2}^{N},U^{N})\leq \epsilon\rightarrow 0
\end{equation}
is guaranteed.

\end{itemize}

Substituting (\ref{appen3.21}), (\ref{appen3.22}), (\ref{appen3.23}), (\ref{appen3.24}) and (\ref{appen3.26xx}) (or (\ref{appen3.28xx}), (\ref{appen3.28xxx}))
into (\ref{appen3.20}),
and using the definition (\ref{e203}), we have
$\lim_{N\rightarrow \infty}\Delta_{1}\geq R_{e1}=\min\{I(X_{1};Z|U,V_{1},V_{2}), I(X_{1};Y)\}+
I(V_{1};Y|U,X_{1})-I(V_{1};V_{2}|U)-I(X_{1},V_{1};Z|U,V_{2})$.

\textbf{Proof of} $\lim_{N\rightarrow \infty}\Delta_{2}\geq R_{e2}=I(V_{2};Z|U)-I(V_{1};V_{2}|U)-I(V_{2};Y|U,X_{1},V_{1})$:

\begin{eqnarray}\label{appen3.30}
H(W_{2}|Y^{N})&\geq& H(W_{2}|Y^{N},V_{1}^{N},U^{N},X_{1}^{N})\nonumber\\
&=&H(W_{20},W_{22}|Y^{N},V_{1}^{N},U^{N},X_{1}^{N})\nonumber\\
&\stackrel{(a)}=&H(W_{22}|Y^{N},V_{1}^{N},U^{N},X_{1}^{N})\nonumber\\
&=&H(W_{22},Y^{N}|V_{1}^{N},U^{N},X_{1}^{N})-H(Y^{N}|V_{1}^{N},U^{N},X_{1}^{N})\nonumber\\
&=&H(W_{22},Y^{N},V_{2}^{N}|V_{1}^{N},U^{N},X_{1}^{N})-H(V_{2}^{N}|W_{22},Y^{N},V_{1}^{N},U^{N},X_{1}^{N})-H(Y^{N}|V_{1}^{N},U^{N},X_{1}^{N})\nonumber\\
&\geq&H(Y^{N},V_{2}^{N}|V_{1}^{N},U^{N},X_{1}^{N})-H(V_{2}^{N}|W_{22},Y^{N},V_{1}^{N},U^{N},X_{1}^{N})-H(Y^{N}|V_{1}^{N},U^{N},X_{1}^{N})\nonumber\\
&=&H(V_{2}^{N}|V_{1}^{N},U^{N},X_{1}^{N})+H(Y^{N}|V_{2}^{N},V_{1}^{N},U^{N},X_{1}^{N})-H(V_{2}^{N}|W_{22},Y^{N},V_{1}^{N},U^{N},X_{1}^{N})\nonumber\\
&&-H(Y^{N}|V_{1}^{N},U^{N},X_{1}^{N})\nonumber\\
&\stackrel{(b)}=&H(V_{2}^{N}|U^{N})-I(V_{1}^{N};V_{2}^{N}|U^{N})-I(Y^{N};V_{2}^{N}|V_{1}^{N},U^{N},X_{1}^{N})\nonumber\\
&&-H(V_{2}^{N}|W_{22},Y^{N},V_{1}^{N},U^{N},X_{1}^{N}),
\end{eqnarray}
where (a) follows from the fact that given $U^{N}$, $W_{20}$ is uniquely determined, and (b) is from that $X_{1}^{N}$ is independent
of $V_{1}^{N}$, $V_{2}^{N}$ and $U^{N}$.

For the first term in (\ref{appen3.30}), we have
\begin{equation}\label{appen3.31}
H(V_{2}^{N}|U^{N})\geq\log2^{N(L_{21}+L_{22}+L_{3})}-\delta_{3}=N(I(V_{2};Z|U)-\epsilon)-\delta_{3},
\end{equation}
where $\delta_{3}$ is small for sufficiently large $N$.

For the second and third terms in (\ref{appen3.30}), using the same approach as that in \cite[Lemma 3]{CK}, we get
\begin{equation}\label{appen3.32}
I(V_{1}^{N};V_{2}^{N}|U^{N})\leq N(I(V_{1};V_{2}|U)+\epsilon^{'}),
\end{equation}
and
\begin{equation}\label{appen3.33}
I(Y^{N};V_{2}^{N}|V_{1}^{N},U^{N},X_{1}^{N})\leq N(I(V_{2};Y|U,V_{1},X_{1})+\epsilon^{''}),
\end{equation}
where $\epsilon^{'}, \epsilon^{''}\rightarrow 0$ as $N\rightarrow\infty$.

Now, we consider the last term of (\ref{appen3.30}).

\begin{itemize}

\item For the case that $R_{22}\leq L_{21}+L_{22}$, given $U^{N}$,
$V_{1}^{N}$ and $W_{22}$, the total number of possible codewords of $V_{2}^{N}$ is
\begin{equation}\label{appen3.34}
N_{3}\leq 2^{NL_{22}}=2^{NI(V_{2};Y|U,X_{1},V_{1})}.
\end{equation}
By using the Fano's inequality and (\ref{appen3.34}), we have
\begin{equation}\label{appen3.35}
H(V_{2}^{N}|W_{22},Y^{N},V_{1}^{N},U^{N},X_{1}^{N})\leq N\epsilon^{'''},
\end{equation}
where $\epsilon^{'''}\rightarrow 0$.

\item For the case that $L_{21}+L_{22}\leq R_{22}\leq L_{21}+L_{22}+L_{3}$, given $U^{N}$,
$V_{1}^{N}$ and $W_{22}$, $V_{2}^{N}$ is totally determined, and therefore
\begin{equation}\label{appen3.36}
H(V_{2}^{N}|W_{22},Y^{N},V_{1}^{N},U^{N},X_{1}^{N})=0.
\end{equation}

\end{itemize}

Substituting (\ref{appen3.31}), (\ref{appen3.32}), (\ref{appen3.33}) and (\ref{appen3.35})
(or (\ref{appen3.36})) into (\ref{appen3.30}),
and using the definition (\ref{e203}), we have
$\lim_{N\rightarrow \infty}\Delta_{2}\geq R_{e2}=I(V_{2};Z|U)-I(V_{1};V_{2}|U)-I(V_{2};Y|U,X_{1},V_{1})$.
This completes the proof for Theorem \ref{T3}.

\section{Proof of Theorem \ref{T9}\label{appen11.1}}

The auxiliary random variables in $\mathcal{R}^{(Co)}$ are defined by
\begin{eqnarray*}
&&Q\triangleq Y_{1}^{J-1}, U\triangleq (Y^{J-1}, W_{0}, Z_{J+1}^{N}, J), V\triangleq (U, W_{1}), Y\triangleq Y_{J}, Z\triangleq Z_{J},
\end{eqnarray*}
where $J$ is a random variable (uniformly distributed over $\{1, 2, ,...,N\}$), and it is independent of
$X^{N}$, $X_{1}^{N}$, $Y^{N}$, $Y_{1}^{N}$, $Z^{N}$, $W_{0}$ and $W_{1}$.
From the above definitions, it is easy to see that the relay $X_{1}$ is represented by the auxiliary random variables $Q$. The common message $W_{0}$ is represented
by $U$, and the confidential message $W_{1}$ is represented by $V$. Now it remains to prove the inequalities of Theorem \ref{T9}, see the followings.

\textbf{Proof of $R_{0}\leq \min\{I(U,Q;Z),I(U;Y_{1}|Q)\}$:}

Note that
\begin{eqnarray}\label{exxc1.1}
&&\frac{1}{N}H(W_{0})\stackrel{(1)}\leq \frac{1}{N}I(W_{0};Z^{N})+\frac{\delta(\epsilon)}{N}\nonumber\\
&&=\frac{1}{N}\sum_{i=1}^{N}I(W_{0};Z_{i}|Z_{i+1}^{N})+\frac{\delta(\epsilon)}{N}\nonumber\\
&&\leq \frac{1}{N}\sum_{i=1}^{N}(H(Z_{i})-H(Z_{i}|Z_{i+1}^{N},W_{0},Y^{i-1},Y_{1}^{i-1}))+\frac{\delta(\epsilon)}{N}\nonumber\\
&&=\frac{1}{N}\sum_{i=1}^{N}(H(Z_{i}|J=i)-H(Z_{i}|Z_{i+1}^{N},W_{0},Y^{i-1},Y_{1}^{i-1},J=i))+\frac{\delta(\epsilon)}{N}\nonumber\\
&&=H(Z_{J}|J)-H(Z_{J}|Z_{J+1}^{N},W_{0},Y^{J-1},Y_{1}^{J-1},J)+\frac{\delta(\epsilon)}{N}\nonumber\\
&&\stackrel{(2)}\leq I(Q,U;Z)+\frac{\delta(\epsilon)}{N},
\end{eqnarray}
where (1) is from Fano's inequality, and (2) is from the above definitions of $U$, $Q$ and $Z$.

Also note that
\begin{eqnarray}\label{exxc1.2}
&&\frac{1}{N}H(W_{0})\stackrel{(3)}\leq \frac{1}{N}I(W_{0};Y_{1}^{N},Z^{N})+\frac{\delta(\epsilon)}{N}\nonumber\\
&&=\frac{1}{N}\sum_{i=1}^{N}I(W_{0};Y_{1,i},Z_{i}|Z^{i-1},Y_{1}^{i-1})+\frac{\delta(\epsilon)}{N}\nonumber\\
&&\leq \frac{1}{N}\sum_{i=1}^{N}(H(Y_{1,i},Z_{i}|Y_{1}^{i-1})-H(Y_{1,i},Z_{i}|Z_{i+1}^{N},W_{0},Y^{i-1},Z^{i-1},Y_{1}^{i-1}))+\frac{\delta(\epsilon)}{N}\nonumber\\
&&\stackrel{(4)}=\frac{1}{N}\sum_{i=1}^{N}(H(Y_{1,i},Z_{i}|Y_{1}^{i-1})-H(Y_{1,i},Z_{i}|Z_{i+1}^{N},W_{0},Y^{i-1},Y_{1}^{i-1}))+\frac{\delta(\epsilon)}{N}\nonumber\\
&&=\frac{1}{N}\sum_{i=1}^{N}(H(Y_{1,i},Z_{i}|Y_{1}^{i-1},J=i)-H(Y_{1,i},Z_{i}|Z_{i+1}^{N},W_{0},Y^{i-1},Y_{1}^{i-1},J=i))+\frac{\delta(\epsilon)}{N}\nonumber\\
&&\leq H(Y_{1,J},Z_{J}|Y_{1}^{J-1})-H(Y_{1,J},Z_{J}|W_{0},Y^{J-1},Z_{J+1}^{N},Y_{1}^{J-1},J)+\frac{\delta(\epsilon)}{N}\nonumber\\
&&\stackrel{(5)}=I(U;Y_{1},Z|Q)+\frac{\delta(\epsilon)}{N}\stackrel{(6)}=I(U;Y_{1}|Q)+\frac{\delta(\epsilon)}{N},
\end{eqnarray}
where (3) is from Fano's inequality, (4) is from the Markov chain $(Y_{1,i},Z_{i})\rightarrow (Z_{i+1}^{N},W_{0},Y^{i-1},Y_{1}^{i-1})\rightarrow Z^{i-1}$,
(5) is from the definitions of $U$, $Q$, $Y_{1}$ and $Z$, and (6) is from the Markov chain $U\rightarrow (Q, Y_{1})\rightarrow Z$.

Letting $N\rightarrow \infty$ and using $\lim_{N\rightarrow \infty}\frac{1}{N}H(W_{0})=R_{0}$, $R_{0}\leq \min\{I(U,Q;Z),I(U;Y_{1}|Q)\}$
is proved.

\textbf{Proof of $R_{0}+R_{1}\leq \min\{I(Q,U,V;Y),I(U,V;Y_{1}|Q)\}$:}

Similar to the proof of $R_{0}\leq \min\{I(U,Q;Z),I(U;Y_{1}|Q)\}$, first, note that
\begin{eqnarray}\label{exxc1.3}
&&\frac{1}{N}H(W_{0},W_{1})\stackrel{(1)}\leq \frac{1}{N}I(W_{0},W_{1};Y^{N})+\frac{\delta(\epsilon)}{N}\nonumber\\
&&=\frac{1}{N}\sum_{i=1}^{N}I(W_{0},W_{1};Y_{i}|Y^{i-1})+\frac{\delta(\epsilon)}{N}\nonumber\\
&&\leq \frac{1}{N}\sum_{i=1}^{N}(H(Y_{i})-H(Y_{i}|Z_{i+1}^{N},W_{0},W_{1},Y^{i-1},Y_{1}^{i-1}))+\frac{\delta(\epsilon)}{N}\nonumber\\
&&=\frac{1}{N}\sum_{i=1}^{N}(H(Y_{i}|J=i)-H(Y_{i}|Z_{i+1}^{N},W_{0},W_{1},Y^{i-1},Y_{1}^{i-1},J=i))+\frac{\delta(\epsilon)}{N}\nonumber\\
&&=H(Y_{J}|J)-H(Y_{J}|Z_{J+1}^{N},W_{0},W_{1},Y^{J-1},Y_{1}^{J-1},J)+\frac{\delta(\epsilon)}{N}\nonumber\\
&&\stackrel{(2)}\leq I(Q,U,V;Y)+\frac{\delta(\epsilon)}{N},
\end{eqnarray}
where (1) is from Fano's inequality, and (2) is from the above definitions of $U$, $V$, $Q$ and $Y$.

Also note that
\begin{eqnarray}\label{exxc1.4}
&&\frac{1}{N}H(W_{0},W_{1})\stackrel{(3)}\leq \frac{1}{N}I(W_{0},W_{1};Y_{1}^{N},Y^{N})+\frac{\delta(\epsilon)}{N}\nonumber\\
&&=\frac{1}{N}\sum_{i=1}^{N}I(W_{0},W_{1};Y_{1,i},Y_{i}|Y^{i-1},Y_{1}^{i-1})+\frac{\delta(\epsilon)}{N}\nonumber\\
&&\leq \frac{1}{N}\sum_{i=1}^{N}(H(Y_{1,i},Y_{i}|Y_{1}^{i-1})-H(Y_{1,i},Y_{i}|Z_{i+1}^{N},W_{0},W_{1},Y^{i-1},Y_{1}^{i-1}))+\frac{\delta(\epsilon)}{N}\nonumber\\
&&=\frac{1}{N}\sum_{i=1}^{N}(H(Y_{1,i},Y_{i}|Y_{1}^{i-1},J=i)-H(Y_{1,i},Y_{i}|Z_{i+1}^{N},W_{0},W_{1},Y^{i-1},Y_{1}^{i-1},J=i))+\frac{\delta(\epsilon)}{N}\nonumber\\
&&\leq H(Y_{1,J},Y_{J}|Y_{1}^{J-1})-H(Y_{1,J},Y_{J}|W_{0},W_{1},Y^{J-1},Z_{J+1}^{N},Y_{1}^{J-1},J)+\frac{\delta(\epsilon)}{N}\nonumber\\
&&\stackrel{(4)}=I(U,V;Y_{1},Y|Q)+\frac{\delta(\epsilon)}{N}\stackrel{(5)}=I(U,V;Y_{1}|Q)+\frac{\delta(\epsilon)}{N},
\end{eqnarray}
where (3) is from Fano's inequality,
(4) is from the definitions of $U$, $V$, $Q$, $Y_{1}$ and $Y$, and (5) is from the Markov chain $(U,V)\rightarrow (Q, Y_{1})\rightarrow Y$.

Letting $N\rightarrow \infty$ and using $\lim_{N\rightarrow \infty}\frac{1}{N}H(W_{0},W_{1})=R_{0}+R_{1}$, $R_{0}+R_{1}\leq \min\{I(Q,U,V;Y),I(U,V;Y_{1}|Q)\}$
is proved.

\textbf{Proof of $R_{e}\leq I(V;Y|U)-I(V;Z|U)$:}

Note that the definitions of $U$ and $V$ are the same as those of \cite{CK}, and therefore, the proof of $R_{e}\leq I(V;Y|U)-I(V;Z|U)$
is the same as that of \cite{CK}. Thus, we omit the proof here.

\textbf{Proof of $R_{e}\leq I(V;Y_{1}|U,Q)-I(V;Z|U,Q)$:}
First, note that
\begin{eqnarray}\label{exxc1}
&&\frac{1}{N}H(W_{1}|Z^{N})=\frac{1}{N}(H(W_{1}|Z^{N},W_{0})+I(W_{1};W_{0}|Z^{N}))\nonumber\\
&&\stackrel{(a)}\leq \frac{1}{N}H(W_{1}|Z^{N},W_{0})+\frac{\delta(\epsilon)}{N}\nonumber\\
&&=\frac{1}{N}(I(W_{1};Y^{N},Y_{1}^{N}|Z^{N},W_{0})+H(W_{1}|Z^{N},W_{0},Y^{N},Y_{1}^{N}))+\frac{\delta(\epsilon)}{N}\nonumber\\
&&\stackrel{(b)}\leq \frac{1}{N}I(W_{1};Y^{N},Y_{1}^{N}|Z^{N},W_{0})+\frac{2\delta(\epsilon)}{N}\nonumber\\
&&=\frac{1}{N}\sum_{i=1}^{N}(I(W_{1};Y_{i},Y_{1,i}|Y^{i-1},Y_{1}^{i-1},Z^{N},W_{0}))+\frac{2\delta(\epsilon)}{N}\nonumber\\
&&\stackrel{(c)}\leq \frac{1}{N}\sum_{i=1}^{N}(H(Y_{i},Y_{1,i}|Z_{i},Z_{i+1}^{N},W_{0},Y^{i-1},Y_{1}^{i-1})
-H(Y_{i},Y_{1,i}|Z_{i},Z_{i+1}^{N},W_{0},Y^{i-1},Y_{1}^{i-1},W_{1}))+\frac{2\delta(\epsilon)}{N}\nonumber\\
&&=\frac{1}{N}\sum_{i=1}^{N}(H(Y_{i},Y_{1,i},Z_{i}|Z_{i+1}^{N},W_{0},Y^{i-1},Y_{1}^{i-1})-H(Z_{i}|Z_{i+1}^{N},W_{0},Y^{i-1},Y_{1}^{i-1})\nonumber\\
&&-H(Y_{i},Y_{1,i},Z_{i}|Z_{i+1}^{N},W_{0},Y^{i-1},Y_{1}^{i-1},W_{1})+H(Z_{i}|Z_{i+1}^{N},W_{0},Y^{i-1},Y_{1}^{i-1},W_{1}))+\frac{2\delta(\epsilon)}{N}\nonumber\\
&&\stackrel{(d)}=\frac{1}{N}\sum_{i=1}^{N}(I(V_{i};Y_{i},Y_{1,i},Z_{i}|U_{i},Q_{i})-I(V_{i};Z_{i}|U_{i},Q_{i}))+\frac{2\delta(\epsilon)}{N}\nonumber\\
&&\stackrel{(e)}=\frac{1}{N}\sum_{i=1}^{N}(I(V_{i};Y_{1,i}|U_{i},Q_{i})-I(V_{i};Z_{i}|U_{i},Q_{i}))+\frac{2\delta(\epsilon)}{N}\nonumber\\
&&\stackrel{(f)}\leq I(V;Y_{1}|U,Q)-I(V;Z|U,Q)+\frac{2\delta(\epsilon)}{N},
\end{eqnarray}
where (a) and (b) are from Fano's inequality, (c) is from the Markov chain
$(Y_{i},Y_{1,i})\rightarrow (Z_{i},Z_{i+1}^{N},W_{0},Y^{i-1},Y_{1}^{i-1},W_{1})\rightarrow Z^{i-1}$,
(d) is from the definitions $Q_{i}=Y_{1}^{i-1}$, $U_{i}=(Y^{i-1}, W_{0}, Z_{i+1}^{N})$, $V_{i}=(U_{i}, W_{1})$,
(e) is from the fact that given $Q_{i}$, $U_{i}$, $Y_{i}$, $Z_{i}$ and $Y_{1,i}$, $V_{i}$ is independent of $X_{1,i}$, and
the Markov chains $V_{i}\rightarrow (X_{1,i}, U_{i},Q_{i}, Y_{1,i})\rightarrow Y_{i}\rightarrow Z_{i}$,
$V_{i}\rightarrow (U_{i},Q_{i})\rightarrow X_{1,i}$ and $V_{i}\rightarrow (U_{i},Q_{i},Y_{1,i})\rightarrow X_{1,i}$, and
(f) is from the above definitions of $U$, $Q$, $V$, $Y_{1}$ and $Z$, and note that $J$ is a time sharing random variable.
Letting $N\rightarrow \infty$ and using $\lim_{N\rightarrow \infty}\frac{1}{N}H(W_{1}|Z^{N})\geq R_{e}$, $R_{e}\leq I(V;Y_{1}|U,Q)-I(V;Z|U,Q)$
is proved.

\end{document}